\documentclass[12pt,epsf]{article}
\usepackage{verbatim}
\usepackage{anyfontsize}
\usepackage{dsfont}
\usepackage{tikz}
\usepackage{sidecap}
\sidecaptionvpos{figure}{t}
\usepackage{subfigure}

\usepackage[english]{babel}
\usepackage{enumitem}  

\usepackage{amssymb,amsmath}
\usepackage{graphicx, xcolor, varwidth}
\usepackage{setspace}
\usepackage[permil]{overpic}
\usepackage{cite}
\usepackage{mathtools}

\DeclareSymbolFont{matha}{OML}{txmi}{m}{it}
\DeclareMathSymbol{v}{\mathord}{matha}{118}

\colorlet{darkblue}{blue!70!black}
\colorlet{darkgreen}{green!70!black}

\usepackage[colorlinks=true,urlcolor=darkblue,linktocpage=true,linkcolor=darkblue,citecolor=darkblue]{hyperref}

\numberwithin{equation}{section}
\DeclareMathSymbol{v}{\mathord}{matha}{118}
\newcommand{\be}{\begin{equation}}
\newcommand{\ee}{\end{equation}}
\newcommand{\bea}{\begin{eqnarray}}
\newcommand{\eea}{\end{eqnarray}}
\newcommand{\bear}{\begin{eqnarray}}
\newcommand{\eear}{\end{eqnarray}}
\newcommand{\beas}{\begin{eqnarray*}}
\newcommand{\p}{\partial}
\newcommand{\eeas}{\end{eqnarray*}}
\newcommand{\ba}{\begin{array}}
\newcommand{\ea}{\end{array}}

\def\ba#1\ea{\begin{align}#1\end{align}}
\def\bs#1\es{\begin{split}#1\end{split}}

\newcommand{\blockS}[5]{
\begin{tikzpicture}[baseline=-3pt]
\coordinate (v1) at (-1,-0.5) {} {};
\coordinate (v2) at (-0.5,0) {} {};
\coordinate (v3) at (-1,0.5) {} {};
\coordinate (v4) at (0.5,0) {} {};
\coordinate (v5) at (1,0.5) {} {};
\coordinate (v6) at (1,-0.5) {} {};
\begin{scope}[very thick]
\draw  (v1) node[left] {$#2$}--(v2);
\draw  (v3) node[left] {$#1$}-- (v2);
\draw  (v4) -- (v2) node[midway, above] {$#5$};
\draw  (v5) node[right] {$#3$}-- (v4);
\draw  (v6) node[right] {$#4$}-- (v4);
\end{scope}
\end{tikzpicture}
}

\renewcommand{\a}{\alpha}
\renewcommand{\b}{\beta}

\newcommand{\ep}{\epsilon}

\newcommand{\pd}[2][1]{\ifnum#1=1 \frac{\partial}{\partial {#2}} \else
  \frac{\partial^#1}{\partial {#2}^{#1}}\fi}
\newcommand{\dpd}[2][1]{\ifnum#1=1 \dfrac{\partial}{\partial {#2}} \else
  \frac{\partial^#1}{\partial {#2}^{#1}}\fi}
\newcommand{\td}[2][1]{\ifnum#1=1 \frac{d}{d{#2}} \else
  \frac{d^#1}{d{#2}^{#1}}\fi}

\renewcommand{\t}{\tilde{t}}
\renewcommand{\a}{\tilde{a}}
\renewcommand{\b}{\tilde{b}}
\renewcommand{\c}{\tilde{c}}
\renewcommand{\ell}{J}

\newcommand{\Deltag}{\Delta_{\text{gap}}}

\newcommand{\e}{\varepsilon}

\newcommand{\x}{\xi}

\newcommand{\nbox}{{\,\lower0.9pt\vbox{\hrule \hbox{\vrule height 0.2 cm \hskip 0.19 cm \vrule height 0.2 cm}\hrule}\,}}

\renewcommand{\v}[1]{\vec{#1}}

\newcommand{\E}{{\mathcal E}}
\def\O{{\cal O}}

\newcommand{\bz}{\bar{z}}



\begin{document}
\begin{spacing}{1.3}
\begin{titlepage}

\begin{center}
{\Large \bf 
A Bound on Massive Higher Spin Particles
}

\vspace*{6mm}

Nima Afkhami-Jeddi,$^m$ Sandipan Kundu,$^J$ and Amirhossein Tajdini$^m$

\vspace*{6mm}

\textit{$^m$Department of Physics, Cornell University, Ithaca, New York, USA\\}

\vspace{3mm}

\textit{$^J$Department of Physics and Astronomy, Johns Hopkins University,
Baltimore, Maryland, USA\\}

\vspace{6mm}

{\tt \small na382@cornell.edu, kundu@jhu.edu, at734@cornell.edu}

\vspace*{6mm}
\end{center}

\begin{abstract}
According to common lore, massive elementary higher spin particles lead to inconsistencies when coupled to gravity. However, this scenario was not completely ruled out by previous arguments. In this paper, we show that in a theory  where the low energy dynamics of the gravitons are governed by the Einstein-Hilbert action, any finite number of massive elementary particles with spin more than two cannot interact with gravitons, even classically, in a way that preserves causality. This is achieved in flat spacetime by studying eikonal scattering of higher spin particles in more than three spacetime dimensions. Our argument is insensitive to the physics above the effective cut-off scale and closes certain loopholes in previous arguments.  Furthermore, it applies to higher spin particles even if they do not contribute to tree-level graviton scattering as a consequence of being charged under a global symmetry such as $\mathds{Z}_2$. We derive analogous bounds in anti-de Sitter spacetime from analyticity properties of correlators of the dual CFT in the Regge limit. We also argue that an infinite tower of fine-tuned higher spin particles can still be consistent with causality. However, they necessarily affect the dynamics of gravitons at an energy scale comparable to the mass of the lightest higher spin particle. Finally, we apply the bound in de Sitter to impose restrictions on the structure of three-point functions in the squeezed limit of the scalar curvature perturbation produced during inflation.

\end{abstract}

\end{titlepage}
\end{spacing}

\vskip 1cm
\setcounter{tocdepth}{2}  
\tableofcontents

\begin{spacing}{1.3}

\section{Introduction}
Weinberg in one of his seminal papers \cite{Weinberg:1964ew} showed that general properties of the S-matrix allow for the presence of the graviton. Not only that, the soft-theorem dictates that at low energies gravitons must interact universally with all particles -- which is the manifestation of the equivalence principle in QFT. This remarkable fact has many far-reaching consequences for theories with higher spin particles. 

Even in the early days of quantum field theory (QFT), it was known that there are restrictions on particles with spin $J>2$ in flat spacetime. For example,  Lorentz invariance of the S-matrix  requires that massless particles interacting with gravity in flat spacetime cannot have spin more than two \cite{Weinberg:1964ew,Weinberg:1980kq,Porrati:2008rm}. Moreover, 
folklore has it that any finite number of massive elementary higher spin particles, however fine-tuned, cannot interact with gravity in a consistent way.  There is ample evidence suggestive of a strict bound  on massive higher spin particles  at least in flat spacetime in dimensions $D\ge 4$ from tree-level unitarity and asymptotic causality \cite{Arkani-Hamed:2017jhn,old-weinberg,Ferrara:1992yc, Porrati:1993in,Cucchieri:1994tx,Camanho:2014apa},\footnote{See comments in section \ref{sec:fornima} for comparison between arguments in the literature and the argument presented in this paper.  } however, to our knowledge there is no concrete argument which completely rules out a finite number of massive particles with spin $J>2$.

Most notably, it was argued in \cite{Camanho:2014apa} that in a theory with finite number of massive particles with spin $J>2$, unless each higher spin particle is charged under a global symmetry such as $\mathds{Z}_2$, they will contribute to eikonal scattering of particles, even with low spin ($J\le 2$), in a way that violates asymptotic causality in flat spacetime. The same statement is true even in anti-de Sitter (AdS) spacetime where the global symmetries of higher spin particles are required by the chaos growth bound of  the dual CFT \cite{Maldacena:2015waa}. In addition, there is no known string compactification which leads to particles with spin $\ell>2$ and masses $M\ll M_s$ in flat spacetime, where $M_s$ is the string scale. Of course, it is well known that higher spin particles do exist in AdS, but they always come in an infinite tower and  these theories become strongly interacting at low energies \cite{Vasiliev:2004qz, Bekaert:2005vh}. All of these observations indicate that there are universal bounds on theories with higher spin massive particles. In this paper, we will prove such a bound from causality. We will show that any finite number of massive elementary  particles with spin $\ell>2$, however fine tuned, cannot interact with gravitons in flat or AdS spacetimes (in $D\ge 4$ dimensions) in a way that is consistent with the QFT equivalence principle and preserves causality. In particular, we will demonstrate that the three-point interaction $J$-$J$-graviton must vanish  for $J>2$. However, this is one interaction that no particle can avoid due to the equivalence principle, implying that elementary particles with spin $J>2$ cannot exist.

For massless higher spin particles, the inconsistencies are even more apparent. The tension between Lorentz invariance of the S-matrix and the existence of massless particles with spin $\ell>2$ was already visible in \cite{Weinberg:1964ew}. Subsequently, the same tension was shown to exist for massless fermions with spin $\ell>3/2$ \cite{Grisaru:1977kk,Grisaru:1976vm}. A concrete manifestation of this tension is an elegant theorem due to Weinberg and Witten which states that any massless particle with spin $\ell>1$ cannot possess a Lorentz covariant and gauge invariant energy-momentum tensor \cite{Weinberg:1980kq}.\footnote{See \cite{Loebbert:2008zz} for a nice review.} Of course, this theorem does not prohibit the existence of gravitons, rather it implies that the graviton must be fundamental. More recently, a generalization of the Weinberg-Witten theorem has been  presented by Porrati which states that massless particles with spin $\ell>2$ cannot be minimally coupled to the graviton in flat spacetime \cite{Porrati:2008rm}. Both of these theorems are completely consistent with various other observations made about interactions of massless higher spin particles in flat spacetime (see \cite{Aragone:1979hx,Berends:1979wu, Aragone:1981yn, Metsaev:2005ar,Boulanger:2006gr,Boulanger:2008tg} and references therein). Furthermore, the generalized Weinberg-Witten theorem and the QFT equivalence principle are sufficient to completely rule out massless particles with spin $\ell>2$ in flat spacetime \cite{Weinberg:1980kq,Porrati:2008rm}. The basic argument is rather simple. The Weinberg-Witten theorem and its generalization by Porrati only allow non-minimal coupling between massless particles with spin $\ell>2$ and the graviton. Whereas, it is well known that particles with low spin can couple minimally with the graviton. Therefore, the QFT equivalence principle requires that massless higher spin particles, if they exist, must couple minimally with the graviton at low energies -- which directly contradicts the Weinberg-Witten/Porrati theorem.

Any well behaved Lorentzian QFT must also be unitary and  causal. Lorentz invariance alone was sufficient to rule out massless higher spin particles in flat spacetime. Whereas, massive elementary particles with spin $\ell>2$ do not lead to any apparent contradiction with Lorentz invariance in flat spacetime. However, any such particle if present, must interact with gravitons. The argument presented in \cite{Camanho:2014apa} implies that finite number of higher spin particles cannot be exchanged in any tree-level scattering. However, this restriction is not sufficient to rule out massive higher spin particles, rather it implies that each massive higher spin particle must be charged under $\mathds{Z}_2$ or some other global symmetry. On the other hand, the equivalence principle requires the coupling between a single graviton and two spin-$\ell$ particles to be non-vanishing. By considering an eikonal scattering experiment between scalars and elementary higher spin particles with spin $\ell$ and mass $m$ in the regime $|s|\gg |t|\gg m$, where $s$ and $t$ are the Mandelstam variables, we will show that any such coupling between the higher spin particle and the graviton in flat spacetime leads to violation of asymptotic causality. This is accomplished by extending the argument of \cite{Camanho:2014apa} to the scattering of higher spin particles which requires the phase shift to be non-negative for all choices of polarization of external particles.

A similar high energy scattering experiment can be designed in AdS to rule out elementary massive higher spin particles. However, we will take a holographic route which has several advantages. We consider a class of large-$N$ CFTs in $d\ge 3$ dimensions with a sparse spectrum. The sparse spectrum condition, to be more precise, implies that the lightest single trace primary operator with spin $\ell >2$ has dimension $\Deltag\gg 1$. It was first conjectured in \cite{Heemskerk:2009pn} that this class of CFTs admit a universal holographic dual description with a low energy description in terms of Einstein gravity coupled to matter fields. The conjecture was based on the observation that there is a one-to-one correspondence between scalar effective field theories in AdS and perturbative solutions of CFT crossing equations in the $1/N$ expansion. The scalar version of this conjecture was further substantiated in \cite{Cornalba:2006xk,Cornalba:2006xm,Cornalba:2007zb,Mack:2009gy,Mack:2009mi,Fitzpatrick:2010zm,Heemskerk:2010ty,Fitzpatrick:2011hu,Fitzpatrick:2011ia,
ElShowk:2011ag,Komargodski:2012ek,Fitzpatrick:2012yx,Fitzpatrick:2012cg,Goncalves:2014rfa,Hijano:2015zsa,Alday:2016htq,Alday:2014tsa,Caron-Huot:2017vep} by using the conformal bootstrap. More recently, the conjecture has been completely proven at the linearized level even for spinning operators including the stress tensor \cite{Afkhami-Jeddi:2016ntf,Kulaxizi:2017ixa,Costa:2017twz,Afkhami-Jeddi:2017rmx,Meltzer:2017rtf,Afkhami-Jeddi:2018own}. In the second half of the paper, we will exploit this connection to constrain massive higher spin particles in AdS by studying large-$N$ CFTs with a sparse spectrum. To this end, we introduced a new non-local operator, capturing the contributions to the Regge limit of the OPE of local operators. This operator is expressed as an integral of a local operator over a ball times a null-ray. It is obtained by generalizing the Regge OPE introduced in \cite{Afkhami-Jeddi:2018own} to non-integer spins, resulting in an operator that is more naturally suited for parametrizing the contribution of Regge trajectories which require analytic continuation in both spin and scaling dimension.

In the holographic CFT side we will ask the dual question: is it possible to add an extra higher spin single trace primary operator with $\ell > 2$ and scaling dimension $\Delta\ll \Delta_{\tiny{\text{gap}}}$ and still get a consistent CFT? A version of this question has already been answered by a theorem in CFT that rules out any finite number of higher spin conserved currents \cite{Maldacena:2011jn,Boulanger:2013zza,Hartman:2015lfa}-- which is the analog of the Weinberg-Witten theorem in AdS. However, ruling out massive higher spin particles in AdS requires a generalization of this theorem for non-conserved single trace primary operators of holographic CFTs.  The chaos (growth) bound of  Maldacena, Shenker, and Stanford \cite{Maldacena:2015waa} partially achieves this by not allowing any finite number of higher spin single trace primary operators to contribute as  exchange operators in CFT four-point functions in the Regge limit. However, this restriction does not rule out the existence of such operators rather it prohibits these higher spin operators to appear in the operator product expansion (OPE) of certain operators. On the other hand, causality (chaos sign bound) imposes stronger constraints on non-conserved single trace primary operators. In particular, by using the holographic null energy condition (HNEC) \cite{Afkhami-Jeddi:2017rmx,Afkhami-Jeddi:2018own} applied to correlators with external higher spin operators, we will show that massive higher spin fields in AdS (in $D\ge 4$ dimensions) lead to causality violation in the dual CFT. This implies that any finite number of massive elementary particles with spin $\ell>2$ in AdS cannot be embedded in a well behaved UV theory in which the dynamics of gravitons at low energies is described by the Einstein-Hilbert action.

\begin{figure}[h!]
\centering
\includegraphics[scale=0.35]{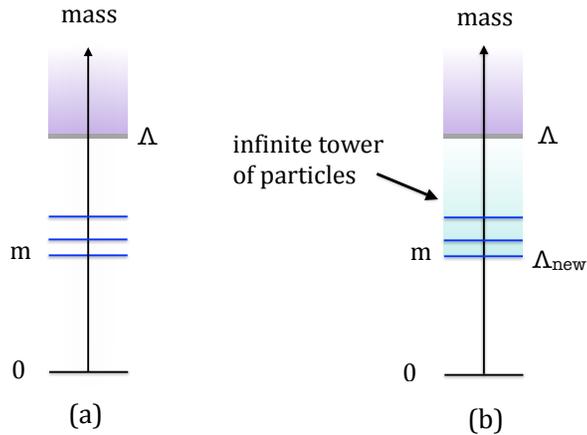}
\caption{\small Spectrum of elementary particles with spin $J>2$ in a theory where the dynamics of gravitons is described by the Einstein-Hilbert action at energy scales $E\ll \Lambda$. The cut-off scale $\Lambda$ can be the string scale and hence there can be an infinite tower of higher spin particles above $\Lambda$. Figure (a) represents a scenario that also contains a finite number of higher spin particles below the cut-off and hence violates causality. Causality can only be restored if these particles are accompanied by an infinite tower of higher spin particles with comparable masses which is shown in figure (b). This necessarily brings down the cut-off scale to $\Lambda_{\text{new}}=m$, where $m$ is the mass of the lightest higher spin particle.}\label{fig:hs}
\end{figure}

One advantage of the holographic approach is that it also provides a possible solution to the causality problem. From the dual CFT side, we will argue  that in a theory where the dynamics of gravitons is described by the Einstein-Hilbert action at energy scales $E\ll \Lambda$ ($\Lambda$ can be the string scale $M_s$), a single elementary particle with spin $\ell>2$ and mass $m\ll \Lambda$ violates causality unless the particle is accompanied by an infinite tower of (finely tuned) higher spin elementary particles with mass $\sim m$. Furthermore, causality also requires that these new higher spin particles (or at least an infinite subset of them) must be able to decay into two gravitons and hence modify the dynamics of gravitons at energy scales $E\sim m$. So, one can have a causal theory without altering the low energy behavior of gravity only if all the higher spin particles are heavier than the cut-off scale $\Lambda$.

Causality of CFT four-point functions in the lightcone limit also places nontrivial constraints on higher spin primary operators. In particular, it generalizes the Maldacena-Zhiboedov theorem of $d = 3$  \cite{Maldacena:2011jn} to higher dimensions by ruling out a finite number of  higher spin conserved currents  \cite{Hartman:2015lfa}. The advantage of the lightcone limit is that the constraints are valid for all CFTs -- both holographic and non-holographic. However, the argument
of \cite{Hartman:2015lfa} is not  applicable when higher spin conserved currents do not contribute to generic CFT four-point functions as exchange operators. We will present an argument in the lightcone limit that closes this loophole by ruling out higher spin conserved currents even when none of the operators are charged under it.\footnote{We should note that we have not ruled out an unlikely scenario in which the OPE coefficients conspire in a non-trivial way to cancel the causality violating contributions. Three-point functions of conserved currents are heavily constrained by conformal invariance and hence this scenario is rather improbable.} For holographic CFTs, this completely rules out a finite number of massless higher spin particles in AdS in $D\ge 4$ dimensions.

The bound on higher spin particles has a natural application in inflation. If higher spin particles are present during inflation, they produce distinct signatures on the late time three-point function of the scalar curvature perturbation in the squeezed limit \cite{Arkani-Hamed:2015bza}. The bounds on higher spin particles in flat space and in AdS were obtained by studying local high energy scattering which is insensitive to the spacetime curvature. This strongly suggests that the same bound should hold even in de Sitter space.\footnote{This argument parallels the argument made by Cordova, Maldacena, and Turiaci in \cite{Cordova:2017zej}. The same point of view was also adopted in our previous paper \cite{Afkhami-Jeddi:2018own}. } Our bound, when applied in de Sitter, immediately implies that contributions of higher spins to the three-point function of the scalar curvature perturbation in the squeezed limit must be Boltzmann suppressed $\sim e^{-2\pi \Lambda/H}$, where $H$ is the Hubble scale. Therefore, if  the higher spin contributions are detected in future experiments, then the scale of new physics must be $\Lambda\sim H$. This necessarily requires the presence of not one but an infinite tower of higher spin particles with spins $\ell >2$ and masses comparable to the Hubble scale. Any such detection can be interpreted as evidence in favor of string theory with the string scale comparable to the Hubble scale.

The rest of the paper is organized as follows. In section \ref{sec:eikonal}, we present an S-matrix based argument to show that massive elementary particles with spin $J>2$ cannot interact with gravitons in a way that preserves asymptotic causality. We derive the same bounds in AdS from analyticity properties of correlators of the dual CFT in section \ref{sec:cft}. In section \ref{sec:causal}, we argue that the only way one can restore causality is by adding an infinite tower of massive higher spin particles. In addition, we also discuss why stringy states in classical string theory are consistent with causality. Finally, in section \ref{sec:cosmo}, we apply our bound in de Sitter to constrain the squeezed limit three-point functions of scalar curvature perturbations produced during inflation.

\section{Higher Spin Fields in Flat Spacetime}\label{sec:eikonal}
\begin{figure}[h!]
\centering
\includegraphics[scale=0.7]{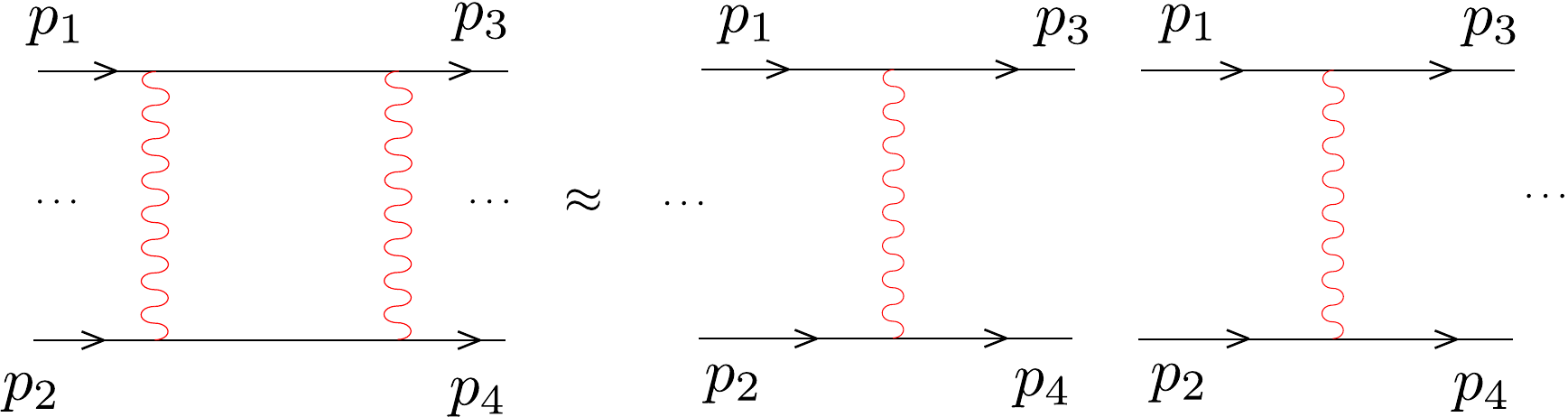}
\caption{Tree-level exchange diagrams are the building blocks of ladder diagrams.}\label{fig:eikonal}
\end{figure}
In this section, we explicitly show that interactions of higher spin particles with gravity lead to causality violation. Eikonal scattering has been used in the literature \cite{ Camanho:2014apa,Hinterbichler:2017qcl,Bonifacio:2017nnt,Levy:1969cr,Camanho:2016opx,Edelstein:2016nml} to impose constraints on interactions of particles with spin. When the center of mass energy is large and transfer momentum is small, the scattering amplitude is captured by the eikonal approximation. Focusing on a specific exchange particle for now, the scattering amplitude is given by a sum of ladder diagrams. These diagrams can be resumed (see figure \ref{fig:eikonal}) and as a result introduce a phase shift in the scattering amplitude \cite{tHooft:1987vrq}.\footnote{We will comment more about the resummation later in the section.} This phase shift produces a Shapiro time delay \cite{Shapiro:1964uw} that particles experience \cite{ Camanho:2014apa}. Asymptotic causality in flat spacetime requires the time delay and hence the phase shift to be non-negative \cite{ Camanho:2014apa ,Gao:2000ga}. Moreover, positivity of the phase shift imposes restrictions on the tree-level exchange diagrams --which are the building blocks of ladder diagrams-- constraining three-point couplings between particles. This method has been utilized to constrain three-point interactions between gravitons, massive spin-2 particles, and massless higher spin particles\cite{Camanho:2014apa,Bonifacio:2017nnt,Hinterbichler:2017qcl}. Here we apply the eikonal scattering method to external massive and massless elementary particles with spin $J>2$.

We will briefly review eikonal scattering in order to explicitly relate the phase shift to the three-point interactions between elementary particles. We will take two of the external particles to be massive or massless higher spin particles ($J>2$) and the other two particles to be scalars. The setup is shown in figure \ref{fig_eik} where particles 1 and 3 are the higher spin particles, whereas particles 2 and 4 are scalars. We will then use on-shell methods  to write down the general three-point interaction between higher spin elementary particles and gravitons \cite{Costa:2011mg}. This allows us to derive the most general form of the amplitude in the eikonal limit. Positivity of the phase shift for all choices of polarization tensors of external particles, constrains the coefficients of three-point vertices. In particular, for both massive and massless particles with spin $J>2$ in space-time dimensions $D\ge 4$, we find that the three-point interaction $J$-$J$-graviton must be zero. However, this is one interaction that no particle can avoid due to the equivalence principle, implying that elementary particles with spin $J>2$ cannot exist.

\subsection{Eikonal Scattering }

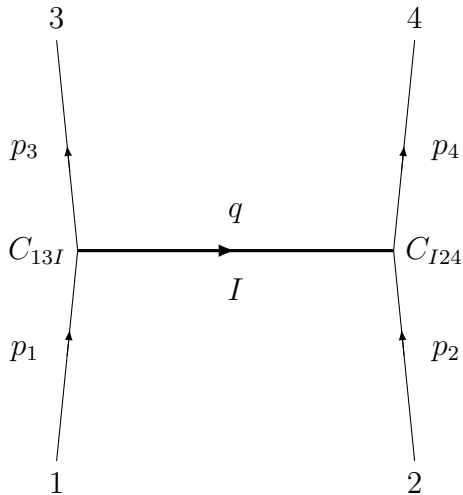
\begin{figure}
\begin{center}
\usetikzlibrary{decorations.markings}    
\usetikzlibrary{decorations.markings}    
\begin{tikzpicture}[baseline=-3pt,scale=0.7]
\begin{scope}[very thick,shift={(4,0)}]
\draw[thin,-latex]  (-3,0) -- (-3.2,2.0);
\draw[thin]  (-3.15,1.5) -- (-3.4,4.0);
\draw[thin]  (-3.2,-2.0) -- (-3,0);
\draw[thin, -latex]  (-3.4,-4)--(-3.15,-1.5) ;
\draw[thin, -latex]  (3,0) -- (3.2,2.0);
\draw[thin]  (3.15,1.5) -- (3.4,4.0);
\draw[thin]  (3.2,-2.0) -- (3,0);
\draw[thin, -latex]  (3.4,-4)--(3.15,-1.5) ;
\draw[very thick, -latex]  (-3,0)--(0,0) ;
\draw[very thick ]  (-0.1,0)--(3,0) ;
\draw(-4,1.5)node[above]{$p_3$};
\draw(-4,-1.5)node[below]{$p_1$};
\draw(4,1.5)node[above]{$p_4$};
\draw(4,-1.5)node[below]{$p_2$};
\draw(0,0.3)node[above]{ $q$};
\draw(0,-0.3)node[below]{ $I$};
\draw(-3.4,-4)node[below]{ $1$};
\draw(-3.4,4)node[above]{ $3$};
\draw(3.4,-4)node[below]{ $2$};
\draw(3.4,4)node[above]{ $4$};
\draw(-3,0)node[left]{ $C_{13I}$};
\draw(3,0)node[right]{ $C_{I24}$};

\end{scope}
\end{tikzpicture}
\end{center}
\caption{\label{fig_eik} \small Eikonal scattering of particles. In this highly boosted kinematics, particles are moving almost in the null directions such that the center of mass energy is large.}
\end{figure}

Let us consider $2 \to 2$ scattering of particles in space-time dimensions $D\ge 4$ as shown in figure \ref{fig_eik}. Coordinates are written in $\mathbb{R}^{1, D-1}$ with the metric
\ba
ds^2  = -du dv + d\vec{x}_{\perp}^2.
\ea
 Denoting the momentum of particles by $p_i, \;$ with $i$ labeling particles $1$ through $4$, the Mandelstam variables are given by 
\ba
s= - (p_1+p_2)^2,  \qquad t= -(p_1 -p_3)^2 = -q^2,
\ea 
where $q$ is the momentum of the particle exchanged which in the eikonal limit has the property $q^2=\vec{q}^2$, where $\vec{q}\ $ has components in directions transverse to the propagation of the external particles.\footnote{See section \ref{sec:kin} for the details of the kinematics.} The tree level amplitude consists of the products of three-point functions\footnote{For a detail discussion about the i$\epsilon$ see \cite{ Camanho:2014apa}. }
\ba\label{eq:factor}
M_{\text{tree}}( s,  \vec{q} ) =\sum_I \frac{C_{13 I}(\vec{q})  C_{I 24}( \vec{q})}{{\vec{q}\;}^2 +m_I^2}\ ,
\ea
where the sum is over all of the states of the exchanged particles with mass $m_I$. In the above expression, $C_{13I}$ and $C_{24I}$ are on-shell three-point amplitudes which are generally functions of the transferred momentum $\vec{q}$, as well as the polarization tensors and the center of mass variables.

In highly boosted kinematics, particles are moving almost in the null directions $u$ and $v$  with momenta $P^u$ and  $P^v$ respectively. The center of mass energy $s$ is large with respect to other dimensionful quantities such as the particle masses. In particular, we have $s \gg |t|=\vec{q}^{~2} $. The total scattering amplitude is given by the sum of all ladder diagrams in t-channel which exponentiates  when it is expressed in terms of the impact parameter  $\vec{b}$ which has components only along the transverse plane,
\ba
i M_{\text{eik}} (s ,  - {\vec{q}\;}^2)  = 2 s \int d^{D-2}\vec{b} e^{- i \vec{q}\cdot \vec{b}} \left( e^{ i \delta (s,\vec{b})} -1\right)\ , 
\ea
where,
\ba
 \delta (s,\vec{b}) = \frac{1}{2 s} \int \frac{d^{D-2}\vec{q}}{(2 \pi)^{D-2}} \; e^{i \vec{q} \cdot \vec{b}} M_{\text{tree}}(s, {\vec{q}\;})\ .
\ea
Before we proceed, let us comment more on the exponentiation since it plays a central role in the positivity argument. We can interpret the phase shift as the Shapiro time-delay only when it exponentiates in the eikonal limit. However, it is known that the eikonal exponentiation fails for the exchange of particles with spin $J<2$ \cite{Tiktopoulos:1971hi,Cheng:1987ga,Kabat:1992pz}. It is also not obvious that the tree level amplitude must exponentiate in the eikonal limit for the exchange of particles with spin $J\ge 2$. A physical argument was presented in \cite{ Camanho:2014apa} which suggests that for higher spin exchanges it is possible to get a final amplitude that is exponential of the tree level exchange diagram. First, let us think of particle 2 as the source of a shockwave and particle 1 to be a probe particle travelling in that background. At tree-level, the amplitude is given by $1+i \delta$, where we ensure that $\delta \ll 1$ by staying in a weakly coupled regime. Let us then send $N$ such shockwaves so that we can treat them as individual shocks and hence the final amplitude, in the limit $\delta\rightarrow 0, N\rightarrow\infty$ with $N\delta=$fixed, is approximately given by $(1+i \delta)^N\approx e^{iN \delta}$. This approximation is valid only if we can view $N$ scattering processes as independent events. Moreover, we want to be in the weakly coupled regime. Both of these conditions can only be satisfied if $\delta$ grows with $s$ -- which is true for the exchange of particles with spin $J\ge 2$ \cite{ Camanho:2014apa}. Therefore, for higher-spin exchanges, we can interpret $\delta$ (or rather $N$ times $\delta$) as the Shapiro time delay of particle 1.

There is one more caveat. The exponentiation also depends on the assumption that $\delta$ is the same for each of the $N$-processes -- in other words, the polarization of particle 3 is the complex conjugate of that of particle 1. In general, particle 3 can have any polarization, however, we can fix the polarization of particle 3 by replacing particle 1 by a coherent state of particles with a fixed polarization. Since we are in the weakly coupled regime, we can make the mean occupation number large without making $\delta$ large. This allows us to fix the polarization of particle 3 to be complex conjugate of that of particle 1 because of  Bose enhancement (see \cite{ Camanho:2014apa} for a detail discussion).

Let us end this discussion by noting that the N-shock interpretation of the eikonal process is also consistent with classical gravity calculations. For example, the Shapiro time delay as obtained in GR from shockwave geometries  is the same as the time delay obtained from  the sum of all ladder diagrams for graviton exchanges -- which indicates that these are the  only important diagrams in the eikonal limit. Thus, it is reasonable to expect that the  exponentiation of the tree-level diagram correctly captures the eikonal process. 

\subsubsection*{Positivity:}
When $\delta (s,\vec{b})$ grows with $s$, we can trust the eikonal exponentiation which allows us to relate the phase shift to  time delay. In particular, for a particle moving in $u$ direction with momentum $P^u>0$, the phase shift $\delta (s,\vec{b})$ is related to the time delay of the particle by
\ba
\delta \left(s, \vec{b}\right)=P^u \Delta v\ .
\ea
Asymptotic causality in flat space requires that particles do not experience a time advance even when they are interacting \cite{Gao:2000ga}. Therefore, $\Delta v \ge 0$, implying that the phase shift must be non-negative as well. 

So far our discussion is very general and it is applicable even when multiple exchanges contribute to the tree level scattering amplitude. From now on, let us restrict to the special case of massless exchanges.\footnote{For non-zero $m_I$, the $\vec{q}$ integral yields $(2\pi)^{\frac{2-D}{2}}(\frac{m_I}{b})^{\frac{D-4}{2}}K_{\frac{D-4}{2}}(m_I b)$, where $K$ is the Bessel-$K$ function.} Using the tree-level amplitude (\ref{eq:factor}), we can write 
\ba\label{positive}
\delta (s,\vec{b})& = \frac{1}{2 s} \sum_I\int \frac{d^{D-2}\vec{q}}{(2 \pi)^{D-2}} \; e^{i \vec{q} \cdot \vec{b}}  \frac{C_{13 I}(\vec{q})  C_{I 24 }(\vec{q})}{q^2} \nonumber\\
&= \frac{\Gamma(\frac{D-4}{2})}{4 \pi^{\frac{D-2}{2}}} \sum_I\frac{C_{I24}(- i \vec{\partial_b}) C_{13 I}(- i \vec{\partial_b}) }{2 s}  \frac{1}{ |\vec{b}|^{D-4}}
\ea
which must be non-negative. Note that $\vec{\partial}_b^2$ annihilates $1/|\vec{b}|^{D-4}$, which is why we can consider the exchange particle to be on-shell.\footnote{The same can be seen from the choice of the integration contour, as described in more detail in \cite{Camanho:2014apa}. By rotating the contour of integration in $\vec{q}$, we cross the pole at $\vec{q}^{~2}= 0$ and hence it is sufficient to consider only three-point functions on-shell. }

\subsection{Higher Spin-graviton Couplings}\label{sec:coupl}

There are Lagrangian formulations of massive higher spin fields in flat spacetime, as well as in AdS \cite{Singh:1974qz,Singh:1974rc,Zinoviev:2001dt}. However, in this section, we present a more general approach that does not require the knowledge of the Lagrangian. We write down all possible local three-point interactions between two higher spin elementary particles with spin $J$ and a graviton. This three-point interaction is of importance for several reasons. First, this is one interaction that no particle can avoid because of the equivalence principle. Therefore the vanishing of this three-point interaction is sufficient to rule out existence of such higher spin particles. Moreover, as we will discuss later, this three-point interaction is sufficient  to compute the full eikonal scattering amplitude between a scalar and a higher spin particle.  

\begin{figure}[h]
\begin{center}
\usetikzlibrary{decorations.markings}    
\usetikzlibrary{decorations.markings}    
\begin{tikzpicture}[baseline=-3pt,scale=0.8]
\begin{scope}[very thick,shift={(4,0)}]
\draw[very thick,-latex]  (0,0) -- (-1.1,1.1);
\draw[very thick]  (-1,1) -- (-2,2);
\draw[very thick]  (-1.1,-1.1) -- (0,0);
\draw[very thick, -latex]  (-2,-2) -- (-1,-1);
\draw[thin,-latex]  (1.7,0.3) -- (2.3,0.3);
\draw  (0,0) circle (1.5pt) ;
\draw [domain=0:3.5, samples=500]  plot (\x, {0.1* sin(5*pi*\x r)});
\draw(-2,2)node[above]{$p_3, z_3$};
\draw(-2,-2)node[below]{$p_1, z_1$};
\draw(2,0.3)node[above]{ $q, z$};
\end{scope}
\end{tikzpicture}
\end{center}
\caption{\label{fig_tpf} \small The three-point interaction between two elementary particles with spin $J$ and a graviton.}
\end{figure}
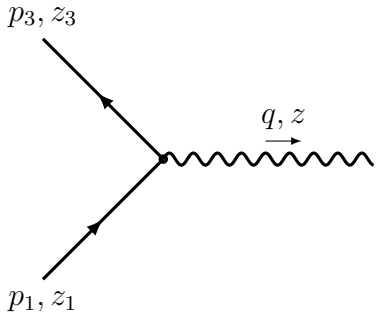

We start with the massive case and consider the massless case later on. Here we use the same method used in \cite{Costa:2011mg, Hinterbichler:2017qcl} for deriving the most general $J-J-2$ interaction. The momenta of higher spin particles are $p_1, p_3$ and the graviton has momentum $q$ (see figure \ref{fig_tpf}). The conservation and the on-shell conditions imply
\be\label{onshell}
p_1=p_3+q\ , \qquad p_1^2= p_3^2 = -m^2\ , \qquad  q^2 =0\ ,
\ee 
where $m$ is the mass of the higher spin particle. It is sufficient for us to consider polarization tensors which are made out of null and transverse polarization vectors $z_1, z_3, z$ satisfying 
\ba\label{eq:pol-tran}
z_1^2=z_3^2=z^2 =0, \qquad z_1 \cdot p_1=z_3 \cdot p_3=z \cdot q=0\ .
\ea
Transverse symmetric polarization tensors can be constructed from null and transverse polarization vectors by substituting  $ z_{i}^{\mu_1} z_{i}^{\mu_2} \cdots z_{i}^{\mu_s} \rightarrow\mathcal{E}_i^{\mu_1 \mu_2 \cdots \mu_s}  - \text{traces}$. In addition, we need to impose gauge invariance for the graviton. This means that each on-shell vertex should be invariant under $z \to z +\alpha q$, where $\alpha$ is an arbitrary number. Using (\ref{onshell}) and (\ref{eq:pol-tran}), we can write down all vertices in terms of only five independent building blocks\footnote{In $D=4$, the collection of momentum and polarization vectors $p_1, p_2, z_j \; i,j=1,2, 3$ are not linearly independent and there are additional relations between the building blocks. }
\ba\label{eq:bilblock}
&z_1\cdot z_3\ ,  \qquad z_1\cdot q\ , \qquad z_3\cdot q\ , \nonumber\\
&z\cdot p_3\ , \qquad  (z\cdot z_3)(z_1 \cdot q) - (z\cdot z_1)(z_3\cdot q) \ .
\ea
In order to list all possible vertices for the interaction $J-J-2$, we must symmetrize the on-shell amplitudes under $1\leftrightarrow 3$. We can then construct the most general form of on-shell three-point amplitude from these building blocks. In particular, for $J\ge 2$, we can write three distinct sets of vertices. The first set contains $J+1$ independent structures all of which are proportional to $(z\cdot p_3)^2$:
\ba\label{set1}
&\mathcal{A}_1  = (z \cdot p_3)^2 (z_1 \cdot z_3)^J\ ,\nonumber \\
&\mathcal{A}_2 =  (z \cdot p_3)^2 (z_1 \cdot z_3)^{J-1} (z_3 \cdot q) (z_1 \cdot q)\ , \nonumber\\
&\vdots \nonumber \\
&\mathcal{A}_{J+1} =  (z \cdot p_3)^2  (z_3 \cdot q)^J (z_1 \cdot q)^J\ .
\ea
The second set contains $J$-independent structures which are proportional to $(z\cdot p_3)$:
\ba\label{set2}
&\mathcal{A}_{J+2} = (z \cdot  p_3) ((z\cdot z_3)(z_1 \cdot q) - (z\cdot z_1)(z_3\cdot q)) (z_1 \cdot z_3)^{J-1}, \nonumber\\
&\mathcal{A}_{J+3} =  (z \cdot  p_3) ((z\cdot z_3)(z_1 \cdot q) - (z\cdot z_1)(z_3\cdot q))  (z_1\cdot z_3)^{J-2}  (z_3 \cdot q) (z_1 \cdot q), \nonumber\\
&\vdots \nonumber\\
& \mathcal{A}_{2J+1} = (z \cdot  p_3) ((z\cdot z_3)(z_1 \cdot q) - (z\cdot z_1)(z_3\cdot q)) (z_3 \cdot q)^{J-1} (z_1 \cdot q)^{J-1}\ .
\ea
Finally the third set consists of $J-1$ independent structures which do not contain $(z\cdot p_3)$:
\ba\label{set3}
&\mathcal{A}_{2J+2} =  ((z\cdot z_3)(z_1 \cdot q) - (z\cdot z_1)(z_3\cdot q))^2  (z_3 \cdot z_1)^{J-2}, \nonumber\\
&\mathcal{A}_{2J+3} =  ((z\cdot z_3)(z_1 \cdot q) - (z\cdot z_1)(z_3\cdot q))^2  (z_3 \cdot z_1)^{J-3} (z_3 \cdot q) (z_1 \cdot q), \nonumber\\
&\vdots \nonumber\\
&\mathcal{A}_{3J} =   ((z\cdot z_3)(z_1 \cdot q) - (z\cdot z_1)(z_3\cdot q))^2  (z_3 \cdot q)^{J-2} (z_1 \cdot q)^{J-2} \ .
\ea
In total there are $3J$ independent structures that contribute to the on-shell three-point amplitude of two higher spin particles with mass $m$ and spin $J$ and a single graviton.  Therefore the most general form of the three-point amplitude for $J\ge 1$, is given by\footnote{Here the propagators of the gravitons are canonically normalized to 1. Therefore we need explicit $G_N$ dependence in (\ref{eq:Css2}) since it couples to the graviton.}
\ba\label{eq:Css2}
C_{JJ2} =  \sqrt{32\pi G_N} \sum_{n= 1}^{3J} a_n \mathcal{A}_n.
\ea 
Note that $3J$ is also the number of independent structures in the three point functions in the CFT side after imposing permutation symmetry between operators $1,3$ and taking conservation of stress-tensor into account.

\subsection{Eikonal Kinematics}\label{sec:kin}
We now study the eikonal scattering of higher spin particles: $1,2\rightarrow 3,4$, where, 1 and 3 label the massive higher spin particles with mass $m$ and spin $J$ and 2, 4 label scalars of mass $m_s$ (see figure \ref{fig_eik}). Let us specify the details of the momentum and polarization tensors. In the eikonal limit, the momentum of particles are parametrized as follows\footnote{Our convention is $p^\mu=(p^u,p^v,\vec{p}).$}
\ba\label{kin4}
&p_1^\mu = \left(P^u,\frac{1}{P^u} \left(\frac{\vec{q} \;^2}{4}+m_1^2\right),  \frac{\vec{q}}{2} \right)\ , \qquad p_3^\mu =  \left(\bar{P}^u,\frac{1}{\bar{P}^u} \left(\frac{\vec{q} \;^2}{4}+m_3^2\right),  -\frac{\vec{q}}{2} \right)\ , \nonumber\\
& p_2^\mu= \left( \frac{1}{P^v}\left(\frac{\vec{q} \;^2}{4}+m_2^2\right), P^v,-\frac{\vec{q}}{2} \right)\ , \qquad p_4^\mu= \left( \frac{1}{\bar{P}^v}\left(\frac{\vec{q} \;^2}{4}+m_4^2\right),\bar{P}^v, \frac{\vec{q}}{2} \right)\ ,
\ea
where, $P^u,\bar{P}^u,P^v,\bar{P}^v>0$ and  $p_1^\mu- p_3^\mu\equiv q$ is the transferred momentum of the exchange particle which is spacelike. The eikonal limit is defined as $P^u, P^v \gg |q|,m_i$. In this limit  $P^u\approx \bar{P}^u, P^v \approx \bar{P}^v$  and the Mandelstam variable $s$ is  given by $s= - (p_1+p_2)^2 \approx  P^u P^v$. Moreover, for our setup we have  $m_1=m_3=m$ and $m_2=m_4=m_s$.

Massless particles have only transverse polarizations but massive higher spin particles can have both transverse and longitudinal polarizations. General polarization tensors can be constructed using the following polarization vectors
\ba\label{vectors}
&\epsilon^\mu_{T,\lambda}(p_1) = \left(0, \frac{\vec{q}\cdot \vec{e}_\lambda^{\;(1)}}{ P^u},  \vec{e}_\lambda^{\;(1)} \right)\ , \qquad \epsilon_L^\mu(p_1) = \left(\frac{P^u}{m},\frac{1}{ m P^u} \left(\frac{\vec{q}\;^2}{4} - m^2 \right),  \frac{\vec{q}}{2m} \right)\ , \nonumber\\
& \epsilon^\mu_{T,\lambda}(p_3) = \left( 0,-\frac{\vec{q}\cdot \vec{e}_\lambda^{\; (3)}}{ P^u},  \vec{e}_\lambda^{\; (3)} \right)\ , \quad \epsilon_L^\mu(p_3) = \left(\frac{P^u}{m},\frac{1}{ m P^u} \left(\frac{\vec{q}\;^2}{4} - m^2 \right),  - \frac{\vec{q}}{2m} \right) \ ,
\ea
where vectors $e_\lambda^{\mu} \equiv (0,0, \vec{e}_{\lambda})$ are complete orthonormal basis in the transverse direction $\vec{x}_\perp$. The longitudinal vectors do not satisfy (\ref{eq:pol-tran}) because $\ep_L \cdot \ep_L \neq 0$. However, they still form a basis for constructing symmetric traceless polarization tensors which are orthogonal to the corresponding momentum.

The polarization tensors constructed from (\ref{vectors}) are further distinguished by their spin under an $SO(D-2)$ rotation group which preserves the longitudinal polarization $\ep_L$ for each particle. We denote this basis of polarization tensors as $\mathcal{E}^{\mu_1 \mu_2 \cdots \mu_J}_{j} (p_i)$ where $j$ labels the spin under $SO(D-2)$. These tensors are basically organized by the number of transverse polarization vectors they contain.  The  most general polarization tensor for a particle with spin $J$ can now be decomposed as
\ba
{\pmb{ \mathcal{E}}}^{\mu_1 \cdots \mu_J}(p) = \sum_{j=0}^{J} r_j \mathcal{E}^{\mu_1 \cdots \mu_J}_j (p), 
\ea
where $r_j$'s are arbitrary complex numbers. However, in order to show that the higher spin particles cannot interact with gravity in a consistent way, we need only to consider a subspace spanned by
\ba\label{eq:3pol}
&\mathcal{E}_J^{\mu_1 \mu_2 \cdots \mu_J} = \ep_{T,\lambda_1}^{\mu_1} \ep_{T,\lambda_2}^{\mu_2} \cdots \ep_{T,\lambda_J}^{\mu_J} ,  \\
&\mathcal{E}_{J-1}^{\mu_1 \mu_2 \cdots \mu_J} = \sqrt{J}  \ep_L^{ ( \mu_1}   \ep_{T,\lambda_2}^{\mu_2} \ep_{T,\lambda_3}^{\mu_3} \cdots \ep_{T,\lambda_J}^{\mu_J)}, \\
&\mathcal{E}_{J-2}^{\mu_1 \mu_2 \cdots \mu_J} = \sqrt{\frac{D-1}{D-2}} \left(\ep_L^{(\mu_1} \ep_L^{\mu_2 } - \frac{\mathcal{P}^{\mu_1 \mu_2}}{D+2J-5}  \right) \ep_{T,\lambda_3}^{\mu_3} \ep_{T,\lambda_4}^{\mu_4} \cdots \ep_{T,\lambda_J}^{\mu_J)}, \quad \mathcal{P}^{\mu \nu} \equiv \eta^{\mu \nu} + \frac{p^{\mu} p^{\nu}}{m^2}, 
\ea
where, after contractions with other tensors we perform the following substitution: $e_{\lambda_1}^{i_1} e_{\lambda_2}^{i_2} \cdots e_{\lambda_j}^{i_j} \to e^{i_1 \cdots i_j}$ in which $e^{i_1 \cdots i_j}$ is a transverse symmetric traceless tensor.\footnote{In other words, whenever we see a combination of transverse polarization vectors: $\ep_{T,\lambda_1}^{\mu_1} \ep_{T,\lambda_2}^{\mu_2} \cdots \ep_{T,\lambda_S}^{\mu_S}$, we will replace that by either of $\ep_{T,+}^{\mu_1} \ep_{T,+}^{\mu_2} \cdots \ep_{T,+}^{\mu_S}\pm \ep_{T,-}^{\mu_1} \ep_{T,-}^{\mu_2} \cdots \ep_{T,-}^{\mu_S}$, where $e_+^{\mu} \equiv (0,0, 1,i,\vec{0})$ and $e_-^{\mu} \equiv (0,0, 1,-i,\vec{0})$. For us, it is sufficient to restrict to these set of polarization tensors.} One can easily continue this construction to generate the remaining polarization tensors. One should add more longitudinal polarization vectors and subtract traces in order to make them traceless.

\subsection{Bounds on Coefficients}
We now have all the tools we need to utilize the positivity condition (\ref{positive}) in the eikonal scattering of a massive higher spin particle and a scalar. The expression (\ref{positive}) requires knowledge of the contributions of all the particles that can be exchanged. However as we explain next, in the eikonal limit the leading contribution is always due to the graviton exchange. Let us explain this by discussing all possible exchanges:
\begin{itemize}
\item{Graviton exchange: Since, gravitons couple to all particles, the scattering amplitude in the eikonal limit will always receive contributions from graviton exchanges. In particular, in the eikonal limit, the contribution of graviton exchange to the phase shift goes as $\delta(s, b)\sim s$.}
\item{Exchange of particles with spin $J<2$: These exchanges are always subleading in the eikonal limit and hence can be ignored.\footnote{We have mentioned before that the eikonal exponentiation fails for the exchange of particles with spin $J<2$. However, we can still ignore them because the exchange of lower spin particles cannot compete with the graviton exchange in the eikonal limit.}}
\item{Exchange of higher spin particles $J>2$: In the eikonal limit, the exchange of a particle with spin $J$ can produce  a phase shift $\delta(s, b)\sim s^{J-1}$. However, it was shown in \cite{Camanho:2014apa}  that a phase shift that grows faster than $s$ leads to additional causality violation. Therefore if higher spin particles are present, their interactions must be tuned in such a way that they cannot be exchanged in eikonal scattering. This happens naturally when each higher spin particle is individually charged under a global symmetry such as $\mathds{Z}_2$. We should note that it is possible to have a scenario in which an infinite tower of higher spin particles can be exchanged without violating causality. However, we will restrict to the case where only a finite number of higher spin particles are present. At this point, let us also note that in AdS, the exchange of a finite number of higher spin particles are ruled out by the chaos growth bound of the dual CFT.}
\item{Exchange of massive spin-2 particles: Massive spin-2 particles can be present in nature. However, the exchange of these particles, as explained in \cite{Camanho:2014apa}, cannot fix the causality violation caused by the graviton exchange. Therefore, without any loss of generality, we can assume that the scalar particles do not interact with any massive spin-2 particle. For now this will allow us to ignore massive spin-2 exchanges. Let us note that it is not obvious that the argument of \cite{Camanho:2014apa} about massive spin-2 exchanges necessarily holds for scattering of higher spin particles. So, at the end of this section, we will present an interference based argument to explain the reason for why even an infinite tower of massive spin-2 exchanges cannot  restore causality. }
\end{itemize}

In summary, in the eikonal limit, it is sufficient to consider only the graviton exchange. In fact, we can safely assume that the scalar interacts with everything, even with itself, only via gravity. Let us also note that we are studying eikonal scattering of higher spin particles with scalars only for simplicity. The calculations as well as the rest of the arguments are almost identical even if we replace the scalar by a graviton. In the graviton case,  the argument of \cite{Camanho:2014apa} about massive spin-2 exchanges holds -- this implies that the presence of massive spin-2 particles will not change our final bounds.

We now use (\ref{positive}) to calculate the phase shift where $C_{13I}$ is given by equation (\ref{eq:Css2}). For scalar-scalar-graviton there is only one vertex, written as 
\ba
C_{I24}\equiv C_{002}=  \sqrt{32 \pi G} (z \cdot p_2)^2 \ .
\ea
Consequently, the sum in (\ref{positive}) is over the polarization of the exchanged graviton. In the eikonal limit, this sum receives a large contribution from only one specific intermediate state corresponding to the polarization tensor of the exchanged graviton appearing in $C_{13I}$ of the form $z^v z^v$ and the polarization tensor appearing in $C_{I24}$ of the form $z^uz^u$.\footnote{In the eikonal limit, the sum over the polarization of the graviton, in general, is given by \cite{Camanho:2014apa}
\be
\sum_I \epsilon^I_{\mu\nu}(q)(\epsilon^I_{\rho\sigma}(q))^* \sim \frac{1}{2} \left(\eta_{\mu\rho}\eta_{\nu\sigma}+\eta_{\nu\rho}\eta_{\mu\sigma} \right)\ .
\ee
}

As discussed earlier, if $\delta(s,\vec{b})$ grows with $s$, causality requires  $\delta(s,\vec{b}) \ge 0$ as a condition which must be true independent of polarization tensors we choose for our external particles. In particular, in the basis ${\pmb{ \mathcal{E}}}$, $\delta(s,\vec{b})$ can be written as 
\ba\label{eq:deltav}
\delta(s,\vec{b})= {\pmb{ \mathcal{E}}}_1^\dag \mathcal{K}(\vec{b}) {\pmb {\mathcal{E}}}_3 ,
\ea
where $\mathcal{K}$ is a Hermitian matrix which is encoding the eikonal amplitude in terms of the structures written in (\ref{eq:Css2}).\footnote{This assumes polarization tensors being properly normalized, i.e. ${\pmb{\mathcal{E}}}_i^{\dagger}  {\pmb{\mathcal{E}}}_i =1$, otherwise (\ref{eq:deltav}) should be divided by ${\pmb{\mathcal{E}}}_1^\dag {\pmb{\mathcal{E}}}_3$.  }
 Causality then requires $\mathcal{K}$ to be a positive semi-definite matrix for any $\vec{b}$. We sketch the argument for constraining three-point interactions here and leave the details to appendices  \ref{sec:transpol} and \ref{sec:laborwork}. 

First, let us discuss $D>4$.\footnote{$D=4$ is more subtle for various reasons and we will discuss it separately.}  We start with the general expressions for on-shell three point amplitudes. The  polarization tensors for both particles 1 and 3 are chosen to be  in the subspace spanned by $\mathcal{E}_J, \mathcal{E}_{J-1}$ and $\mathcal{E}_{J-2}$:
\be
{\pmb{ \mathcal{E}}}= r_J \mathcal{E}_J+ r_{J-1} \mathcal{E}_{J-1}+ r_{J-2}\mathcal{E}_{J-2}\ ,
\ee
where, $r_J, r_{J-1}$ and $r_{J-2}$ are real numbers. Using eikonal scattering we organize the phase shift in the small $b$ limit in terms of the highest negative powers of the impact parameter $b$. We start by setting $r_{J-2}=0$. We then demand $\mathcal{K}(\vec{b})$ to have non-negative eigenvalues order by order in $1/b$ for transverse polarization  $e^\oplus$ (or  $e^\otimes$) for all directions of the impact parameter $\vec{b}$.\footnote{Transverse polarizations $e^\otimes, e^\oplus$ are given  explicitly in appendix \ref{sec:transpol}.} This imposes the following constraints on the coefficients
\ba
a_{i}= 0\ , \qquad i \in \left\{ 2, 3, \cdots 3J \right\} \setminus \{J+2, 2J+2\} \ ,
\ea
where, $a_i$ is defined in (\ref{eq:Css2}). In other words, we find that all vertices with more than two derivatives must vanish. Moreover, the coefficients $a_{1}, a_{ J+2}, a_{2J+2}$ are related and the interaction $C_{JJ2}$ can be reduced to the following vertex
\ba\label{eq:minimal}
C_{JJ2}=&  a_1 (z_1\cdot z_3)^{J-2} \Big((z_1 \cdot z_3)^2 (z\cdot p_3)^2+ J \big((z\cdot z_3)(z_1 \cdot q) - (z\cdot z_1)(z_3\cdot q)\big) (z_1 \cdot z_3)(z \cdot p_3)  \nonumber\\
&  + \frac{J(J-1)}{2} \big((z\cdot z_3)(z_1 \cdot q) - (z\cdot z_1)(z_3\cdot q)\big)^2  \Big)\ .
\ea
When $J=2$, no further constraints can be obtained using any other choice of polarization tensors. On the other hand, for $J>2$  we can use the polarization tensor $\mathcal{E}_{J-2}$ (which always exists for $J \ge 2$) yielding
\ba
a_1 = 0\ ,
\ea
implying that $C_{JJ2}=0$. Therefore,  there is no consistent way of coupling higher spin elementary particles with gravity in flat spacetime in $D>4$ dimensions.\footnote{There are parity odd structures in $D=5$ for massive particles of any spin. As we show in appendix \ref{odd5}, These interactions also violate causality for $J>2$ as well as $J\le 2$. }

\subsection{$D=4$}
The $D=4$ case is special for several reasons. First of all, the $3J$ structures of on-shell three-point amplitude of two higher spin particles with mass $m$ and spin $J$ and a single graviton are not independent in $D=4$. These structures are built out of 5 vectors, however, in $D=4$, any 5 vectors are necessarily linearly dependent.  In particular, one can show that 
\be
m^2 B^2+2AB (q\cdot z_3)(q\cdot z_1)+2A^2 (q\cdot z_3)(q\cdot z_1)(z_1\cdot z_3)=0\ ,
\ee
where, $A= (z \cdot  p_3)$ and $B=(z\cdot z_3)(z_1 \cdot q) - (z\cdot z_1)(z_3\cdot q)$ are two of the building blocks of on-shell three-point amplitudes. The above relation implies that structures in the set (\ref{set3}) in $D=4$ are not independent since they can be written as structures from set (\ref{set1}) and (\ref{set2}). Therefore, for spin $J$ in $D=4$, there are $2J+1$ independent structures which is in agreement with the number of independent structures in the CFT three point function of the stress tensor and two spin-$J$ non-conserved primary operators. The $D=4$ case is special for one more reason -- there are parity odd structures for any spin $J$. In order to list all possible parity odd vertices for the interaction $J-J-2$, we introduce the following building block that does not preserve parity :
\begin{align}
\mathcal{B}= \epsilon^{\mu_1 \mu_2 \mu_3 \mu_4}{z_{1}}_{ \mu_1}{z_{3}}_{\mu_2}z_{\mu_3}q_{\mu_4}\ .
\end{align}
The parity odd on-shell three-point amplitude can be constructed using this building block. In particular, we can write two distinct sets of vertices with $\mathcal{B}$. The first set contains $J$  independent structures:
\ba\label{set1odd}
&\mathcal{A}_1^{odd}  =\mathcal{B} (z \cdot p_3) (z_1 \cdot z_3)^{J-1}\ ,\nonumber \\
&\mathcal{A}_2 ^{odd}= \mathcal{B} (z \cdot p_3) (z_1 \cdot z_3)^{J-2} (z_3 \cdot q) (z_1 \cdot q)\ , \nonumber\\
&\vdots \nonumber \\
&\mathcal{A}_{J}^{odd} =\mathcal{B}  (z \cdot p_3)  (z_3 \cdot q)^{J-1} (z_1 \cdot q)^{J-1}\ .
\ea
The second set contains $J-1$ independent structures:
\ba\label{set2odd}
&\mathcal{A}_{J+1}^{odd} =\mathcal{B} ((z\cdot z_3)(z_1 \cdot q) - (z\cdot z_1)(z_3\cdot q)) (z_1 \cdot z_3)^{J-2}, \nonumber\\
&\mathcal{A}_{J+2}^{odd} =  \mathcal{B}  ((z\cdot z_3)(z_1 \cdot q) - (z\cdot z_1)(z_3\cdot q))  (z_1\cdot z_3)^{J-3}  (z_3 \cdot q) (z_1 \cdot q), \nonumber\\
&\vdots \nonumber\\
& \mathcal{A}_{2J-1}^{odd} = \mathcal{B}  ((z\cdot z_3)(z_1 \cdot q) - (z\cdot z_1)(z_3\cdot q)) (z_3 \cdot q)^{J-2} (z_1 \cdot q)^{J-2}\ .
\ea
In $d=4$, there is another parity odd structure which is not related to the above structures and hence should be considered independent\footnote{We would like to thank J. Bonifacio for pointing this out.} 
\be
\mathcal{A}_{2J}^{odd} = \epsilon^{\mu_1 \mu_2 \mu_3 \mu_4}{z_{1}}_{ \mu_1}{p_1}_{\mu_3}{z_{3}}_{\mu_3}{p_3}_{\mu_4} (z \cdot p_3)^2 (z_3 \cdot q)^{J-1} (z_1 \cdot q)^{J-1}\ .
\ee
Therefore, the most general form of the three-point amplitude for $J\ge 1$ is given by
\ba
C_{JJ2} =  \sqrt{32 \pi G_N}\left( \sum_{n= 1}^{2J+1} a_n \mathcal{A}_n + \sum_{n= 1}^{2J} \bar{a}_n \mathcal{A}_n^{odd}\right)\ .
\ea 

We can again use the polarization tensors (\ref{eq:3pol}) to derive constraints. However, for $D=4$ the setup of this section is not adequate to completely rule out particles with $J>2$. In $D=4$, the transverse space is only two-dimensional and therefore does not provide enough freedom to derive optimal bounds. In particular, we find that a specific non-minimal coupling is consistent with the positivity of the phase shift. We eliminate this remaining non-minimal coupling by considering interference between the graviton and the higher spin particle.  

In $D=4$, the use of the polarization tensors (\ref{eq:3pol}) leads to the following bounds: $\bar{a}_n=0$ and $a_2,\cdots, a_{2J+1}$ are fixed by $a_1$ (see (\ref{bounds4})). The same set of bounds can also be obtained by using a simple null polarization vector
\be\label{pol4d}
\epsilon^\mu(p_1) =i \epsilon^\mu_{L}(p_1) +\epsilon^\mu_{T, \hat{x}}(p_1)\ , \qquad \epsilon^\mu(p_3) =-i \epsilon^\mu_{L}(p_3) +\epsilon^\mu_{T, \hat{x}}(p_3)\ ,
\ee
where the transverse and longitudinal vectors are defined in (\ref{vectors}) and the vector $\hat{x}$ is given by $\hat{x}=(0,0,1,0)$. The phase-shift in $D=4$ is 
\ba
\delta(s, \vec{b}) =  \frac{1}{4 \pi s} \sum_I C_{I24}(- i \vec{\partial_b}) C_{13 I}(- i \vec{\partial_b}) \ln\left( \frac{L}{b}\right)\ ,
\ea
where, $L$ is the IR regulator. Introduction of the IR regulator is necessary because of the presence of IR divergences in $D=4$. Using the polarization (\ref{pol4d}) we obtain
\be\label{ps4d}
\delta(s, \vec{b}) \sim s a_1  \ln\left( \frac{L}{b}\right)+ s \sum_{n=0}^{2J-1}\frac{1}{b^{2J-n}} \left(f_n \cos((2J-n)\theta)+\bar{f}_n \sin ((2J-n)\theta) \right)\ ,
\ee
where, $\cos\theta=\hat{b}\cdot \hat{x}$. Coefficients $f_n$ and $\bar{f}_n$ are linear combinations of parity even and parity odd coupling constants respectively. Requiring the phase shift to be positive order by order in $1/b$ in the limit $b \ll 1/m$ imposes the condition $f_n=\bar{f}_n=0$. This implies that all the parity odd couplings must vanish and all the parity even couplings are completely fixed once we specify $a_1$ (full set of constraints for spin $J$ are shown in (\ref{bounds4}).) Therefore, positivity of the phase shift (\ref{ps4d}) is consistent with a specific non-minimal coupling of higher spin particles in $D=4$. In order to rule out this specific interaction, we now consider interference between the graviton and the higher spin particle.

\subsubsection*{Bound from Interference}
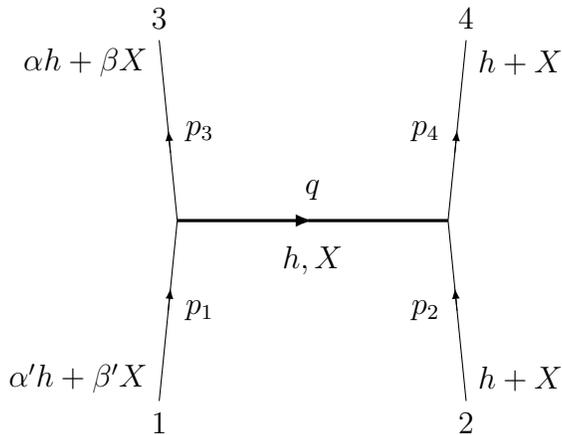
\begin{figure}
\begin{center}
\usetikzlibrary{decorations.markings}    
\usetikzlibrary{decorations.markings}    
\begin{tikzpicture}[baseline=-3pt,scale=0.6]
\begin{scope}[very thick,shift={(4,0)}]

\draw[thin,-latex]  (-3,0) -- (-3.2,2.0);
\draw[thin]  (-3.15,1.5) -- (-3.4,4.0);
\draw[thin]  (-3.2,-2.0) -- (-3,0);
\draw[thin, -latex]  (-3.4,-4)--(-3.15,-1.5) ;

\draw[thin, -latex]  (3,0) -- (3.2,2.0);
\draw[thin]  (3.15,1.5) -- (3.4,4.0);
\draw[thin]  (3.2,-2.0) -- (3,0);
\draw[thin, -latex]  (3.4,-4)--(3.15,-1.5) ;

\draw[very thick, -latex]  (-3,0)--(0,0) ;
\draw[very thick ]  (-0.1,0)--(3,0) ;

\draw(-2.5,1.5)node[above]{$p_3$};
\draw(-2.5,-1.5)node[below]{$p_1$};
\draw(2.5,1.5)node[above]{$p_4$};
\draw(2.5,-1.5)node[below]{$p_2$};
\draw(0,0.2)node[above]{ $q$};
\draw(0,-0.3)node[below]{ $h, X$};
\draw(-3.4,-4)node[below]{ $1$};
\draw(-3.4,4)node[above]{ $3$};
\draw(3.4,-4)node[below]{ $2$};
\draw(3.4,4)node[above]{ $4$};

\draw(3.4,3.5)node[right]{ $h+X$};
\draw(3.4,-3.5)node[right]{ $h+X$};
\draw(-3.4,3.5)node[left]{ $\alpha h +\beta X$};
\draw(-3.4,-3.5)node[left]{ $\alpha' h +\beta' X$};


\end{scope}
\end{tikzpicture}
\end{center}
\caption{\label{fig_int} \small Bounds from interference in $D=4$. In-states are linear combinations of massive higher spin particle $X$ and the graviton $h$. }
\end{figure}

We now consider eikonal scattering of gravitons and  massive higher spin particles: $1,2\rightarrow 3,4$. In this setup, 1 and 3 are linear combinations of massive higher spin particle $X$ and the graviton: $\alpha h +\beta X$ and $\alpha' h+ \beta' X$ respectively, where $\alpha,\alpha',\beta,\beta'$ are arbitrary real coefficients. While 2 and 4 are a fixed combination of $X$ and the graviton: $ h + X$. We will treat 2 as the source and 1 as the probe (see figure \ref{fig_int}). This setup is very similar to the setup of \cite{Bonifacio:2017nnt}.

Positivity of the phase-shift can now be expressed as semi-definiteness of the following matrix
\begin{align}\label{matrix}
\left(
\begin{array}{cc}
\delta_{hh} & \delta_{hX}\\ 
\delta_{Xh} & \delta_{XX}
\end{array} 
\right)\succeq 0\ ,
\end{align}
where, $\delta_{Xh}$ represents phase-shift when particle 1 is a higher spin particle of mass $m$ and spin $J$ and particle 3 is a graviton.\footnote{similar notation is used for other elements of the phase-shift matrix.} The above condition can also be restated as an interference bound
\be\label{intf_bound}
|\delta_{Xh}|^2 \le \delta_{hh}\delta_{XX}\ ,
\ee
where we have used the fact that $\delta_{Xh}=\delta_{hX}^*$. In the eikonal limit, the dominant contribution to both $\delta_{hh}$ and $\delta_{XX}$ comes from the graviton exchange and hence $\delta_{hh}, \delta_{XX}\sim s$, where $s$ is the Mandelstam variable. Therefore, asymptotic causality requires that $\delta_{Xh}$ should not grow faster than $s$.  

Let us now compute $\delta_{Xh}$ for a specific configuration. Momenta of the particles are again given by (\ref{kin4}) with appropriate masses. Moreover, we will use the following null polarization vectors for various particles:
\ba
&\epsilon_X^\mu(p_1) =i \epsilon^\mu_{L}(p_1) +\epsilon^\mu_{T, \hat{x}}(p_1)\ , \qquad \epsilon_h^\mu(p_3) = \epsilon^\mu_{T, \hat{x}}(p_3)+i \epsilon^\mu_{T, \hat{y}}(p_3)\ ,\nonumber\\
&\epsilon_X^\mu(p_2) =i \epsilon^\mu_{L}(p_2) +\epsilon^\mu_{T, \hat{x}}(p_2)\ , \qquad \epsilon_h^\mu(p_2) = \epsilon^\mu_{T, \hat{x}}(p_2)-i \epsilon^\mu_{T, \hat{y}}(p_2)\ ,
\nonumber\\
&\epsilon_X^\mu(p_4) =-i \epsilon^\mu_{L}(p_4) +\epsilon^\mu_{T, \hat{x}}(p_4)\ , \qquad \epsilon_h^\mu(p_4) = \epsilon^\mu_{T, \hat{x}}(p_4)+i \epsilon^\mu_{T, \hat{y}}(p_4)\ ,
\ea
where $\hat{x}=(0,0,1,0)$ and $\hat{y}=(0,0,0,1)$. In the eikonal limit the dominant contribution to $\delta_{Xh}$ comes from X-exchange. In particular, after imposing constraints (\ref{bounds4}), we find that 
\be
\delta_{Xh} \sim a_1 s^{J-1} \frac{e^{-2i(J-2)\theta}}{b^{2(J-2)}m^{4(J-2)}}\ ,
\ee
where $\cos\theta=\hat{b}\cdot \hat{x}$. The above phase-shift violates causality for $J>2$ implying 
\be
a_1=0 \qquad \text{for} \qquad J>2\ .
\ee
Therefore there is no consistent way of coupling higher spin elementary particles with gravity even in four dimensional flat spacetime.

\subsection{Comments}

\subsubsection*{Comparison with other arguments}\label{sec:fornima}

As mentioned in the introduction, there are qualitative arguments in the literature in $D=4$ suggesting that elementary massive higher spin particles cannot exist. The idea originally advocated by Weinberg, is to require physical theories for elementary particles to have a well behaved high energy limit or equivalently to demand a smooth limit for the amplitude as $m_X \to 0$ \cite{old-weinberg, Ferrara:1992yc }.  However, for minimal coupling with spin $J>2$ particles, the amplitude grows with powers of $   \left( \frac{s}{m_X^2} \right)$ as $m_X \to 0$  \cite{Arkani-Hamed:2017jhn}. Therefore, given a fixed and finite cutoff scale $\Lambda$ and a mass $m_X$, the amplitude can become $\mathcal{O}(1)$ for $ m_X \ll \sqrt{s} \ll \Lambda$. For instance, it was shown in \cite{ Porrati:1993in} by considering only the minimal coupling of spin $\frac{5}{2}$ to gravity, that tree-level unitarity breaks down at the energy $\sqrt{s} \sim \sqrt{ m_X M_{pl}} \ll M_{pl}$.  Moreover, the break-down scale for a particle of spin $J$ was conjectured to be even lower  $\sim \left(m_X^{2J-2} M_{pl}\right)^{\frac{1}{2J-1}}$ \cite{Rahman:2009zz}. This was shown to be true for massive spin $J=2$ particles \cite{Bonifacio:2018aon}. The existence of this scale implies that this particle cannot exist if tree-level unitarity is required to persist for scales up to $M_{pl}$. This seems natural if we require the theory of higher spin fields to be renormalizable. However, from an effective field theory point of view, the smooth $ m_X\to 0$ requirement, determines only the range of masses and cut-off scales over which the low energy tree level amplitude is a good description of this massive higher spin scattering experiment. Note that even within the tree level unitarity arguments, one still needs to consider all possible non-minimal couplings as well as all contact interactions in order to ensure that they do not conspire to change the singular behavior of the amplitude in the $m_X \to 0$ limit. In fact, \cite{Porrati:1993in,Cucchieri:1994tx} demonstrates examples in which adding non-minimal couplings can change the high energy singular behavior of the amplitude for longitudinal part of polarizations.   

By contrast, the causality arguments used here, require only the cut-off to be parametrically larger than the mass of the higher spin particle, $\Lambda \gg m_X$. Then, given an impact parameter $ b  \ll m_X^{-1}$, the desired bounds are obtained even if the amplitude or phase shift $M(s,t)$ , $\delta(s,b) \ll 1 $ (unlike the violation of tree-level unitarity requiring the amplitude to be $\mathcal{O}(1)$) since even the slightest time advance is forbidden by causality.   Moreover, in the eikonal experiment, the two incoming particles do not overlap and hence contributions from the other channel and contact diagrams can be ignored \cite{Camanho:2014apa}.

\subsubsection*{An Interference Argument for $D>4$}
A generalization of the interference argument of $D=4$ to higher dimensions also suggests that there is tension between massive higher spin particles and asymptotic causality. In fact, it might be possible to derive the bounds of this section by demanding that the phase shift $\delta_{Xh}$ does not grow faster than $s$, however, we have not checked this explicitly. This argument has one immediate advantage. For a particle with spin $J$, $\delta_{Xh}\sim s^{J-1}$ and therefore it is obvious that even an infinite tower of massive spin-2 exchanges cannot restore causality. The only way causality can be restored is if we add an infinite tower of massive higher spin particles. We should note that this arguments rely on the additional assumption that the eikonal approximation is valid for spin-$J$ exchange with $J>2$. The $N$-shocks argument of \cite{Camanho:2014apa} is also applicable here which strongly suggests that  the eikonal exponentiation holds even for $J>2$, however, a rigorous proof is still absent.

\subsubsection*{Massless Case}
Higher spin massless particles are already ruled out by the Weinberg-Witten theorem. Nonetheless, we can rederive this fact using the eikonal scattering setup. If the higher spin particles are massless, then gauge invariance requires that each vertex is invariant under the shift $z_i \to z_i + \alpha_i \; p_i$, where $\alpha_i$'s are arbitrary real numbers. In this case only the three following structures are allowed for $J\ge 2$
\ba
& \mathcal{D}_1 = (z\cdot p_3)^2 (z_1\cdot q)^{J} (z_3 \cdot q)^{J}, \\
&\mathcal{D}_2 = ((z_3\cdot q) (z_1\cdot z)-(z\cdot z_3) (z_1 \cdot q) - (z\cdot p_3) (z_1\cdot z_3)) (z\cdot p_3) (z_1\cdot q)^{J-1} (z_3 \cdot q)^{J-1},\nonumber\\
&\mathcal{D}_3 =((z_3\cdot q) (z_1\cdot z)-(z\cdot z_3) (z_1 \cdot q) - (z\cdot p_3) (z_1\cdot z_3))^2 (z_1\cdot q)^{J-2} (z_3 \cdot q)^{J-2}\ .\nonumber
\ea
This is again, as we will see in the next section, in agreement with the three structures appearing in the CFT three point function once we impose conservation constraints for all three operators. The general form of the three-point function for $J\ge 2$ is now given by
\ba
C_{JJ2} =  \sqrt{32\pi G_N} \sum_{n= 1}^{3} d_n \mathcal{D}_n\ .
\ea
For massless particles, $\mathcal{E}_J$ is the only polarization tensor. As before, by requiring asymptotic causality we find
\ba
d_n = 0 \qquad n= 1, 2, 3\ 
\ea
for $J>2$.

\subsubsection*{Parity Violating Interactions of Massive Spin-2 in $D=4$}
The argument presented in this section can also be applied to $J=2$ in $D\ge 4$. Of course, our argument does not rule out massive spin-2 particles. Rather it restricts the coupling between two massive spin-2 particles and a graviton to be minimal (\ref{eq:minimal}) which agrees with \cite{Bonifacio:2017nnt}. However, for $D=4$ our argument does rule out parity violating interactions between massive spin-2 particles and the graviton. Moreover, the same conclusion about parity violating interactions holds even for massive spin-1.

\subsubsection*{Restoration of Causality}
Let us now discuss the possible ways of bypassing the arguments presented in this section. Our arguments utilized the eikonal limit $m,q\ll \sqrt{s}\ll \Lambda$, where $\Lambda$ is the UV cut-off of the theory. Hence, our argument breaks down if the mass of the higher spin particle $m\sim \Lambda$. 

There is another interesting possibility. One can have a massive higher spin particle with mass $m\ll \Lambda$ and causality is restored by adding one or more additional particles. The contribution to the phase shift  for a tree level exchange of a particle of mass $M\gg m, \frac{1}{b}$  is exponentially suppressed $\sim e^{-bM}$. Hence, these additional contributions can be significant enough if the masses of these particles are not much larger than $m$. In addition, exchange of these additional particles can only restore causality if they have spin $J>2$. However, exchange of any finite number of such particles will lead to additional causality violation. Hence, the only possible way causality can be restored is by adding an infinite tower of fine-tuned higher spin particles with masses comparable to $ m$. Furthermore, causality for the scattering $J$+graviton$\rightarrow$ $J$+graviton also requires that an infinite subset of these new higher spin particles must be able to decay into two gravitons which implies that this infinite tower does affect the dynamics of gravitons at energies $\sim m$.\footnote{Note that we ignored loops of the higher spin tower. From the scattering $J$+graviton$\rightarrow$ $J$+graviton, it is clear that an infinite tower of higher spin particles with mass $M\gg m$ cannot restore causality even if we consider loops.} We will discuss this in more detail in section \ref{sec:causal}.

\subsubsection*{Composite Higher Spin Particles}
The argument of this section is applicable to elementary massive higher spin particles. However, whether a particle is elementary or not must be understood from the perspective of effective field theory. Hence, the argument of this section is also applicable to composite higher spin particles as long as they look elementary enough at a certain energy scale. In particular, if the mass of a composite particle is $m$ but it effectively behaves like an elementary particle up to some energy scale $\Lambda$ which is parametrically higher than $m$, then the argument of this section is still applicable. More generally, argument of this section rules out any composite higher spin particle which is isolated enough such that it does not decay to other particles after interacting with high energy gravitons $q\gg m$.

\subsubsection*{Validity of the Causality Condition}
Let us end this section by mentioning a possible caveat of our argument. In this section, we have shown that presence of massive higher spin particles is inconsistent with asymptotic causality which requires that particles do not experience a time advance even when they interact with each other. It is believed that any Lorentzian QFT must obey this requirement. However, there is no rigorous S-matrix based argument that shows that positivity of the time delay is a necessary requirement of any UV complete theory. A physical argument was presented in \cite{Camanho:2014apa} which relates positivity of the phase shift to unitarity but it would be nice to have a more direct derivation. In the next section, we present a CFT-based derivation of the same bounds in anti-de Sitter spacetime which allows us to circumvent this technical loophole.

\section{Higher Spin Fields in AdS$_D$}\label{sec:cft}

Let us now consider large-$N$ CFTs  in dimensions $d\ge 3$ with a sparse spectrum. CFTs in this class are special because at low energies they exhibit universal, gravity-like behavior. This duality allows us to pose a question in the CFT in $d$-dimensions which is dual to the question about higher spin fields in AdS in $D=d+1$ dimensions. Is it possible to have additional higher spin single trace primary operators $X_\ell$ with $\ell >2$ and scaling dimension $\Delta\ll \Delta_{\tiny{\text{gap}}}$ in a holographic CFT?

In general, any such operator $X_\ell$ will appear as an exchange operator in a four-point function of even low spin operators. In the Regge limit $\sigma \rightarrow 0$,\footnote{In terms of the conformal cross-ratios, $z\sim \sigma$ and $\bz \sim \eta \sigma$. The Regge limit is defined as $\sigma\rightarrow 0$ with $\eta=$ fixed after we analytically continue $\bz$ around the singularity at $1$ (see \cite{Afkhami-Jeddi:2016ntf,Afkhami-Jeddi:2017rmx,Afkhami-Jeddi:2018own}). } the contribution to the four-point function from the $X_\ell$-exchange goes as $\sim 1/\sigma^{\ell-1}$  which violates the chaos growth bound of \cite{Maldacena:2015waa} for $\ell>2$ and hence all CFT three-point functions $\langle X_\ell O O\rangle$ must vanish for any low spin operator $O$. In the gravity side, this rules out all bulk couplings of the form $\O \O {\cal X}_\ell$ in AdS, where ${\cal X}_\ell$ is a higher spin bulk field (massive or massless) and $\O$ is any other bulk field with or without spin. For example, this immediately implies that in a theory of quantum gravity where the dynamics of gravitons at low energies is described by Einstein gravity, decay of a higher spin particle  into two gravitons is not allowed. 

The above condition is not sufficient to completely rule out the existence of higher spin operators. In particular, we can still have higher spin operators without violating the chaos growth bound if the higher spin operator $X_\ell$ does not appear in the OPE of any two identical single trace primary operators. For example, if each higher spin operator  has a $\mathds{Z}_2$ symmetry,  they will be prohibited from appearing in the OPE of identical operators. However, a priori we can still have non-vanishing $\langle X_\ell X_\ell O\rangle$. In fact, the Ward identity dictates that the three-point function $\langle X_\ell X_\ell T\rangle$ must be non-zero where $T$ is the CFT stress tensor. In this section, we will utilize the holographic null energy condition to show that $\langle X_\ell X_\ell T\rangle$ must vanish for CFTs (in $d\ge 3$) with large $N$ and a sparse spectrum, or else causality (the chaos sign bound) will be violated. The Ward identity then requires that the two-point function $\langle X_\ell X_\ell \rangle$ must vanish as well. However, the two-point function $\langle X_\ell X_\ell \rangle$ is a measure of the norm of a state created by acting $X_\ell$ on the vacuum and therefore must be strictly positive in a unitary CFT. Vanishing of the norm necessarily requires that the operator $X_\ell$ itself is zero.
 
In the gravity language, this forbids the bulk interaction ${\cal X}_\ell$-${\cal X}_\ell$-graviton  -- which directly contradicts the equivalence principle. Therefore, a finite number of higher spin elementary particles, massless or massive, cannot interact with gravity in a consistent way even in AdS spacetime (in $D\ge 4$).

\subsection{Causality and Conformal Regge Theory}
We start with a general discussion about the Regge limit in generic CFTs and then review the holographic null energy condition (HNEC) in holographic CFTs which we will use to rule out higher spin single trace primary operators. The HNEC was derived in \cite{Afkhami-Jeddi:2017rmx,Afkhami-Jeddi:2018own}, however, let us provide a more general discussion of the HNEC here. The advantage of the new approach is that it can be applied to more general CFTs. However, that makes this subsection more technical, so casual readers can safely skip this subsection. 

As discussed in \cite{Cornalba:2006xk,Cornalba:2007zb,Afkhami-Jeddi:2018own} the relevant kinematic regime of the CFT 4-point function for accessing the physics of deep inside the bulk interior is the Regge limit. In terms of the familiar cross-ratios, in our conventions this limit corresponds to analytically continuing $\bz$ around the singularity at $1$ followed by taking the limit $z,\bz\rightarrow 0$ with $z/\bz$ held fixed. Unlike the more familiar euclidean OPE limit, the contributions to the correlation function in this limit are not easily organized in terms of local CFT operators. In fact contributions of individual local operators become increasingly singular with increasing spin. Using conformal Regge theory \cite{Costa:2012cb}, these contributions may be resummed into finite contributions by rewriting the sum over spins as a contour integral using the Sommerfeld-Watson transform. This formalism relied on the fact that the coefficients in the conformal block expansion are well defined analytic functions of $J$ away from integer values which was later justified in \cite{Caron-Huot:2017vep}. This allows one to rewrite the sum over spins in the conformal block expansion as a deformed contour integral over $J$, reorganizing the contributions to a sum over Regge trajectories. We will not discuss the derivation here as the details are well reviewed in \cite{Costa:2012cb,Kulaxizi:2017ixa,Costa:2017twz,Afkhami-Jeddi:2017rmx}.
We will instead derive an expression for the contribution of a Regge trajectory directly to the OPE of two local operators in terms of a non-local operator $\mathbb{E}_{\Delta,J}$ described below.

We will first derive an expression for the contribution to the OPE of scalar operators $\psi \psi$ by an operator of spin $J$ and scaling dimension $\Delta$. To this end, we will utilize the methods introduced in \cite{Costa:2011mg} to encode primary symmetric traceless tensor operators into polynomials of degree $J$ by contracting them with null polarization vectors $z^\mu$ :
\begin{align}
\O(x;z)\equiv z^{\mu_1}...z^{\mu_J}\O(x)_{\mu_1...\mu_J}.
\end{align}
It was shown in \cite{Costa:2011mg} that the tensor may be recovered from this polynomial by using the Thomas/Todorov operator. We are however interested in the case where the spin $J$ is not necessarily an integer. Therefore we will employ the procedure introduced in \cite{Kravchuk:2018htv} to generalize this expression to continuous spin by dropping the requirement that $\O(x;z)$ be a polynomial in $z$. With this definition, the expression for the contribution to the OPE by a continuous spin operators is given by a simple generalization of the expression appearing in \cite{Afkhami-Jeddi:2017rmx}. We will then use the shadow representation\cite{Ferrara:1972uq,SimmonsDuffin:2012uy,Dolan:2000ut} for the OPE in Lorentzian signature\cite{Czech:2016xec,deBoer:2016pqk}:
\begin{align}
\left.\frac{\psi(x_1)\psi(x_2)}{\langle \psi(x_1)\psi(x_2)\rangle}\right|_{\Delta,J}=\mathcal{N}&\int_{\text{\large{$\diamond$}}_{12}} d^dx_3 \int D^{d-2}zD^{d-2}z'\nonumber\\
&\times \frac{(-2z.z')^{2-d-J}\langle \psi(x_1)\psi(x_2)\tilde{\O}(x_3;z)\rangle}{\langle \psi(x_1)\psi(x_2)\rangle} \O(x_3,z').
\end{align}
where we let points $x_1$ and $x_2$ to be time-like separated and the integration of $x_3$ is performed over the intersection of causal future of $x_1$ and the causal past of $x_2$, $\mathcal{N}$ is a normalization constant and
\begin{align}
D^{d-2}z\equiv \frac{d^dz\delta(z^2)\theta(z_0)}{\text{vol}~{\mathbb{R}_+}}\ .
\end{align}
The integrals over $z$ and $z'$ replace the contraction over tensor indices that would appear for integer $J$ using the inner product for Lorentzian principal series introduced in \cite{Kravchuk:2018htv}. These are manifestly conformal integrals and the integration can be performed using the methods described in \cite{SimmonsDuffin:2012uy}.

In order to obtain the contribution to the Regge limit we will set $x_1=-x_2=(u,v,\vec{0})$ and analytically continue the points to space-like separations resulting in integration over a complexified Lorentzian diamond. We will then take the Regge limit by sending $v\rightarrow 0$ and $u\rightarrow \infty$ with $uv$ held fixed. The resulting expression is an integral over a complexified ball times a null ray along the $u$ direction:
\begin{align}\label{OPEdj}
&\left.\frac{\psi(u,v,\vec{0})\psi(-u,-v,\vec{0})}{\langle \psi(u,v,\vec{0})\psi(-u,-v,\vec{0})\rangle}\right|_{\Delta,J}=(-1)^{\frac{\Delta-1}{2}}\pi^{\frac{1-d}{2}}2^\Delta 
\frac{\Gamma\left(\frac{\Delta+J+1}{2}\right)\Gamma(\Delta-d/2+1)}{\Gamma\left(\frac{\Delta+J}{2}\right)\Gamma(\Delta-d+2)}\frac{C_{\psi\psi\O_{\Delta,J}}}{C_{\O_{\Delta,J}}}\notag\\ 
&\hspace{.9in}\times \frac{(uv)^{\frac{d-\Delta-J}{2}}}{u^{1-J}}\int_{-\infty}^{\infty} d\tilde{u}\int_{\vec{x}^2\leq uv}d^{d-2}\vec{x}(uv-\vec{x}^2)^{\Delta-d+1}\O((\tilde{u},0,i \vec{x});(0,1,0))\notag\\
&\hspace{.9in}\equiv u^{J-1} \mathbb{E}_{\Delta,J}\ ,
\end{align}
where $C_{\psi\psi\O_{\Delta,J}}$ is the OPE coefficient, $C_{\O_{\Delta,J}}$ is the normalization of $\langle \O\O\rangle$ and we have used $(u,v,\vec{x}_\perp)$ to express coordinates. This operator captures the contribution to OPE of $\psi\psi$ in the Regge limit. Therefore, analytically continued conformal blocks can be computed by inserting $\mathbb{E}_{\Delta,J}$ inside a three-point function. For example, in the case of external scalars we find
\begin{align}
\frac{\langle \phi(x_3)\phi(x_4) \mathbb{E}_{\Delta,J}\rangle}{\langle \phi(x_3)\phi(x_4)\rangle} u^{J-1} &\sim\lim_{\overset{z,\bz\rightarrow0}{z/\bz~\text{fixed}}}G_{\Delta,J}^\circlearrowleft(z,\bz)\notag\\
&=\frac{i (-1)^J 2^{2 \Delta +3 J-2} \Gamma \left(\frac{ J+\Delta -1}{2}\right) \Gamma \left(\frac{J+\Delta +1}{2}\right)}{ \Gamma \left(\frac{J+\Delta }{2}\right)^2}\frac{z^{\frac{\Delta -J}{2}} \bz^{-\frac{\Delta }{2}-\frac{J}{2}+2}}{(z-\bz)}\ ,
\end{align}
where $G_{\Delta,J}^\circlearrowleft(z,\bz)$ is obtained from the conformal block by taking $\bz$ around $1$ while holding $z$ fixed. In (\ref{OPEdj}) this analytic continuation corresponds to the choice of contour in performing the $\tilde{u}$ integral. The integrand encounters singularities in $\tilde{u}$ as the points become null separated from $x_3$ or $x_4$. Different analytic continuations of the conformal block can be obtained by choosing appropriate contours. The choice of contour in the $\tilde{u}$ plane was discussed in \cite{Afkhami-Jeddi:2018own} in greater detail. By an identical Sommerfeld-Watson transform and contour deformation argument as in\cite{Costa:2012cb}, the expression for the Regge OPE can now be used to capture the contribution of Regge trajectories 
\begin{align}\label{CRT}
&\left.\frac{\psi(u,v,\vec{0})\psi(-u,-v,\vec{0})}{\langle \psi(u,v,\vec{0})\psi(-u,-v,\vec{0})\rangle}\right|_{J(\nu)}=\int d\nu  u^{J(\nu)-1} a(\nu)\mathbb{E}_{\Delta(J(\nu)),J(\nu)}\ ,
\end{align}
where the coefficient $a(\nu)$ encodes the dynamical information about the spectrum of the CFT for the Regge trajectory parametrized by $J(\nu)$. 

The operator $\mathbb{E}_{\Delta,J}$ can be contrasted with the light-ray operator $\bold{L}[\O]$ introduced in \cite{Kravchuk:2018htv}. Although both correspond to non-local contributions to the OPE in the Regge limit, they do not compute the same quantity. As mentioned above $\mathbb{E}_{\Delta,J}$ computes the analytic continuation of the conformal block, whereas $\bold{L}[\O]$ computes the analytic continuation of conformal partial wave which is the sum of the block and its shadow which is proportional to $G_{1-J,1-\Delta}(z,\bz)$. However, because of the symmetry of the coefficient $a(\nu)$ under $\nu\rightarrow -\nu$ using either operator in the Regge limit will yield the same results after integration.

\subsubsection*{Holographic CFT: Holographic Null Energy Condition}
As described in more detail in \cite{Brower:2006ea,Kulaxizi:2017ixa,Costa:2012cb,Costa:2017twz,Meltzer:2017rtf,Afkhami-Jeddi:2017rmx} the leading Regge trajectory in a holographic theory with a large $\Delta_{gap}$ can be parametrized as
\ba
J(\nu) =2 -\frac{1}{\Delta_{gap}^2}\left( \frac{d^2}{4}+ \nu^2\right) + \mathcal{O} \left(\frac{1}{\Delta_{gap}^4} \right)\ .
 \ea
Using this expression for the trajectory we find that at leading order in $\Delta_{gap}$ the coefficient $a(\nu)$ will have single poles corresponding to the stress-tensor exchange as well as an infinite set of double-trace operators. As shown in \cite{Afkhami-Jeddi:2018own,Afkhami-Jeddi:2017rmx}, in the class of states in which we are interested, the dominant contribution to this OPE is given by the stress-tensor and the double-trace operators will not contribute. This contribution is captured by the holographic null energy operator  
\be\label{hne}
 \E_{r}(v)= \int_{-\infty}^{+\infty}  du' \int_{\vec{x}^2 \le r^2} d^{d-2}\vec{x}\left( 1-\frac{ \vec{x}^2}{ r^2}\right) T_{uu} \left(u' ,v , i \vec{x} \right)\ 
 \ee
which is a generalization of the averaged null energy operator \cite{Afkhami-Jeddi:2017rmx} and a special case of the operator $\mathbb{E}_{\Delta,J}$ described above with $\Delta=d$ and $J=2$.\footnote{We are using the following convention for points $x \in \mathbb{R}^{1,d-1}$ in CFT$_d$:
\be
x = (t,x^1,\vec{x})\equiv (u,v,\vec{x})\ , \qquad \text{where},  \qquad u=t-x^1\ , \qquad v=t+x^1\ .
\ee} In particular, in the limit $r\rightarrow 0$, this operator is equivalent to the averaged null energy operator.  

Causality in CFT implies that the four-point function obeys certain analyticity properties \cite{Hartman:2015lfa,Hartman:2016dxc,Hofman:2016awc,Hartman:2016lgu}. For generic CFTs in $d\ge 3$, these analyticity conditions dictate that the averaged null energy operator must be non-negative \cite{Hartman:2016lgu}. However, for holographic CFTs, causality leads to stronger constraints. In particular, causality of CFT four-point functions in the Regge limit implies that the expectation value of the holographic null energy operator is positive in a subspace of the total Hilbert space of holographic CFTs  \cite{Afkhami-Jeddi:2017rmx,Afkhami-Jeddi:2018own}:
\be\label{hnec}
\E(\rho)\equiv\lim_{B\rightarrow \infty}\langle \Psi| \E_{\sqrt{\rho}B}(B) |\Psi  \rangle \ge 0\ , 
\ee
where, $0< \rho < 1$. The class of states $|\Psi\rangle$ are created by inserting an arbitrary operator $O$ near the origin
\be\label{psi}
|\Psi\rangle=\int dy^1 d^{d-2}\vec{y}\, \epsilon.O(-i\delta,y^1,\vec{y})|0 \rangle\ , \qquad \langle \Psi|=\int dy^1 d^{d-2}\vec{y}\, \langle 0| \epsilon^*.O(i\delta,y^1,\vec{y})\ ,
\ee
where, $\epsilon$ is the polarization of the operator $O$ with 
\be
\epsilon.O\equiv \epsilon_{\mu \nu...}O^{\mu\nu...}
\ee 
and $\delta>0$. The state $|\Psi\rangle$ is equivalent to the Hofman-Maldacena state of the original conformal collider \cite{Hofman:2008ar} which was created by acting local operators, smeared with Gaussian wave-packets, on the CFT vacuum. 

\begin{figure}
\begin{center}
\includegraphics[width=0.65\textwidth]{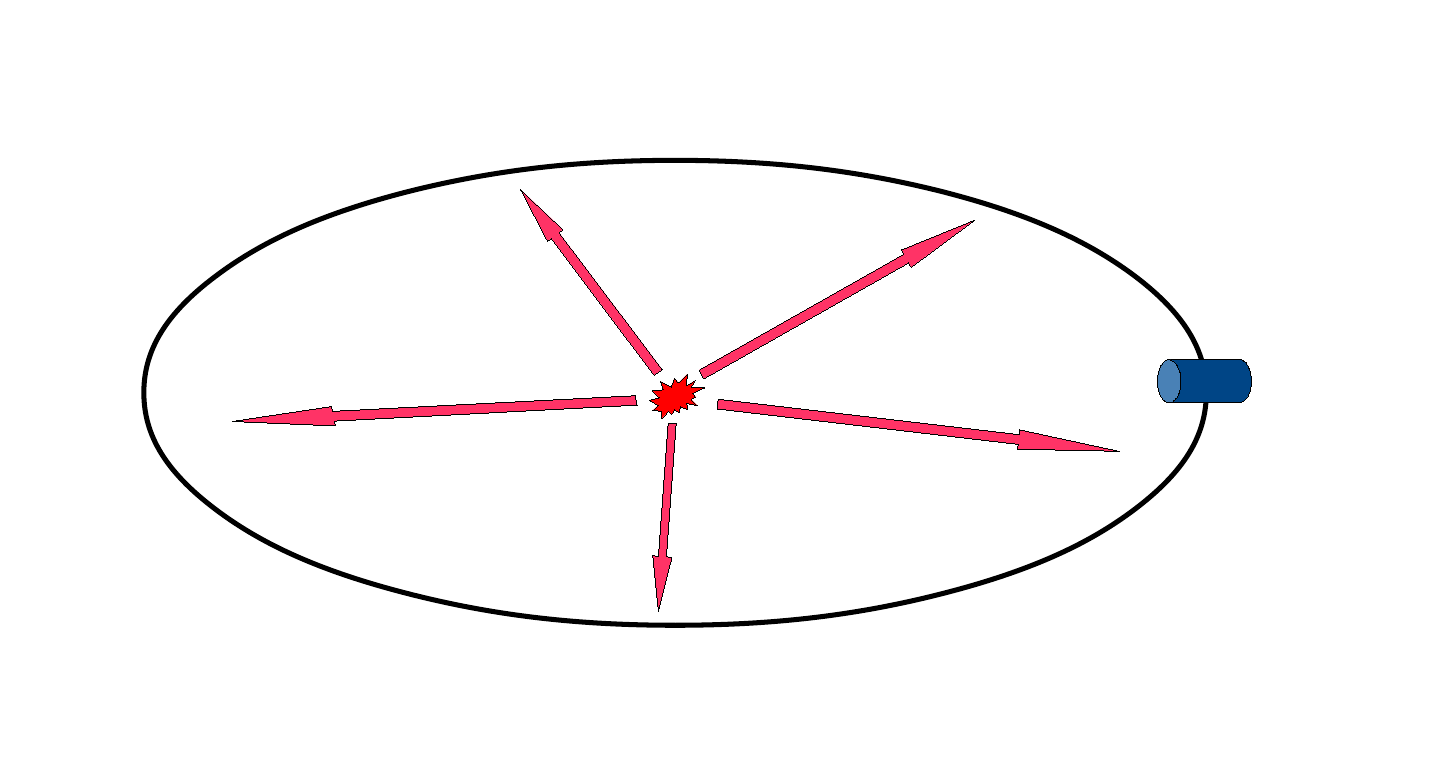}
\caption{Holographic null energy condition (HNEC): A holographic CFT is prepared in an excited state $|\Psi  \rangle$ by inserting an operator $O$ near the origin and an instrument which is shown in blue, measures the holographic null energy $\E_r$ far away from the excitation.}\label{coll_fig}
\end{center}
\end{figure}

The HNEC is practically a conformal collider experiment for holographic CFTs (in $d\ge 3$) in which the CFT is prepared in an excited state $|\Psi  \rangle$ by inserting an operator $O$ near the origin and an instrument measures $\E(\rho)$ far away from the excitation, as shown in figure \ref{coll_fig}. Then, causality implies that the measured value $\E(\rho)$ must be non-negative for large-$N$ CFTs with a sparse spectrum. Next, creating the state $|\Psi\rangle$ by inserting the higher spin operator $X_\ell$, we show that the inequality (\ref{hnec})  leads to surprising equalities among various OPE coefficients that appear in $\langle X_\ell X_\ell T \rangle$.

\subsection{$D>4$}
We will use the HNEC to derive bounds on higher spin single trace primary operators in $d\ge 4$ (or AdS$_D$ with $D\ge 5$). We will explicitly show that spin 3 and 4 operators are completely ruled out and then argue that the same must be true even for $J>4$. The case of $D=4$ is more subtle and will be discussed separately. 

\subsubsection{Spin-3 Operators}
Let us start with an operator  $X_\ell$ with $\ell= 3$ which does not violate the chaos growth bound because it has $\mathds{Z}_2$ or some other symmetry which sets $\langle OOX_{J=3}\rangle=0$ for all $O$. Consequently, this operator does not contribute as an exchange operator in any four-point function in the Regge limit and the leading contribution to the Regge four-point function still comes from the exchange of spin-2 single trace (stress tensor) and double trace operators. Therefore, the HNEC is still valid and we can use it with states created by smeared $X_{\ell=3}$ to derive constraints on $\langle X_{\ell=3}X_{\ell=3}T\rangle$.  

The CFT three-point function $\langle X_{\ell=3}X_{\ell=3}T\rangle$, is completely fixed by conformal symmetry up to a finite number of OPE coefficients (see appendix \ref{CFT_corr}). After imposing permutation symmetry and conservation equation, the three-point function $\langle X_{\ell=3}X_{\ell=3}T\rangle$ has $9$ independent OPE coefficients. We now compute the expectation value of the holographic null energy operator $\E(\rho)$ in states created by smeared $X_{\ell=3}$:
\be\label{state3}
|\Psi\rangle=\int dy^1 d^{d-2}\vec{y}\, \epsilon^{\mu_1}\epsilon^{\mu_2}\epsilon^{\mu_3}X_{\mu_1 \mu_2 \mu_3}(-i\delta,y^1,\vec{y})|0 \rangle\ , 
\ee
 where, $\epsilon^{\mu}$ is a null polarization vector:
 \begin{align}\label{polarization}
\epsilon^{\mu}=(-i \xi,-i ,\vec{\e}_{\perp})\  ,
\end{align}
with $\xi=\pm 1$ and $\vec{\e}_{\perp}{}^2=0$.\footnote{Note that in $d=3$ this choice of polarization vector does not work. In this case, one needs to use a general polarization tensor to derive constraints. 
 } 
 Following the procedure outlined in \cite{Afkhami-Jeddi:2018own}, we can compute $\E(\rho)$ in state (\ref{state3}). The result has the following form
\be
\E(\rho)=  \frac{1}{(1-\rho)^{d+3}}\sum_{n=0}^\infty I^{(n)}_\xi(\lambda^2)(1-\rho)^n\ ,
\ee
where, $I^{(n)}_\xi(\lambda^2)$ are polynomials in $\lambda^2$ which in general have  terms up to order $\lambda^6$, where
\be\label{lambda}
\lambda^2=\frac{1}{2}\vec{\e}_{\perp}\cdot {\vec{\e}_{\perp}}^*\ge 0\ .
\ee
Given our choice of polarization, different powers of $\lambda^2$ correspond to independent spinning structures and decomposition of $SO(d-1,1)^3$ to representations under $SO(d-2)$. Therefore positivity of $\E(\rho)$ implies that the coefficients of each power of $\lambda^2$ must individually satisfy positivity, for $\xi=+1$ as well as $\xi=-1$. Now, applying the HNEC order by order in the limit $\rho\rightarrow 1$, the inequalities lead to $9$ equalities among the $9$ OPE coefficients. We find that the 9 OPE coefficients cannot be consistently chosen to satisfy these equalities. Hence, causality implies that
\be
\langle X_{\ell=3}X_{\ell=3}T\rangle=0\ .
\ee
Moreover, the Ward identity relates $C_{X_3}$, coefficient of the two-point function $\langle X_{\ell=3}X_{\ell=3}\rangle$ (see eq \ref{2ptfn}), to a particular linear combination of the OPE coefficients $C_{i,j,k}$ and hence the two-point function $\langle X_{\ell=3}X_{\ell=3}\rangle$ must vanish as well.  This implies that we cannot have individual spin-3 single trace primary operators in the spectrum. The detail of the calculation are rather long and not very illuminating, so we relegate them to appendix \ref{app3}.

\subsubsection{Spin-4 Operators}
We can perform a similar analysis with a spin-4 operator which leads to the same conclusion, however, the details are little different. The three-point function $\langle X_{\ell=4}X_{\ell=4}T\rangle$, after imposing permutation symmetry and conservation equation, has $12$ independent OPE coefficients (see appendix \ref{app4}). But the HNEC leads to stronger constraints as we increase the spin of $X$ and these 12 OPE coefficients cannot be consistently chosen to satisfy all the positivity constraints. In fact, as we will show, it is easier to rule out spin-4 operators using the HNEC  than spin-3 operators.

We again perform a conformal collider experiment for holographic CFTs (in $d\ge 3$) in which the CFT is prepared in an excited state
\be\label{state4}
|\Psi\rangle=\int dy^1 d^{d-2}\vec{y}\, \epsilon^{\mu_1}\epsilon^{\mu_2}\epsilon^{\mu_3}\epsilon^{\mu_4}X_{\mu_1 \mu_2 \mu_3\mu_4}(-i\delta,y^1,\vec{y})|0 \rangle\ , 
\ee
where, $\epsilon^{\mu}$ is the null polarization vector (\ref{polarization}). The expectation value of the holographic null energy operator $\E(\rho)$ in states created by smeared $X_{\ell=4}$ can be computed using methods used in \cite{Afkhami-Jeddi:2018own}
\be \label{eq:comp}
\E(\rho)=  \frac{1}{(1-\rho)^{d+5}}\sum_{n=0}^\infty \tilde{I}^{(n)}_\xi(\lambda)(1-\rho)^n\ ,
\ee
where, $\tilde{I}^{(n)}_\xi(\lambda^2)$ are polynomials in $\lambda^2$ (\ref{lambda}) with terms up to $\lambda^8$ in general. Causality implies that different powers of $\lambda^2$ must satisfy positivity individually, for $\xi=+1$ as well as $\xi=-1$. We find that the 12 OPE coefficients cannot be consistently chosen to satisfy all the positivity constraints implying (see appendix \ref{app4})
\be
\langle X_{\ell=4}X_{\ell=4}T\rangle=0\ .
\ee
Consequently, the Ward identity dictates that the two-point function of $X_{\ell=4}$ must vanish as well. This rules out single trace spin-4 operators with scaling dimensions below $\Delta_{\tiny{\text{gap}}}$ in the spectrum of a holographic CFT. As shown in the appendix \ref{app4}, we ruled out spin-4 operators even without considering $\E_{\xi=-1}(\rho)$. This is because as we increase the spin of $X$, the number of constraint equations increases faster than the number of independent OPE coefficients. This is also apparent from the fact that for spin-3, we had to go to order $\frac{1}{(1-\rho)^{d-2}}$ to derive all constraints. Whereas, for spin-4, the full set of constraints were obtained at the order $\frac{1}{(1-\rho)^{d-1}}$.

\subsubsection{Spin $\ell>4$}
For operators with spin $\ell\ge 5$, the argument is exactly the same. In fact, it is easier to rule them out because the HNEC leads to stronger constraints at higher spins. For example, for $\ell=1$, there are 3 independent OPE coefficients but the HNEC yields 2 linear relations among them. Consequently, the three-point function  $\langle X_{\ell=1}X_{\ell=1}T\rangle$ is fixed up to one coefficient. The same is true for $\ell=2$ -- there are 6 independent OPE coefficients and 5 constraints from the HNEC. Furthermore, in both of these cases, constraint equations ensure that the expectation value of the holographic null energy operator behaves exactly like that of the scalars: $\E(\rho)\sim \frac{1}{(1-\rho)^{d-3}}$ for $d \ge 4$. In fact, this is true for all low spin operators of holographic CFTs.

The HNEC barely rules out operators with $\ell=3$. There are 9 independent OPE coefficients. Using the positivity conditions all the way up to order $\frac{1}{(1-\rho)^{d-2}}$ for $\xi=\pm 1$, we showed that the OPE coefficients cannot be consistently chosen to satisfy all the positivity constraints. Whereas, the HNEC rules out $\ell=4$ operators quite comfortably.  We only needed to consider positivity conditions up to order $\frac{1}{(1-\rho)^{d-1}}$ and only for $\xi=+1$ to rule them out. The same pattern persists even for operators with spins $\ell\ge 5$ so we will not repeat our argument for each spin. Instead, we present a general discussion about the structure of $\E(\rho)$ at each order in the limit $\rho\to 1$ for general $\Delta$ and $J$ (in $d\ge 4$ dimensions). This enables us to count the number of constraint equations at each order. A simple counting immediately suggests that a non-vanishing $\langle X_{\ell}X_{\ell}T\rangle$ cannot be consistent with the HNEC even for spins higher than 4. By studying various examples with specific values of $\ell$, $\Delta$ and $d$, we have explicitly checked that our simple counting argument is indeed true.

The three point function $\langle X_{\ell}X_{\ell}T\rangle$ has $5+6(\ell-1)$ OPE coefficients to begin with, however not all of them are independent. Permutation symmetry implies that only $4\ell$ OPE coefficients can be independent. In addition, conservation of the stress-tensor operator $T$ imposes $\ell$ additional constraints among the remaining $4\ell$ OPE coefficients. Therefore, the three-point function $\langle X_\ell X_\ell T\rangle$ is fixed by conformal invariance up to $3\ell$ truly independent OPE coefficients.\footnote{The number of independent OPE coefficients is different in $d=3$.} Furthermore, the Ward identity leads to a relation between these OPE coefficients and the coefficient of the two-point function $C_{X_\ell}$.

We again perform a conformal collider experiment for holographic CFTs (in $d\ge 4$) in which the CFT is prepared in an excited state created by smeared $X_{\ell}$. In the limit $\rho\rightarrow 1$, the leading contribution to $\E(\rho)$ goes as
\be
\E(\rho)\sim  \frac{1}{(1-\rho)^{d+2\ell-3}}\ ,
\ee
where only a single structure contributes with an overall factor that depends on a specific linear combination of OPE coefficients. Just like before, the structure changes sign for different powers of $\lambda^2$ and hence in the 1st order, the HNEC produces only one constraint. It is clear from \cite{Afkhami-Jeddi:2017rmx,Afkhami-Jeddi:2018own} that the coefficient of the term $\E(\rho)\sim  \frac{1}{(1-\rho)^{d-3}}$ is fixed by the Ward identity and hence automatically positive. On the other hand, the HNEC in general can lead to constraints up to the $2\ell$-th order, i.e. the order $\E(\rho)\sim  \frac{1}{(1-\rho)^{d-2}}$. But for $\ell>3$, one gets $3\ell$ independent constraints from the HNEC even before the $2\ell$-th order. 

It is easier to rule out operators with higher and higher spins. A simple counting clearly shows why this is not at all surprising. First, let us assume that the HNEC rules out any operator with some particular spin $\ell=\ell_*>2$. That means for spin $\ell_*$ the HNEC generates  $3\ell_*$ independent relations among the OPE coefficients. If we increase the spin by 1: $\ell=\ell_*+1$, we get 3 more independent OPE coefficients. However, the $(2\ell_*+1)$-th and  $(2\ell_*+2)$-th orders in $\E(\rho)$ produce new constraints and at each new order there can be $\ell_*+1$ new equalities. Moreover, the $\lambda^2$ polynomials at each order now has a $\lambda^{2(\ell_*+1)}$ term with its own positivity condition -- this means that there can be $2\ell_*$ additional equalities from the first $2\ell_*$ orders. Therefore, for spin $\ell_*+1$, there are $3$ new OPE coefficients, whereas there can be $2(2\ell_*+1)$ new constraints among them. Of course, this is not exactly true because some of $2(2\ell_*+1)$ constraints are  not independent. However, for $\ell_* \ge 4$, the number of new constraints  $2(2\ell_*+1)\gg 3$ and hence this simple counting suggests that the HNEC must rule out operators with spin $\ell \ge 5$.

Let us now demonstrate that this simple counting argument is indeed correct. First, consider $\ell=1$. This is the simplest possible case which was studied in \cite{Afkhami-Jeddi:2018own}. For $\ell=1$, there are 3 independent OPE coefficients. The number of constraints (equality) from the HNEC at each order is given by $\{1,1\}$.\footnote{The $n$-th element of the sequence $\{c_1,\cdots,c_n, \cdots, c_{2\ell}\}$ represents the number of independent constraints at the order $n$.} After imposing these constraints the expectation value of the holographic null energy operator goes as $\sim\frac{1}{(1-\rho)^{d-3}}$. Similarly, for $\ell=2$ the number of constraints from the HNEC at each order is given by \{1,1,2,1\} and the total number of constraints is still less than the number of independent OPE coefficients \cite{Afkhami-Jeddi:2018own}.

For $\ell=3$, the sequence is  $\{1,1,2,2,2,1\}$ (see appendix \ref{app3}) and hence spin-3 operators were completely ruled out at the order $\frac{1}{(1-\rho)^{d-2}}$. If we increase the spin by 1, we find that the number of constraints from the HNEC at each order is $\{1,1,2,2,3,2,1,0\}$ (see appendix \ref{app4}). The zero at the end indicates that spin-4 operators were already ruled out at the order $\frac{1}{(1-\rho)^{d-1}}$. Our simple counting suggests that the number of zeroes should increase as we go to higher spins. Explicit computation agrees with this expectation. In particular, for $\ell=5$,  there are $15$ independent OPE coefficients and the number of constraints at each order is $\{1,1,3,3,5,2,0,0,0,0\}$. Therefore the spin-5 operators are ruled out at the order $\frac{1}{(1-\rho)^{d+2}}$. Similarly, for $\ell=6$,  there are $18$ independent OPE coefficients. Explicit calculation shows that the number of constraints at each order is $\{1,1,3,3,5,5,0,0,0,0,0,0\}$. Therefore, spin-6 operators can be ruled out even at the order $\frac{1}{(1-\rho)^{d+4}}$. All of these results imply that the presence of any single trace primary operator with spin $\ell>2$ is not compatible with causality.

\subsection{AdS$_4$/CFT$_3$}
Similar to the $D=4$ case on the gravity side, CFTs in $d=3$ are special. Of course, large-$N$ CFTs with a sparse spectrum in $(2+1)$-dimensions are still holographic and the HNEC once again implies that higher spin single trace operators with $\Delta \ll \Delta_{gap}$ are ruled out. However, there are several aspects of the $d=3$ CFTs which are  different from the higher dimensional case. 

First of all, in CFT$_3$ the three-point functions $\langle X_\ell X_\ell T\rangle$ have both parity even and parity odd structures for any $\ell$
\be
\langle X_\ell X_\ell T\rangle = \langle X_\ell X_\ell T\rangle_+ + \langle X_\ell X_\ell T\rangle_-\ .
\ee
Furthermore, the number of independent parity even structures at $d=3$ is different from the higher dimensional case. The general three-point function (\ref{general3pt}) implies that after imposing permutation symmetry and conservation equation, similar to the higher dimensional case $ \langle X_\ell X_\ell T\rangle_+$ should contain $3\ell$ independent structures. However, for $d=3$, not all of these structures are independent. In particular, this overcounting should be corrected by setting OPE coefficients $C_{1,1,k}=0$ for $k\ge 1$ in  (\ref{general3pt}) \cite{Costa:2011mg}. Therefore, in $d=3$, the parity even part $\langle X_\ell X_\ell T\rangle_+$ has $2\ell+1$ independent OPE coefficients. Whereas, the parity odd part $\langle X_\ell X_\ell T\rangle_-$ has $2\ell$ independent OPE coefficients. Note that this is exactly what is expected from interactions of gravitons with higher spin fields in 4d gravity. 

There is another aspect of $d=3$ which is different from the higher dimensional case. The choice of polarization (\ref{polarization})  in $d=3$ implies that $\vec{\varepsilon}_\perp=0$ and hence the $\lambda$-trick does not work. However, the full set of bounds can be obtained by considering the full polarization tensor for $ X_\ell$. This can be achieved by using the projection operator of \cite{Costa:2011mg} which makes the analysis more complicated. However the final conclusion remains unchanged.

Since we expect that the HNEC imposes stronger constraints as we increase the spin, it is sufficient to only rule out $X_{\ell=3}$. The steps are exactly the same but details are little different. After imposing permutation symmetry and conservation equation, the three-point function $\langle X_{\ell=3}X_{\ell=3}T\rangle$ has $7$ parity even and 6 parity odd independent OPE coefficients. We again compute the expectation value of the holographic null energy operator $\E(\rho)$ in states created by smeared $X_{\ell=3}$:
\be
|\Psi\rangle=\int dy^1 dy^2\, \epsilon^{\mu_1\mu_2\mu_3}X_{\mu_1 \mu_2 \mu_3}(-i\delta,y^1,y^2)|0 \rangle\ , 
\ee
 where $\epsilon^{\mu_1\mu_2\mu_3}$ is the traceless symmetric polarization tensor. Using the techniques developed in \cite{Afkhami-Jeddi:2018own}, we now compute the expectation value of the holographic null energy operator $\E(\rho)$ in this state which can be schematically expressed in the following form
\be
\E(\rho)= \sum_{n=1}^6 \frac{j_n(\epsilon^{\mu_1\mu_2\mu_3},C_{i,j,k})}{(1-\rho)^{n}}+j_0(\epsilon^{\mu_1\mu_2\mu_3},C_{i,j,k}) \ln(1-\rho)+\cdots\ ,
\ee
where $j_n(\epsilon^{\mu_1\mu_2\mu_3},C_{i,j,k})$ are specific functions of the the polarization tensors and the OPE coefficients. The dots in the above expression represent terms that vanish in the limit $\rho\rightarrow 1$. The $\ln(1-\rho)$ term is unique to the 3d case and is a manifestation of soft graviton effects in the IR. 

By applying the HNEC order by order in the limit $\rho\rightarrow 1$, we again find that the HNEC can only be satisfied for all polarizations if and only if all the OPE coefficients vanish. Consequently, the Ward identity implies that we cannot have individual spin-3 operators in the spectrum.\footnote{As explained in appendix \ref{3d} it is still possible to use the $\lambda$-trick to derive constraints in dimension $d=3$. This implies that individual spin-4 single trace operators (at least the parity even part) are also ruled out.} Moreover, a simple counting again suggests that the same is true even for $\ell>3$. In $d=3$, as we increase the spin by one, the number of parity even OPE coefficients increases by 2. However, now there are two more orders perturbatively in $(1-\rho)$ that generate new relations among the OPE coefficients. Each new order produces at least one new constraint suggesting that if the HNEC rules out parity even operators with some particular spin $\ell$, it will also rule out all parity even operators with spin $\ell+1$. In addition, it is straightforward to extend this argument to include parity odd structures, however, we will not do so in this paper.

\subsection{Maldacena-Zhiboedov Theorem and Massless Higher Spin Fields }
In this section we argued that in holographic CFTs, any higher spin single trace non-conserved primary operator violates causality. On the gravity side, this rules out any higher spin massive field with mass  below the cut-off scale (for example the string scale). But what about massless higher spin fields? In asymptotically flat spacetime, this question has already been answered by the Weinberg-Witten/Porrati theorem \cite{Weinberg:1980kq,Porrati:2008rm}. The same statement can be proven in AdS by using the argument of this section but for conserved $X_\ell\equiv \mathcal{J}$. Conservation of $\mathcal{J}$ leads to additional relations among the OPE coefficients $C_{i,j,k}$'s in $\langle \mathcal{J}\mathcal{J} T\rangle$. Even before we impose these additional conservation relations, the HNEC implies $C_{i,j,k}=0$ for $\ell>2$, which is obviously consistent with these new  relations from conservation. Hence, our argument is valid even for higher spin conserved current $\mathcal{J}$. 

Causality of CFT four-point functions in the lightcone limit also rules out a finite number of conserved higher spin currents in any CFT \cite{Hartman:2015lfa}. This is a partial generalization of the Maldacena-Zhiboedov theorem \cite{Maldacena:2011jn}, from $d=3$ to higher dimensions. The argument which was used in \cite{Hartman:2015lfa} to rule out higher spin conserved current is not applicable here since $\mathcal{J}$ does not   contribute to generic CFT four-point functions  as exchange operators.\footnote{Let us recall that none of the operators are charged under $\mathcal{J}$ and hence one can tune $\langle \mathcal{J} O O\rangle=0$ for any $O$. Consequently, $\mathcal{J}$ does not   contribute as an exchange operator.} However, we can repeat the argument of \cite{Hartman:2015lfa} for a mixed correlator $\langle \O\O\O\O \rangle$ in the lightcone limit where, $\O\equiv T+\mathcal{J}$. For this mixed correlator, $\mathcal{J}$ does contribute as an exchange operator in the lightcone limit. In particular, we can schematically write
\be\label{mixed}
\langle \O\O\O\O \rangle =
\blockS{\O}{\O}{\O}{\O}{1}
+
\blockS{\O}{\O}{\O}{\O}{T}
 +
\blockS{T}{\mathcal{J}}{T}{\mathcal{J}}{\mathcal{J}}+\cdots\ ,
\ee
where each diagram represents a spinning conformal block and dots represent contributions suppressed by the lightcone limit. The argument of \cite{Hartman:2015lfa}, now applied to the correlator $\langle \O\O\O\O \rangle$, implies that this correlator is causal if and only if the last term in (\ref{mixed}) is identically zero. The $\mathcal{J}$-exchange conformal blocks, for $\ell>2$,  in the lightcone grow faster than allowed by causality. This necessarily requires that the three-point function  $\langle \mathcal{J}\mathcal{J} T\rangle$ must vanish -- which is sufficient to rule out $\mathcal{J}$ for $\ell>2$. This generalizes the argument of \cite{Hartman:2015lfa} ruling out higher spin conserved currents even when none of the operators are charged under it. We should note that technically it might be plausible for the OPE coefficients to conspire in a non-trivial way such that a conserved current $\mathcal{J}$ cannot contribute as an exchange operator (for all polarizations of the external operators) but still has a non-vanishing $\langle \mathcal{J}\mathcal{J} T\rangle$. However, it is very unlikely that such a cancellation is possible since the three-point function  $\langle \mathcal{J}\mathcal{J} T\rangle$ can only have three independent OPE coefficients. This unlikely scenario can be ruled out by explicit calculations. 

The above argument is applicable only because $\mathcal{J}$ is conserved. However, one might expect that a similar argument in the Regge limit should rule out even non-conserved $X_\ell$ for holographic CFTs. This is probably true but the argument is more subtle in the Regge limit because an infinite tower of double trace operators also contribute to the correlator $\langle \O\O\O\O \rangle$. Hence, one needs to smear all four operators appropriately, in a way similar to \cite{Afkhami-Jeddi:2016ntf, Afkhami-Jeddi:2017rmx}, such that the double trace contributions are projected out. One might then use causality/chaos bounds to rule out the three-point function  $\langle X_\ell X_\ell T\rangle$. However, it is possible that the smearing procedure sets  contributions from certain spinning structures in $\langle X_\ell X_\ell T\rangle$ to zero as well. In that case, this argument will not be sufficient. A proof along this line requires the computation of a completely smeared spinning Regge correlator which is technically challenging even in the holographic limit.

\subsection{Comments}

\subsubsection*{Small Deviation from the Holographic Conditions}

Large-$N$ CFTs with a sparse spectrum are indeed special because at low energies they exhibit gravity-like behavior. This immediately poses a question about the assumptions of large-$N$ and sparse spectrum: how rigid are these conditions? In other words, do we still get a consistent CFT if we allow small deviations away from these conditions?

In this section, we answered a version of this question for the sparseness condition. The sparseness condition requires that any single trace primary operator with spin $\ell > 2$, must necessarily have dimension $\Delta\ge \Delta_{gap}\gg 1$. This condition ensures that the dual  gravity theory has a low energy description given by Einstein gravity. However, we can imagine a small deviation from this condition by allowing a finite number of additional higher spin single trace primary operators $X_\ell$ with $\ell >2$ and scaling dimension $\Delta\ll \Delta_{\tiny{\text{gap}}}$. As we have shown in this section, these new operators violate the HNEC implying the resulting CFTs are acausal.

\subsubsection*{Minkowski vs AdS}
It is rather apparent that the technical details of the flat spacetime argument and the AdS argument are very similar. For example, the number of independent structures for a particular spin is the same in both cases. In flat spacetime as well as in AdS, we start with inequalities which can be interpreted as some kind of time-delay. In addition, these inequalities when applied order by order, lead to equalities among various structures. These equalities eventually rule out higher spin particles. However, the AdS argument has one conceptual advantage,  namely, it does not require any additional assumption about the exponentiation of the leading contribution. The CFT-based argument relies on the HNEC. The derivation of the HNEC utilized the causality of a CFT correlator which was designed to probe high energy scattering deep into the AdS bulk. It is therefore not a coincidence that the technical details of the AdS and the flat space arguments are so similar. Since the local high energy scattering is insensitive to the spacetime curvature, it is not very surprising that the bounds in flat space and in AdS are identical. This also suggests that the same bound should hold even in de Sitter.

\subsubsection*{Higher  Spin Operators in Generic CFTs}
The argument of this section does not rule out higher spin non-conserved operators in non-holographic CFTs. However, the HNEC in certain limits can be utilized to constrain interactions of higher spin operators even in generic CFTs. In particular, the limit $\rho\rightarrow 0$ in (\ref{hnec}) corresponds to the lightcone limit and in this limit, the HNEC becomes the averaged null energy condition (ANEC). The proof of the ANEC \cite{Hartman:2016lgu,Faulkner:2016mzt} implies that in the limit $\rho\rightarrow 0$, the inequality $\E(\rho)\ge 0$ must be true for any interacting CFT in $d\ge 3$. Moreover in this limit, the HNEC is equivalent to the conformal collider setup of \cite{Hofman:2008ar} which is known to yield optimal bounds.  Therefore, the same computation performed in the limit $\rho\rightarrow 0$ can be used to derive non-trivial but weaker constraints on the three-point functions $\langle X_\ell X_\ell T\rangle$ which are true for any interacting CFT in $d\ge 3$. These constraints, even though easy to obtain from our calculations of $\E(\rho)$, are rather long and complicated and we will not transcribe them here.  

\subsubsection*{Other Applications of the Regge OPE}
In this note we specialized $\mathbb{E}_{\Delta,J}$ to the case of $\Delta=d$ and $J=2$ to arrive at the HNEC operator in order to make use of the universality of the stress-tensor Regge trajectory in holographic theories. However $\mathbb{E}_{\Delta,J}$ more generally describes the contribution of any operator to the Regge OPE of identical scalar operators. It would be interesting to find the actual spectrum of these operators contributing to the Regge limit of the OPE in specific theories. It would also be worthwhile to try and understand the subleading contributions to the Regge OPE in holographic theories. Although these contributions are not universal, we expect that causality will impose constraints on these contributions as well.

We have explored the Regge limit of the OPE of two identical scalars. Generalization to other representations is straightforward as it only requires knowledge of the CFT three-point functions whose functional form is fixed by symmetry. Positivity of these generalized Regge OPE operators will likely lead to new constraints since they allow access to more general representations. Furthermore decomposition of the additional Lorentz indices under the little group will result on more constraint equations which need to be satisfied to preserve causality. 
\section{Restoring Causality}\label{sec:causal}
\subsection{Make CFT Causal Again}
In the previous section, we considered large-$N$ CFTs in $d\ge 3$ dimensions with the property that the lightest single trace operator with spin $\ell >2$ has dimension $\Delta\equiv \Deltag\gg 1$.  These holographic conditions are equivalent to the statement that in the gravity side the low energy behavior is governed by the Einstein gravity.
Moreover, $ \Deltag$ corresponds to the scale of new physics $\Lambda$ in the effective action in AdS (for example it can be the string scale $M_s$). In any sensible theory of quantum gravity it is expected that the Einstein-Hilbert action should receive higher derivative corrections which are suppressed by the scale $\Lambda$. On the CFT side, this translates into the fact that there is an infinite tower of higher spin operators with dimensions above the $\Deltag$. All of these higher spin operators must appear as exchange operators in CFT four-point functions in order to restore causality at high energies \cite{Afkhami-Jeddi:2016ntf}. Furthermore, in this paper we showed that the sparseness condition is very rigid and we are not allowed to add an additional higher spin operator $X_\ell$ with spin $\ell>2$ and $\Delta\ll \Deltag$ if causality is to be preserved. Let us consider adding an additional higher spin primary single trace operator $X_\ell$ with dimension $\Delta=\Delta_0\ll \Deltag$ (or on the gravity side a higher spin particle with mass $M_0\ll \Lambda$) and ask whether it is possible to restore causality  by adding one or more primary operators (or new particles) that cancel the causality violating contributions? In this section, we answer this question from the CFT side.

The bound obtained in the previous section from the HNEC is expected to be exact strictly in the limit  $\Deltag \rightarrow \infty$. However, it is easy to see that the same conclusion is true even when $\Deltag$ is large but finite, as long as $\Delta_0\ll \Deltag$. In this case, one might expect that the OPE coefficients are no longer exactly zero but receive corrections $C_{i,j,k}/C_{X_\ell}\sim\frac{1}{\Deltag^a}$, where $a$ is some positive number.\footnote{$C_{X_\ell}$ is the coefficient of the two-point function of $X_\ell$ and $C_{i,j,k}$ are the OPE coefficients for $\langle X_\ell X_\ell T\rangle$ (see appendix \ref{CFT_corr}).} However, this is inconsistent with the Ward identity which requires that at least some of $C_{i,j,k}/C_{X_\ell}\sim \O(1)$. Therefore, even for large but finite $\Deltag$, the operator $X_\ell$ is ruled out as long as $\Delta_0\ll \Deltag$. In addition, this also implies that if we want to add $X_\ell$, it will not be possible to save causality by changing the spectrum above $\Deltag$. Let us add extra operators at dimensions $\sim\Deltag'\ll \Deltag$ in order to restore causality. Note that if $\Deltag'\gg\Delta_0$, then contributions of these extra operators are expected to be suppressed by $\Deltag'$ and hence we can again make the above argument. Therefore, contributions of these extra operators can be significant enough to restore causality if and only if $\Deltag'\sim\Delta_0$.

The above argument also implies that perturbative $1/N$ effects are not sufficient to save causality either. Any such correction must be suppressed by positive powers of $1/N$ and hence inconsistent with the Ward identity. This is also clear from the gravity side, both in flat space and in AdS. Causality requires that the tree level higher spin-higher spin-graviton amplitude must vanish. One might expect that loop effects can generate a non-vanishing amplitude without violating causality, however, these effects must be $1/N$ suppressed. Hence, this scenario is in tension with the universality of gravitational interactions dictated by the equivalence principle. 

The behavior of four-point functions in the Regge limit makes it obvious that these extra operators at $\Deltag'$ must have spin $\ell\ge 2$ so that they can contribute significantly in the Regge limit to restore causality. Furthermore, causality imposes strong restrictions on what  higher spin operators can be added at $\Deltag'$. The simplest possibility is to add a finite or infinite set of higher spin operators at $\Deltag'$ which do not contribute as exchange operators in any four-point functions. However, this scenario makes the causality problem even worse. The causality of the Regge four point functions still leads to the HNEC and one can rule out even an infinite set of such operators by applying the HNEC to individual higher spin operators. The only other possibility is to add a set of higher spin operators at $\Deltag'$ which do contribute as exchange operators in the four-point function $\langle  X_\ell X_\ell \psi \psi\rangle$, where $\psi$ is a heavy scalar operator. In this case2, the HNEC is no longer applicable and hence the argument of the previous section breaks down. However, a finite number of higher spin primaries ($\ell>2$) that contribute as exchange operators violate chaos/causality bound \cite{Maldacena:2015waa,Afkhami-Jeddi:2016ntf} and consequently this scenario necessarily requires an infinite tower of higher spin operators.\footnote{Note that the chaos bound does not directly rule out spin-2 exchange operators. Therefore, one might expect that the causality problem may be resolved by adding a finite number of spin-2 non-conserved single trace primaries. However, it was shown in \cite{Camanho:2014apa}  that non-conserved spin-2 primaries when contribute as exchange operators lead to additional causality violation and hence we will not consider this scenario.} Therefore, the only way causality can be restored is to add an infinite tower of finely tuned higher spin primaries with $\Delta\sim \Deltag'\sim \Delta_0$. In other words, addition of a single higher spin operator with $\Delta=\Delta_0$ necessarily brings down the gap to $\Delta_0$.

Let us note that the above argument did not require that this new tower of operators contribute to the $TT$ OPE. For this reason, one might hope that it is possible to fine-tune the higher spin operators such that causality is restored and the gap is still at $\Deltag$ when considering states created by the stress tensor. However, this scenario is also not allowed as we explain next. In this case, one can still prove the HNEC starting from the Regge OPE of $TT$ when both operators are smeared appropriately (see \cite{Afkhami-Jeddi:2018own}). One can then repeat the argument of the previous section to rule out $X_\ell$, as well as the entire tower of operators at $\Deltag'$. Therefore, the only way  the tower at $\Deltag'\sim\Delta_0$ can lead to a causal CFT is if they also contribute to the $TT$ OPE. In particular, an infinite subset of all higher spin operators must appear in the OPE of the stress tensor (and all low spin operators) 
\be
TT \sim \sum_J X_J\ .
\ee

Let us end this section by summarizing in the gravity language. At the energy scale $E\ll \Lambda$, the dynamics of gravitons is completely determined by the Einstein-Hilbert action. If we wish to add even one higher spin elementary particle ($\ell>2$) with mass $M_0\ll \Lambda$, the only way for the theory to remain causal is if we also add an infinite tower of higher spin particles with mass $\sim M_0$. Causality also requires that an infinite subset of these new higher spin particles should be able to decay into two gravitons. As a result, the dynamics of graviton can now be approximated by the the Einstein-Hilbert action only in the energy scale $E\ll M_0$ and hence $M_0$ is the new cut-off even if we only consider external states created by gravitons.

\subsection{Stringy Operators above the Gap}

We concluded from both gravity and CFT arguments that finitely many higher spin fields with scaling dimensions $\Delta \ll \Delta_{gap}$ are inconsistent with causality even as external operators. We can ask how this result may be modified if we consider external operator $X$ to be  a heavy state above the gap, analogous to stringy states in classical string theory.

Let us consider the expectation value of the generalized HNEC operator (\ref{CRT})  in the Hofman-Maldacena states created by a  heavy single-trace higher spin operator with spin $l$.  Following  \cite{Costa:2017twz} we parametrize the leading Regge trajectory as
\ba
 j(\nu) =2 -\frac{1}{\Delta_{gap}^2}\left( \frac{d^2}{4}+ \nu^2\right) + \mathcal{O} \left(\frac{1}{\Delta_{gap}^4} \right)\ .
 \ea
 The external operator has the scaling dimension $\Delta_X \ge \Delta_{gap}$. Consequently, we cannot take the $\Delta_{gap}\to \infty$ limit as before. Instead we must take $\Delta_{gap}$ to be large but finite and keep track of terms that may grow in this limit.
In the Regge limit $u \to \infty$, with $1- \rho \gtrsim \frac{\log(u)}{\Delta_{gap}^2} $, we expect the leading trajectory to be nearly flat and integration over the spectral density (\ref{CRT}) to be approximated by the stress-tensor contribution at $\nu=-i \frac{d}{2}$ up to $\frac{1}{\Deltag^2}$ corrections. This limit is similar to the discussion in section 5.5 in \cite{Kulaxizi:2017ixa} for bounds on real part of phase shift for scattering in AdS. See also discussion about imaginary part of phase shift for AdS scattering in \cite{Kulaxizi:2017ixa, Costa:2017twz, Meltzer:2017rtf}.

 Therefore the operator with a positive expectation value is given by\footnote{    The second line follows from the fact that at large $\Delta_{gap}$ the saddle point is dominated by the stress-tensor. Here we have assumed that the OPE coefficients do not scale exponentially with increasing $\Delta_{gap}$ and hence will not affect the saddle-point.}
\ba
u  \langle \mathbb{E}_{\Delta(J=2),2}(\rho) \rangle_X & =  u\sum_{i=0}^{2l} \frac{t^{(i)}}{ (1-\rho)^{d-3+ i}} + \cdots\ ,
\ea
where the dots denote terms which are subleading in $\Delta_{gap}$, $t^{(i)}$'s consist of certain combination of OPE coefficients and polarization tensors. The OPE coefficients $t^{(i)}$, are analytic continuation of original OPE coefficients. We have already seen that if the OPE coefficients do not grow with $\Delta_{gap}$, the existence of the operator $X$ is inconsistent with causality. One way in which causality may be restored, is to impose the following gap dependence on the OPE coefficients between heavy operators and the exchange operator\footnote{In fact, in the case of stress-tensor exchange, Ward identities forces at least one combination of OPE coefficients to grow with $\Delta_X\sim \Delta_{gap}$.}:
\ba\label{eq:OPEX}
 \frac{t^{(i)}  }{t^{(0)}  }  \lesssim  \frac{1}{\Delta_{{gap}}^{i}}\ .
\ea
The dependence of OPE coefficients on $\Delta_{gap}$ is chosen in (\ref{eq:OPEX}) such that  higher negative powers of $1-\rho$ would be multiplied by higher powers of $\frac{1}{\Delta_{gap}}$ and consequently become more suppressed in the regime of validity of stress-tensor exchange. This means that we would not get the previous constraints by sending $\rho \to 1$ and as a result, there is no inconsistency with Ward identity or causality for higher spin operators above the gap. 

Based on our CFT arguments, (\ref{eq:OPEX}) is not fixed to be the unique choice which restores causality. However, this behaviour is very similar to how the scattering amplitude in classical string theory is consistent with causality. The high energy limit of scattering amplitudes in string theory are explored in \cite{  Amati:1993tb,Amati:1992zb,Amati:1988tn,Amati:1987uf,Amati:1987wq}. In addition, generating functions of three point and four point amplitudes for strings on the leading Regge trajectory with arbitrary spin are constructed in \cite{ Taronna:2010qq,Sagnotti:2010at}. Here we focus on a high energy limit of a two to two scattering between closed higher spin strings and tachyons in bosonic string theory. Using the results of \cite{ Taronna:2010qq,Sagnotti:2010at}, the string amplitude is given by the compact expression
\ba\label{eq:Ven}
M (s,t ) = (\text{POL})  \frac{\Gamma(- \frac{\alpha^\prime s}{4}) \Gamma(-\frac{\alpha^\prime t}{4}) \Gamma( - \frac{\alpha^\prime u}{4}) }{\Gamma(1+\frac{\alpha^\prime s}{4}) \Gamma(1+ \frac{\alpha^\prime t}{4}) \Gamma(1+ \frac{\alpha^\prime u}{4})  }\ ,
\ea
where the Mandelstam variables satisfy $s+t+u = \frac{4}{\alpha^\prime} ( l -4) $ for closed strings. Here, (POL) represents the tensor structures and polynomials of different momenta. The Gamma functions poles in the numerator  of (\ref{eq:Ven}) correspond to the exchange of infinitely many higher spin particles with even spins and the mass relation $m(J)^2 =\frac{2}{\alpha^\prime} (J-2)$.  In the Regge limit, $s \to \infty$ with $t $ held fixed, the amplitude simplifies to
\ba
M(s,t) \approx   (\text{POL}) \frac{\Gamma( - \frac{\alpha^\prime t}{4})}{\Gamma( 1+ \frac{\alpha^\prime t}{4}) } \left( - i \frac{s \alpha^\prime}{4} \right)^{-2 + \frac{\alpha^\prime t}{2}}\ .
\ea
Note that the Mandelstam variable $s$ plays the same role as $u$ in the CFT analogue.  Therefore, to make gravity the dominant force we can either take $\alpha^\prime \to 0$ which corresponds to $\Delta_{\text{gap}} \to \infty$ in the CFT, or take $ t \to 0$ which in CFT language is  the lightcone limit $\rho \to 0$.  In both cases, the polarization part, $(\text{POL})$ becomes
\ba\label{eq:limit}
\lim_{\alpha^\prime \to 0 \; \text{or}\;  t \to 0} \text{(POL)} \propto s^4  {\mathcal{E}_1}_{\mu_1 \mu_2 \cdots \mu_l} {\mathcal{E}_3}^{\mu_1 \mu_2 \cdots \mu_l}\ ,
\ea
where powers of $s$ are dictated by consistency with the gravity result in limits mentioned above. Note that the tensor structure in (\ref{eq:limit}) is  independent of the  momenta and does not change sign even if we perform the eikonal experiment in this limit. Thus, in the limit that gravity is dominant, possible causality violating structures are also vanishing and there is no problem with causality.  This happens naturally in string theory since there is only one scale $\alpha^\prime$, controlling coefficients in  tensor structures, interactions between particles and their masses. As a result, vertices or tensor structures which have higher powers of momentum $\vec{q}$ (analogous to powers of $\frac{1}{1-\rho}$ in CFT) should be accompanied with higher powers of $\sqrt{\alpha^\prime}$ (analogous to powers of $\frac{1}{\Delta_{gap}}$) on dimensional grounds. See also \cite{Camanho:2014apa,DAppollonio:2015fly} for interesting details of eikonal experiment in string theory.

\section{Cosmological Implications }\label{sec:cosmo}
The bound on higher spin particles has a natural application in inflation. The epoch of inflation is a quasi de Sitter expansion of the universe, immediately after the big bang. The primordial cosmological fluctuations produced during inflation naturally explains the observed temperature fluctuations of cosmic microwave background (CMB) and the large-scale structures of the universe. If higher spin particles were present during inflation, they would affect the behavior of primordial cosmological fluctuations. In particular, higher spin particles would produce distinct signatures on the three-point function of scalar perturbations in the squeezed limit. Hence, the bound on higher spin particles imposes rather strong constraints on these three-point functions.

Consider one or  more higher spin particles during inflation. The approximate de Sitter symmetry during inflation dictates that mass of any such particle, even before we impose our causality constraints, must satisfy the Higuchi bound \cite{Higuchi:1986py,Deser:2001us}
\be
m^2>\ell (\ell-1)H^2\ ,
\ee
where, $H$ is the Hubble rate during inflation. Particles with masses that violate the Higuchi bound correspond to non-unitary representations in de Sitter space, so the Higuchi bound is analogous to the unitarity bound in CFT.\footnote{We should note that certain discrete values of mass below the Higuchi bound are also allowed. See \cite{Baumann:2018muz} for a nice review.}  The bound on higher spin particles obtained in this paper are valid in flat and AdS spacetime. We will not attempt to derive similar bounds directly in de Sitter. Instead, we will adopt the point of view of \cite{Camanho:2014apa, Cordova:2017zej} and assume that the same bounds hold even in de Sitter spacetime. This is indeed a reasonable assumption since these bounds were obtained by studying local high energy scattering which is insensitive to the spacetime curvature.  Therefore, in de Sitter spacetime in  Einstein gravity, any additional elementary particle with spin $\ell>2$  cannot have a mass $m \lesssim \Lambda$, where $\Lambda$ is the scale of new physics in the original effective action. In any sensible low energy theory we must have $H\ll \Lambda$ and hence the causality bound is stronger than the Higuchi bound. Furthermore, the causality bound also implies that all elementary higher spin particles must belong to the principal series of unitary representation of the de Sitter isometry group.

\begin{figure}[h]
\begin{center}
\includegraphics[width=0.45\textwidth]{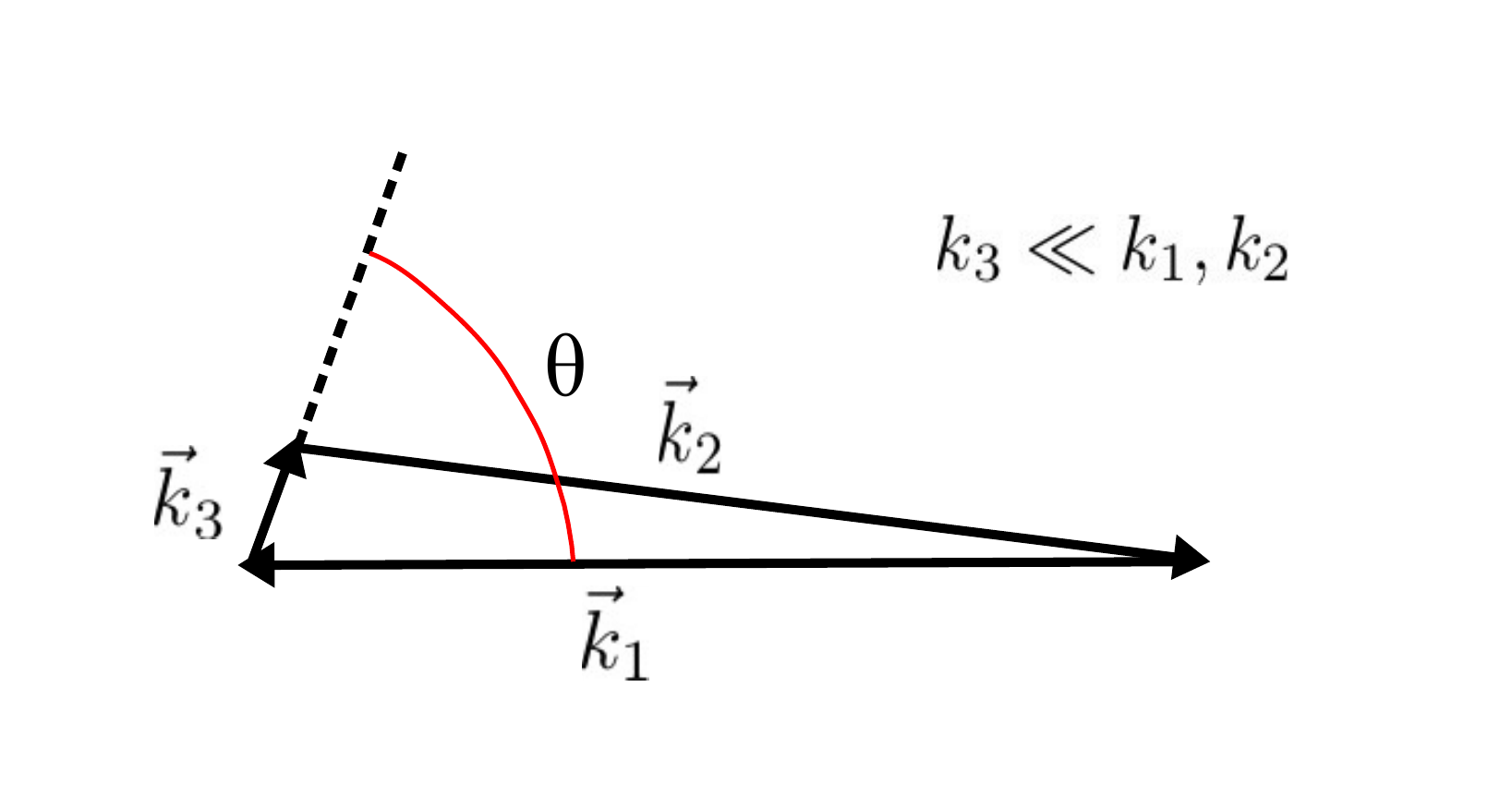}
\caption{The squeezed limit of three-point functions.}\label{ng_fig}
\end{center}
\end{figure}

Inflation naturally predicts that the scalar curvature perturbation $\zeta$ produced during inflation is nearly scale invariant and Gaussian. The momentum space three-point function of the scalar curvature perturbation $\langle \zeta(\v{k}_1)\zeta(\v{k}_2)\zeta(\v{k}_3)\rangle$ is a good measure of the deviation from exact Gaussianity. Higher spin particles affect the three-point function of scalar perturbations in a unique way. In an inflating universe, the massive higher spin particles can be spontaneously created.  It was shown in \cite{Arkani-Hamed:2015bza} that the spontaneous creation of higher spin particles produces characteristic signatures on the late time three-point function of scalar fluctuations. In particular, in the squeezed limit $k_1,k_2\gg k_3$ (see figure \ref{ng_fig}), the late time scalar three-point function admits an expansion in spin of the new particles present during inflation:\footnote{For simplicity of notation, we are omitting the Dirac delta functions.}
\be\label{AHM}
\frac{\langle \zeta(\v{k}_1)\zeta(\v{k}_2)\zeta(\v{k}_3)\rangle}{\langle \zeta(\v{k}_1)\zeta(-\v{k}_1)\rangle\langle \zeta(\v{k}_3)\zeta(-\v{k}_3)\rangle}\sim \epsilon M_{Pl}^2\sum_\ell \lambda_\ell^2\ I_\ell \left(\frac{m_\ell}{H}, \frac{k_3}{k_1}\right)P_\ell (\cos\theta)\ ,
\ee 
where $\epsilon$ is one of the slow roll parameters and $\lambda_\ell$ is the coupling between $\zeta$ and the higher spin particle with mass $m_\ell$ and spin $\ell$. $P_\ell (\cos\theta)$ is the Legendre polynomial whose index is fixed by the spin of the particle and $\theta$ is the angle between vectors $\v{k}_1$ and $\v{k}_3$. The exact form of the function $I_\ell \left(\frac{m_\ell}{H}, \frac{k_3}{k_1}\right)$ can be found in \cite{Arkani-Hamed:2015bza}. The bound  on higher spin particles from causality implies that $m_\ell\sim \Lambda \gg H$ for $\ell>2$ and hence
\be
I_\ell \left(\frac{m_\ell}{H}, \frac{k_3}{k_1}\right)\sim - \pi^2 e^{-\frac{2\pi \Lambda}{H}}\left(\frac{\Lambda}{H} \right)^{2\ell-3}\left(\frac{k_3}{k_1} \right)^{3/2} \text{Re}\left[e^{\frac{i \pi}{4}} \left(\frac{k_3}{4k_1} \right)^{i \frac{\Lambda}{H}}\right]\ .
\ee
The oscillatory behavior of the above expression is a consequence of a quantum interference effect between two different processes \cite{Arkani-Hamed:2015bza}. Moreover, the above expression also implies that contributions of higher spins to the three-point function in the squeezed limit must be exponentially suppressed. The exponential suppression can be understood as the probability for the spontaneous production of massive higher spin particles in the principal series at de Sitter temperature $T_{dS}=H/2\pi$.

Now, if  $I_\ell$ with $\ell>2$  is detected in future experiments, then the scale of new physics must be $\Lambda\sim H$. This necessarily requires the presence of not one but an infinite tower of higher spin particles with spins $\ell >2$ and masses comparable to the Hubble scale. This scenario is very similar to string theory. Any detection of $I_\ell$ with $\ell>2$ can be interpreted as evidence in favor of string theory with the string scale comparable to the Hubble scale and a very weak coupling which explains small $H/M_{pl}$. 

It is obvious from (\ref{AHM}) that the effects of higher spin particles are always suppressed by the slow roll parameter and hence not observable in the near future. The derivation of (\ref{AHM}) relied heavily on the approximate conformal invariance of the inflationary background. This approximate conformal invariance is also responsible for the slow roll suppression. However, if we allow for a large breaking of conformal invariance, the signatures of massive higher spin particles can be large enough to be detected by  future experiments. In particular, using the framework of effective field theory of inflation it was shown in \cite{Lee:2016vti} that there are interesting scenarios in which higher spin particles contribute significantly to the scalar non-Gaussanity. Furthermore, it was shown in \cite{Lee:2016vti} that higher spin particles can also produce   detectable as well as distinctive signatures on the scalar-scalar-graviton three-point function in the squeezed limit. Experimental exploration of this form of non-Gaussanity through the measurement of the $\langle B TT\rangle$ correlator of CMB anisotropies can actually be a reality in the near future \cite{Lee:2016vti}. In fact, in the most optimistic scenario, the proposed CMB Stage IV experiments \cite{Abazajian:2013oma} will be sensitive enough to detect massive higher spin particles, providing indirect evidence in favor of a theory which is very similar to low scale string theory.


\section*{Acknowledgements}
It is our pleasure to thank Tom Hartman for several helpful discussions as well as comments on a draft. We would also like to thank Nima Arkani-Hamed, Ibou Bah, Brando Bellazzini, James Bonifacio, Ted Jacobson, Marc Kamionkowski, David Kaplan, Jared Kaplan, Petr Kravchuk, David Meltzer, Joao Penedones, Eric Perlmutter, and David Simmons-Duffin  for discussions. The work of NAJ and AT is supported by Simons Foundation grant 488643. SK is supported by the Simons Collaboration Grant on the Non-Perturbative Bootstrap. We are grateful to Caltech and the Simons Bootstrap Collaboration for hospitality and support during the Bootstrap 2018 workshop where part of this work was completed.

\appendix
\section{Transverse Polarizations}\label{sec:transpol}
We construct the transverse polarization tensors used in section \ref{sec:eikonal} explicitly. These polarization tensors have only component in transverse directions $x-y$ so they can be used in $D \ge 4$. Let us define 
\ba
x^+= x+ i y\ , \qquad {x^-} =x -i y\ .
\ea
Let us consider following basis vectors
\ba
&e^+ =  \frac{1}{\sqrt{2}} (\p_x - i \p_y)\ , \qquad  e_{^-} =  \frac{1}{\sqrt{2}} (\p_x + i \p_y)\ , \nonumber\\
& {e^+}^\mu \p_{b^\mu} = \frac{1}{\sqrt{2}} \p_{b^+}\ , \qquad {e^-}^\mu \p_{b^\mu} = \frac{1}{\sqrt{2}} \p_{b^-}\ ,
\ea
where both of them are null vector.  Also we have $e^+\cdot e^- = 1$. Hence they can be used for constructing the transverse traceless polarization tensor $e^{\mu_1 \mu_2 \cdots \mu_s}$:
\ba
{e^{(+)}}^{\mu_1 \mu_2 \cdots \mu_s} = {e^{+}}^{\mu_1} {e^{+}}^{\mu_2} \cdots {e^{+}}^{\mu_s}\ , \qquad {e^{(-)}}^{\mu_1 \mu_2 \cdots \mu_s} = {e^-}^{\mu_1} {e^-}^{\mu_2} \cdots {e^-}^{\mu_s}.
\ea 
These polarization tensors are not orthogonal to each other. They can be made orthogonal by taking the following linear combinations
\ba
&{e^{\oplus}}^{\mu_1 \mu_2 \cdots \mu_s} = \frac{1}{\sqrt{2}} \left( {e^{(+)}}^{\mu_1 \mu_2 \cdots \mu_s} + {e^{(-)}}^{\mu_1 \mu_2 \cdots \mu_s} \right)\ , \nonumber \\
&{e^{\otimes}}^{\mu_1 \mu_2 \cdots \mu_s } = \frac{i}{\sqrt{2}} \left( {e^{(+)}}^{\mu_1 \mu_2 \cdots \mu_s} - {e^{(-)}}^{\mu_1 \mu_2 \cdots \mu_s} \right) \ ,
\ea
where they satisfy
\ba
&{e^{\oplus}}^{\mu_1 \mu_2 \cdots \mu_s} {e^{\otimes}}_{\mu_1 \mu_2 \cdots \mu_s}=0\ ,  \qquad {e^{\oplus}}^{\mu_1 \mu_2 \cdots \mu_s} {e^{\oplus}}_{\mu_1 \mu_2 \cdots \mu_s} =  {e^{\otimes}}^{\mu_1 \mu_2 \cdots \mu_s} {e^{\otimes}}_{\mu_1 \mu_2 \cdots \mu_s} = 1\ ,  \nonumber\\
&{e^{\oplus}}^{\mu_1 \mu_2 \cdots \mu_s} \; e^{\oplus}_{\mu_1 \mu_2 \cdots \mu_s \mu_{s+1}  \cdots \mu_{s+j} }  = \frac{1}{ \sqrt{2}} {e^{\oplus}}_{\mu_{s+1} \mu_{s+2} \cdots \mu_{s+j}}\ , \nonumber\\
& {e^{\otimes}}^{\mu_1 \mu_2 \cdots \mu_s} \; e^{\otimes}_{\mu_1 \mu_2 \cdots \mu_s \mu_{s+1}  \cdots \mu_{s+j} }  = \frac{1}{\sqrt{2}} {e^{\oplus}}_{\mu_{s+1} \mu_{s+2} \cdots \mu_{s+j}}\ ,
\ea
where $s$ and $j$ are positive numbers.
\section{Phase Shift Computations }\label{sec:laborwork}
\subsection*{A Lemma}\label{sec:trick}

In order to get the bounds in the transverse plane, we can use a trick that will be used many times in this appendix. After plugging the polarization tensors for particles, we always find the following equation
\ba
I =e^{\mu_1 \mu_2 \cdots \mu_i \mu_{i+1} \cdots \mu_J} {e_{\mu_1 \mu_2 \cdots \mu_i}}^{\nu_{i+1} \nu_{i+2} \cdots \nu_{J^\prime}} \p_{b^{\mu_{i+1}}} \cdots \p_{b^{\mu_J}} \p_{b^{\nu_{i+1}}} \cdots \p_{b^{\nu_{J^\prime}}} \frac{1}{b^{D-4}}\ .
\ea
We would like to show that sign of $I$ alternates by choosing different directions for $\vec{b}$ in the transverse plane.  \\

Let us first consider $J \neq J^\prime$, $J^\prime = J+ K$. We specify $x^+,x^-$ to be two arbitrary directions in the transverse plane and the direction of the impact parameter $\vec{b}$ is picked in the same plane spanned by $x^+, x^-$. By using $e = e^\oplus$ we find
\ba
I=& 2^{-1 + J+ K/2} \left( \p_{b^+}^K + \p_{b^-}^K  \right) (\p_{b^+} \p_{b^-})^J \frac{1 }{b^{D-4}} \nonumber \\
&=  2^{  J+K/2} (-1)^K  \left[\left(\frac{D-4}{2}\right)_J \right]^2 \left(\frac{D-4}{2}+J\right)_K  \; \frac{ \cos( K \theta)}{ b^{D-4 + 2J + K} } 
\ea
where $(a)_b \equiv \frac{\Gamma(a+b)}{\Gamma(a)}$ and $\theta$ is the angle between the vector $\vec{b}$ and the $x$-axis, where $x=\frac{1}{\sqrt{2}}( x^+ + x^-)$. This implies that rotating $\vec{b}$ with respect to x-axis changes the sign of $I$ for $K \neq 0$.\\

	If $K=0$, both $e^\oplus$ and $e^\otimes$ yield the same sign for $I$, and we need to use polarizations having components in other transverse directions, therefore the following argument could not be applied to $D=4$. For $D \ge 5$, we can separate  another transverse coordinate $z$ from $x^+,x^-$ and after taking derivative we place the impact parameter $\vec{b}$ in $x, y, z$ plane. These coordinates are enough for getting the bounds and we do not have to consider other transverse directions in for $D \ge 6$.  Again by plugging $e=e^\oplus$, we find

\ba\label{eq:hyper}
I= 2^{-1+ J}  \frac{\cos{(\theta)}^{2 J} }{b^{D-4+2 J} } \left(\frac{\Gamma(\frac{D-4}{2}+ J)}{\Gamma( \frac{D-4}{2})}\right)^2 {_2} F_1 \left(- J, - J, \frac{D-4}{2}, - \tan(\theta)^2 \right) \ ,
\ea
where $\theta$ is the angle between $\hat{z}$ and $\vec{b}$. For any integer value of $J$ and $D$, the hypergeometric function in (\ref{eq:hyper}) is a polynomial in its variable, changing sign for both even and odd $J$.

\subsection*{Diagonal Element Between  $\mathcal{E}_J$}
We set ${\mathcal{E}^{(3)}}^{\mu_1 \mu_2 \cdots \mu_J} = {z_3}_T^{\mu_1}   {z_3}_T^{\mu_1} \cdots  {z_3}_T^{\mu_J}, {\mathcal{E}^{(1)}}^{\mu_1 \mu_2 \cdots \mu_J} = {z_1}_T^{\mu_1}   {z_1}_T^{\mu_1} \cdots  {z_1}_T^{\mu_J} $ and send $ e^{\mu_1} e^{\mu_2} \cdots e^{\mu_J} \to  {e}^{\mu_1 \mu_2 \cdots \mu_J}$. We also need to impose  $e_3^{\mu_1 \mu_2 \cdots \mu_J}= (e_1^{\mu_1 \mu_2 \cdots \mu_J})^\dagger$ to have positivity. With this choice of polarization, only $\mathcal{A}_1,\cdots, \mathcal{A}_{J+1}$ contribute to phase shift and we write down the contribution of  each vertex to the phase shift. Let us define $\tilde{\delta}(s,\vec{b}) = \frac{\pi^{D/2 -2}} {\Gamma \left( \frac{D-4}{2} \right) G_N s}   \delta(s,\vec{b})$,
\ba
\left. \tilde{\delta}(s,\vec{b}) \right|_{\mathcal{A}_i} =(-1)^{(i-1)} a_i e^{\mu_1  \cdots \mu_{i-1} \mu_i \mu_{i+1} \cdots \mu_J} {e^{ \nu_1 \cdots \nu_{i-1}}}_{\mu_i \mu_{i+1} \cdots \mu_J} \p_{b^{\mu_1}} \cdots \p_{b^{\mu_{i-1}}}  \p_{b^{\nu_1}} \cdots \p_{b^{\nu_{i-1}}} \frac{1}{|b|^{D-4}}.
\ea
In the small impact parameter limit, the term with the most negative powers of $b$ dominates over other terms. As explained in the lemma \ref{sec:trick}, choosing different direction for $\vec{b}$ for $D \ge 5$ changes the sign for each of these terms. Therefore by applying the argument successively, we find
\ba
a_i = 0 \qquad   2 \le i \le J+1.
\ea
 Note  that for $a_1$, there is no derivative and hence rotating direction of $\vec{b}$ does not change the sign of this term. Choosing $ e$ to be either $e^\otimes$ or $e^\oplus$ we find for $\mathcal{A}_1$ a manifestly positive contribution
\ba
\left. \tilde{\delta}(s,\vec{b})^{\oplus} \right|_{\mathcal{A}_1} =\left. \tilde{\delta}(s,\vec{b})^{\otimes} \right|_{\mathcal{A}_1} =  \frac{a_1}{|b|^{D-4}} \ .
\ea

\subsection*{$\mathcal{E}_{J-1}$}
We again set ${\mathcal{E}^{(3)}}^{\mu_1 \mu_2 \cdots \mu_J} = {\mathcal{E}^{(1)}}^{\mu_1 \mu_2 \cdots \mu_J} =  \ep_L^{ ( \mu_1}   \ep_T^{\mu_2} \ep_T^{\mu_2} \cdots \ep_T^{\mu_J )} $. In this case all the remaining vertices contribute to the phase shift and each vertex contribution is as follows
\ba
\left. \tilde{\delta}(s,\vec{b}) \right|_{\mathcal{A}_{2J+1+K}} =& \frac{2 (-1)^{i-1}}{m^2 J^2} (a_{2J+1+K} - (J-K) a_{J+K+1})  \nonumber \\
&\times e^{\mu_1  \cdots \mu_{i} \mu_{i+1} \cdots \mu_J} {e^{ \nu_1 \cdots \nu_{i}}}_{\mu_{i+1}  \cdots \mu_J} \p_{b^{\mu_1}} \cdots \p_{b^{\mu_{i}}}  \p_{b^{\nu_1}} \cdots \p_{b^{\nu_{i}}} \frac{1}{|b|^{D-4}}\ ,
\ea
 which by taking $b$ small and using the trick discussed in \ref{sec:trick} yields 
\ba
a_{2J+1+K} = (J-K) a_{J+K+1} \qquad 2 \le K \le J-1\ .
\ea
While at the $\frac{1}{b^{D-2}}$ order, $\mathcal{A}_1$ contributes and we find
\ba
 a_{2J+2}- (J-1) a_{J+2}  = - a_{1} \frac{J(J-1)}{2} \ .
 \ea

\subsection*{Off-diagonal Components of $\mathcal{E}_J$ and $\mathcal{E}_{J-1}$}
In order to impose constraints on $\mathcal{A}_{J+2}, \mathcal{A}_{J+3}, \cdots \mathcal{A}_{2J+1}$, we use $\mathcal{E}^{(1)} = \mathcal{E}_J$, $\mathcal{E}^{(3)} = \mathcal{E}_{J-1}$. Subsequently, we find the contribution due to each of remaining vertices
\ba
\left. \tilde{\delta}(s,\vec{b}) \right|_{\mathcal{A}_{J+1+i}} = \frac{2(-1)^i}{J m} a_{J+1+i} e^{\mu_1  \cdots \mu_{i} \mu_{i+1} \cdots \mu_J} {e^{ \nu_2 \cdots \nu_{i}}}_{\mu_{i+1}  \cdots \mu_J} \p_{b^{\mu_1}} \cdots \p_{b^{\mu_{i}}}  \p_{b^{\nu_2}} \cdots \p_{b^{\nu_{i}}} \frac{1}{|b|^{D-4}}
\ea
impling that $a_{J+1+i}=0$. Using the diagonal elements in $\mathcal{E}_{J-1}$ 
 we find
\ba
&a_{J+1+i} = 0 \qquad i = 2, \cdots, J,  \\
&a_{2J+1+i}=0 \qquad i= 2, \cdots, J-1.
\ea
 However the contribution from $\mathcal{A}_1$ is given by
\ba
&\left. \tilde{\delta}(s,\vec{b}) \right|_{\mathcal{A}_{1}} = \frac{2(-i)}{m} a_1 e^{\mu_1 \mu_2 \cdots \mu_{J-1} \mu_J} e_{\mu_1 \mu_2 \cdots \mu_{J-1}} \p_{b^{\mu_J}} \frac{1}{|b|^{D-4}}\ .
\ea
Therefore, we find $ a_{J+2} =J \; a_{1}\ ,\ a_{2J+2}= \frac{J(J-1)}{2} a_1$. This proves (\ref{eq:minimal}).

\subsection*{Diagonal Elements of $\mathcal{E}_{J-2}$ }
For constraining $a_1$ we used the diagonal elements in $\mathcal{E}_{J-2}$ for both particles. Computing $C_{JJ2}$ after imposing all the other constraints, we find for $J \ge 4$
\ba\label{eq:j-2}
 \tilde{\delta}(s,\vec{b})  = a_1 \frac{3(J-2) (J-3) }{m^4 J(J-1) }\left(\frac{D+2J- 6}{D+2J -5}\right)^2   e^{\mu_1 \mu_2 \mu_3 \cdots \mu_{J-2}} {e_ {\mu_3 \cdots \mu_{J-2}}}^{ \nu_1 \nu_2} \p_{b^{\mu_{1}}}  \p_{b^{\mu_2}} \p_{b^{\nu_1}} \p_{b^{\nu_2}} \frac{1}{|b|^{D-4}}
\ea
and hence $a_1=0$ due to the trick used in \ref{sec:trick}. The equation \ref{eq:j-2} is valid for $J \ge 4 $. For $J=3$, we used interference between $\mathcal{E}^{(1)} = \mathcal{E}_0$ and $\mathcal{E}^{(3)} = \mathcal{E}_3$ to set $a_1=0$. 

\subsection*{Bounds for $D=4$}
Positivity of the phase shift (\ref{ps4d}) leads to the following constraints in $D=4$:
\begin{align}\label{bounds4}
&\bar{a}_n=0\ , \qquad n=1,\cdots,2J\ ,\nonumber\\
&\frac{a_{n+1}}{a_n}=\frac{(n-J)(n+J-1)}{n(2n-1)}\frac{1}{m^2}\ , \qquad n=1,\cdots,J\ , \nonumber\\
&\frac{a_{J+n+2}}{a_{J+n+1}}=\frac{n^2-J^2}{n(2n+1)}\frac{1}{m^2}\ , \qquad n=1,\cdots,J-1\ ,
\end{align}
with $a_{J+2}=J a_1$.

\section{Parity Violating Interactions in $D=5$}\label{odd5}
Only in $D=4$ and $5$, the massive higher spin particles can interact with gravity in a way that violates parity. We already discussed the case of $D=4$. Let us now discuss the parity odd interactions in $D=5$. Unlike $D=4$, only massive particles are allowed to couple to gravity in a way that does not preserve parity. In order to list all possible parity odd vertices for the interaction $J-J-2$, we introduce the following parity odd building block:
\begin{align}
\mathcal{B}= \epsilon^{\mu_1 \mu_2 \mu_3 \mu_4 \mu_5}z_{1; \mu_1}z_{3; \mu_2}z_{\mu_3}q_{\mu_4}p_{3;\mu_5}\ .
\end{align}
The most general form of parity odd on-shell three-point amplitude can then be constructed using this building block. In particular, we can write two distinct sets of vertices. The first set contains $J$  independent structures:
\ba\label{set1odd}
&\mathcal{A}_1^{odd}  =\mathcal{B} (z \cdot p_3) (z_1 \cdot z_3)^{J-1}\ ,\nonumber \\
&\mathcal{A}_2 ^{odd}= \mathcal{B} (z \cdot p_3) (z_1 \cdot z_3)^{J-2} (z_3 \cdot q) (z_1 \cdot q)\ , \nonumber\\
&\vdots \nonumber \\
&\mathcal{A}_{J}^{odd} =\mathcal{B}  (z \cdot p_3)  (z_3 \cdot q)^{J-1} (z_1 \cdot q)^{J-1}\ .
\ea
While the second set contains $J-1$ independent structures:
\ba\label{set2odd}
&\mathcal{A}_{J+1}^{odd} =\mathcal{B} ((z\cdot z_3)(z_1 \cdot q) - (z\cdot z_1)(z_3\cdot q)) (z_1 \cdot z_3)^{J-2}, \nonumber\\
&\mathcal{A}_{J+2}^{odd} =  \mathcal{B}  ((z\cdot z_3)(z_1 \cdot q) - (z\cdot z_1)(z_3\cdot q))  (z_1\cdot z_3)^{J-3}  (z_3 \cdot q) (z_1 \cdot q), \nonumber\\
&\vdots \nonumber\\
& \mathcal{A}_{2J-1}^{odd} = \mathcal{B}  ((z\cdot z_3)(z_1 \cdot q) - (z\cdot z_1)(z_3\cdot q)) (z_3 \cdot q)^{J-2} (z_1 \cdot q)^{J-2}\ .
\ea
The most general form of the parity violating three-point amplitude is given by
\ba
C_{JJ2} =  \sqrt{32\pi G_N} \sum_{n= 1}^{2J-1} \bar{a}_n \mathcal{A}_n^{odd}\ .
\ea 
Bounds on parity violating interactions can be obtained by using a simple null polarization vector
\be\label{pol5d}
\epsilon^\mu(p_1) =i \epsilon^\mu_{L}(p_1) -i \epsilon^\mu_{T, \hat{x}}(p_1)+ \sqrt{2}\epsilon^\mu_{T, \hat{y}}(p_1) \ , \qquad \epsilon^\mu(p_3) =-i \epsilon^\mu_{L}(p_3) +i \epsilon^\mu_{T, \hat{x}}(p_3)+\sqrt{2}\epsilon^\mu_{T, \hat{y}}(p_3)\ ,
\ee
where the transverse and longitudinal vectors are defined in (\ref{vectors}). The vectors $\hat{x}$ and $\hat{y}$ are given by $\hat{x}=(0,0,1,0,0)$ and $\hat{y}=(0,0,0,1,0)$. Positivity of the phase shift for this polarization leads to
\be
\bar{a}_n=0\ , \qquad n=1,\cdots, 2J-1\ 
\ee
for any spin $J$. Note that this bound holds even for $J=1$ and $2$.

\section{Correlators of Higher Spin Operators in CFT}\label{CFT_corr}
Let us first define the building blocks
\begin{align}\label{HV}
 H_{ij}&\equiv x_{ij}^2\e_i\cdot\e_j- 2 (x_{ij}\cdot\e_i)(x_{ij}\cdot\e_j),\qquad V_{i,jk}\equiv \frac{x_{ij}^2 x_{ik}\cdot \e_i-x_{ik}^2 x_{ij}\cdot \e_i}{x_{jk}^2}\ ,
 \end{align}
 where, $x_{ij}^\mu=(x_i-x_j)^\mu$ \qquad .

\subsection*{Two-point function}
\be\label{2ptfn}
\langle \varepsilon_1 . X_\ell(x_1) \varepsilon_2 . X_\ell(x_2) \rangle=C_{X_\ell} \frac{H_{12}^\ell}{x_{12}^{2(\Delta+\ell)}}\ ,
\ee
where, $\Delta$ is the dimension of the operator $X_\ell$ and $C_{X_\ell}$ is a positive constant. $\varepsilon_1$ and $\varepsilon_2$ are null polarization vectors contracted with the indices of $X_\ell$ in the following way
\be
 \left(\e^{\mu}\e^\nu \cdots\right)X_{\mu \nu \cdots}\equiv \e.X\ .
 \ee 
 
 \subsection*{Three-point Function}
 Let us now discuss the three-point function $\langle \varepsilon_1 . X_\ell(x_1) \varepsilon_2 . X_\ell(x_2)  \varepsilon_3.T(x_3)\rangle$:
 \begin{align}\label{general3pt}
 \langle \varepsilon_1 . X_\ell(x_1) &\varepsilon_2 . X_\ell(x_2)  \varepsilon_3.T(x_3)\rangle\nonumber\\
& =\sum_{\{n_{23},n_{13},n_{12}\}}C_{n_{23},n_{13},n_{12}}\frac{V_{1,23}^{\ell-n_{12}-n_{13}}V_{2,13}^{\ell-n_{12}-n_{23}}V_{3,12}^{2-n_{13}-n_{23}}H_{12}^{n_{12}}H_{13}^{n_{13}}H_{23}^{n_{23}}}{x_{12}^{(2h-d-2)}x_{13}^{(d+2)}x_{23}^{(d+2)}},
 \end{align} 
where $C_{n_{23},n_{13},n_{12}}$ are OPE coefficients and $h \equiv \Delta +\ell$.  In the above expression all of the polarization vectors are null, however polarizations $\e^{\mu}\e^\nu \cdots$ can be converted into an arbitrary polarization tensor $\e^{\mu \nu \cdots}$ by using projection operators from \cite{Costa:2011mg}.

The sum in (\ref{general3pt}) is over all triplets of non-negative integers $\{n_{23},n_{13},n_{12}\}$ satisfying
 \begin{align}
	\ell-n_{12}-n_{13} \geq 0\ , \qquad 	\ell-n_{12}-n_{23} \geq 0\ , \qquad 	2-n_{13}-n_{23} \geq 0\ .
 \end{align}
To begin with, there are $5+6(\ell-1)$ OPE coefficients $C_{n_{23},n_{13},n_{12}}$, however, not all of them are independent. The three-point function (\ref{general3pt}) must be symmetric with respect to the exchange $(x_1,\varepsilon_1)\leftrightarrow (x_2,\varepsilon_2)$ which implies that only $4\ell$ OPE coefficients can be independent in general. Moreover, conservation of the stress-tensor operator $T$ will impose additional restrictions on  the remaining OPE coefficients $C_{n_{23},n_{13},n_{12}}$.

\subsection*{Conservation Equation}
Relations between the OPE coefficients from conservation of the stress-tensor operator $T$ can be obtained by imposing the vanishing of $\frac{\partial}{\partial x^\mu}\langle T(x)\cdots\rangle$ up to contact terms. For $\langle X_\ell X_\ell T\rangle$, the conservation equation leads to $\ell$ additional constraint amongst the remaining $4\ell$ OPE coefficients. Therefore, the three-point function $\langle X_\ell X_\ell T\rangle$ is fixed by conformal invariance up to $3\ell$ independent OPE coefficients. Furthermore, the Ward identity leads to a relation between these OPE coefficients and the coefficient of the two-point function $C_{X_\ell}$.

 \section{Details of Spin-3 Calculation in $D>4$}\label{app3}
 \subsection*{Constraints from Conservation Equation}
 Conservation equation leads to $3$ relations among the OPE coefficients 
 \begin{align}
& C_{0,0,0}= -\frac{1}{3} \left(d^2+4 d\right) C_{0,2,0}-\frac{1}{6} \left(-d^2-4 d+12\right) C_{1,1,0}+2 C_{0,1,0}, \\
& C_{0,0,1}= -\frac{1}{2} \left(d^2+2 d\right) C_{0,2,1}-\frac{1}{4} \left(-d^2-2 d+8\right) C_{1,1,1}-\frac{3}{2} d C_{0,2,0}\nonumber\\
&~~~~~~~~~~-\frac{1}{2} (2-d) C_{1,1,0}+2 C_{0,1,1},\\
& C_{0,0,2}= -\frac{1}{2} \left(4-d^2\right) C_{1,1,2}-2 d C_{0,2,1}-\frac{1}{2} (2-d) C_{1,1,1}+2 C_{0,1,2}\ .
 \end{align}
 \subsection*{Deriving Constraints from the HNEC}
 Let us first start with $\xi=+1$. In the limit $\rho\rightarrow 1$, the leading contribution to $\E(\rho)$ goes  as $(1-\rho)^{-(d+3)}$, in particular
\be
\E_+(\rho)=  \frac{d (-4 + d^2) - 18 d (2 + d) \lambda^2 + 72 (2 + d) \lambda^4 - 48 \lambda^6 }{(1-\rho)^{d+3}}t_1 +\cdots
\ee 
up to some overall positive coefficient. $t_1$ in the above expression is a particular linear combination of all the OPE coefficients. Positivity of coefficients of each powers of $\lambda^2$ leads to the constraint
\be
t_1=0\ .
\ee
After imposing this constraint, the next leading term becomes 
\be
\E_+(\rho)=  \frac{(d-2) d-12 d \lambda ^2+24 \lambda ^4 }{(1-\rho)^{d+2}}t_2+\cdots\ ,
\ee 
where, $t_2$ is another linear combination of all the OPE coefficients. Positivity now implies 
\be
t_2=0\ .
\ee
After imposing both these constraints the next leading contribution can be written in terms of two new linear combinations $t_3$ and $t_4$ of OPE coefficients:
\be
\E_+(\rho)=  \frac{t_3-(a_3 t_3+a_4 t_4) \lambda ^2+ t_4\lambda ^4 -(b_3 t_3+b_4 t_4)\lambda^6}{(1-\rho)^{d+1}}+\cdots\ ,
\ee 
where, $a_3,a_4,b_3,b_4$ are numerical factors shown later in this appendix. The exact values of these numerical factors are not important, but note that $a_3,a_4,b_4>0$ for $d>3$. Positivity of coefficients of $\lambda^0$  and $\lambda^4$ imply that $t_3, t_4\ge 0$. Then, positivity of coefficients of $\lambda^2$ dictates that
\be
t_3=t_4=0\ .
\ee
After imposing these constraints, we get something very similar
\be
\E_+(\rho)=  \frac{t_5-(a_5 t_5+a_6 t_6) \lambda ^2+ t_6\lambda ^4 }{(1-\rho)^{d}}+\cdots\ ,
\ee 
where, $t_5$ and $t_6$ are two new linear combinations  of OPE coefficients and $a_5, a_6$ are positive numerical factors shown at the end of this appendix. Note that there is no $\lambda^6$ term in this order. However, positivity of coefficients of $\lambda^0$, $\lambda^2$  and $\lambda^4$ still produces two equalities:
\be
t_5=t_6=0\ .
\ee
Repeating the same procedure for the next order, we obtain
\be
\E_+(\rho)=  \frac{t_7-(a_7 t_7+a_8 t_8) \lambda ^2+ t_8\lambda ^4 -(b_7 t_7+b_8 t_8)\lambda^6}{(1-\rho)^{d-1}}+\cdots\ ,
\ee
where, $a$ and $b$ coefficients are shown at the end of this appendix. A similar argument in $d\ge 4$ leads to constraints 
\be
t_7=t_8=0\ .
\ee
After imposing all these constraints, finally we obtain
\be
\E_+(\rho)=\frac{t_9}{(1-\rho)^{d-2}} \left(1+\frac{4 \Delta  \lambda^2}{-d+2 \Delta -2}+\frac{4 \Delta  (\Delta +1) \lambda^4}{(d-2 \Delta ) (d-2 \Delta +2)} \right)\ ,
\ee
where, coefficients of $\lambda^0$, $\lambda^2$  and $\lambda^4$ are now all positive. Hence, the holographic null energy condition now leads to $t_9 \ge 0$. We can now choose $\xi=-1$ and calculate $\E_-(\rho)$. After imposing $t_i=0$ for $i=1,\cdots,8$, we get  
\be
\E_-(\rho)=-\frac{t_9}{(1-\rho)^{d-2}} \left(1+\frac{4 \Delta  \lambda^2}{-d+2 \Delta -2}+\frac{4 \Delta  (\Delta +1) \lambda^4}{(d-2 \Delta ) (d-2 \Delta +2)} \right)\ 
\ee
and hence $t_9 \le 0$. Therefore, combining both these inequalities, we finally get
\be
t_9=0\ .
\ee
From the definitions of $t_i$'s it is apparent that $t_1,\cdots,t_9$ are independent linear combinations of the OPE coefficients. Therefore, irrespective of their exact structures, $\{t_1,\cdots,t_9\}$ forms a complete basis in the space of OPE coefficients. As a consequence, the constraints $t_1,\cdots,t_9=0$ necessarily require that all OPE coefficients $C_{i,j,k}$ must vanish.

 \subsection*{$a$ and $b$ Coefficients}
 $a$ and $b$ coefficients are given by
 \begin{align*}
 a_3 &=\frac{2 d (13 \Delta +9)-8 (\Delta +3)}{(d-2) (d (4 \Delta +3)-2 (\Delta +2))}\ , \qquad b_3=-\frac{16 \Delta }{(d-2) d (d (4 \Delta +3)-2 (\Delta +2))}\ ,\\
a_4 &= \frac{(d-3) d (\Delta +1)}{8 d \Delta +6 d-4 \Delta -8} \ , ~~~~~~~~~~~~~~~~~~~~ b_4=\frac{4 \Delta +2}{4 d \Delta +3 d-2 \Delta -4}\ ,\\
a_5 &=\frac{6 (\Delta -1)}{(d-2) (2 \Delta -1)}\ , ~~~~~~~~~~~~~~~~~~~~~~~~ a_6=\frac{(d-3) \Delta }{2(2 \Delta -1)}\ ,\\
a_7 &=\frac{2 d^2 (2 \Delta  (\Delta +1) (\Delta +2)-3)-4 d (\Delta  (\Delta +2) (7 \Delta +1)-3)+48 (\Delta +1) \Delta ^2}{(d-2) (d-2 \Delta ) \left(2 (d-5) \Delta ^2+4 (d-1) \Delta -3 d+6\right)}\ ,\\
b_7 &=-\frac{8 \Delta ^2 (\Delta +1)}{(d-2) (d-2 \Delta ) \left(2 (d-5) \Delta ^2+4 (d-1) \Delta -3 d+6\right)}\ ,\\
a_8 &=\frac{(d-3) (\Delta -1) (d-2 (\Delta +1))}{2 (d-5) \Delta ^2+4 (d-1) \Delta -3 d+6}\ , ~~~b_8=\frac{2 \left(2 \Delta ^2+\Delta -1\right)}{2 (d-5) \Delta ^2+4 (d-1) \Delta -3 d+6}\ .
 \end{align*}
 
 \subsection*{$t$-basis in $d=4$}
 For the purpose of illustration, let us transcribe $t_1,\cdots,t_9$ for $d=4$. We will not show the general $d$ expressions because the exact structures of $t_1,\cdots,t_9$ are not important. The fact that $t_1,\cdots,t_9$ are independent linear combinations of $C_{i,j,k}$ is sufficient to rule out the existence of spin-3 operators.
 {\scriptsize 
 \begin{align*}
 t_1&= -\frac{5 \pi ^{7/2} 4^{1-\Delta } \Gamma \left(\Delta -\frac{1}{2}\right)}{\Delta  \left(\Delta ^2-1\right) \Gamma (\Delta +4)}\{ -(2 \Delta +5) ((\Delta +5) ((\Delta +5) \Delta +28) \Delta +168) C_{0,1,0}+24 (2 \Delta +5) ((\Delta +5) \Delta +10) C_{0,1,1}\\
 &+\Delta  (2 (((((\Delta +17) \Delta +119) \Delta +471) \Delta +1044) \Delta +1156) C_{0,2,0}-24 (((3 \Delta +34) \Delta +121) \Delta +170) C_{0,2,1}\\
 &-\Delta  (((((\Delta +13) \Delta +91) \Delta +379) \Delta +964) C_{1,1,0}-12 ((3 \Delta +26) \Delta +103) C_{1,1,1}+864 C_{1,1,2})-576 C_{0,1,2}\\
 &-8 (173 C_{1,1,0}-300 C_{1,1,1}+468 C_{1,1,2}))+864 C_{0,0,3}-48 (30 C_{0,1,2}-17 C_{0,2,0}+22 C_{0,2,1}+18 C_{1,1,0}-39 C_{1,1,1}+114 C_{1,1,2})\}\ ,\\
 t_2&=\frac{5 \pi ^{7/2} 2^{1-2 \Delta } (2 \Delta -3) \Gamma \left(\Delta -\frac{3}{2}\right)}{3 (\Delta -1) \Delta  \left(3 \Delta ^4+26 \Delta ^3+103 \Delta ^2+200 \Delta +156\right) \Gamma (\Delta +3)}\{-6 \Delta ^9 C_{0,2,0}+3 \Delta ^9 C_{1,1,0}-102 \Delta ^8 C_{0,2,0}\\
 &+45 \Delta ^8 C_{1,1,0}-828 \Delta ^7 C_{0,2,0}+334 \Delta ^7 C_{1,1,0}+72 \Delta ^6 C_{0,1,1}-4156 \Delta ^6 C_{0,2,0}-288 \Delta ^6 C_{0,2,1}+1562 \Delta ^6 C_{1,1,0}\\
 &+864 \Delta ^5 C_{0,1,1}-14446 \Delta ^5 C_{0,2,0}-432 \Delta ^5 C_{0,2,1}+5067 \Delta ^5 C_{1,1,0}-2592 \Delta ^5 C_{1,1,2}+4584 \Delta ^4 C_{0,1,1}-2592 \Delta ^4 C_{0,1,2}\\
 &-36662 \Delta ^4 C_{0,2,0}+9888 \Delta ^4 C_{0,2,1}+11773 \Delta ^4 C_{1,1,0}-21600 \Delta ^4 C_{1,1,2}+13632 \Delta ^3 C_{0,1,1}-18432 \Delta ^3 C_{0,1,2}\\
 &-67616 \Delta ^3 C_{0,2,0}+55920 \Delta ^3 C_{0,2,1}+19292 \Delta ^3 C_{1,1,0}-79200 \Delta ^3 C_{1,1,2}+24816 \Delta ^2 C_{0,1,1}-53856 \Delta ^2 C_{0,1,2}\\
 &-85464 \Delta ^2 C_{0,2,0}+129408 \Delta ^2 C_{0,2,1}+21108 \Delta ^2 C_{1,1,0}-156960 \Delta ^2 C_{1,1,2}+1728 \left(3 \Delta ^3+15 \Delta ^2+35 \Delta +30\right) C_{0,0,3}\\
 &+\left(3 \Delta ^8+40 \Delta ^7+236 \Delta ^6+762 \Delta ^5+1393 \Delta ^4+1190 \Delta ^3-720 \Delta ^2-3024 \Delta -1728\right) C_{0,1,0}+27072 \Delta  C_{0,1,1}\\
 &-77760 \Delta  C_{0,1,2}-67392 \Delta  C_{0,2,0}+157824 \Delta  C_{0,2,1}+13968 \Delta  C_{1,1,0}-184896 \Delta  C_{1,1,2}+12096 C_{0,1,1}-41472 C_{0,1,2}\\
 &-25920 C_{0,2,0}+86400 C_{0,2,1}+4320 C_{1,1,0}-103680 C_{1,1,2} \}\ ,
 \end{align*}
\begin{align*}
t_3&=\frac{-\pi ^{7/2} 4^{-\Delta } (2 \Delta -3) \Gamma \left(\Delta -\frac{3}{2}\right)}{\left.3 (\Delta -1) \left(3 \Delta ^9+51 \Delta ^8+414 \Delta ^7+2078 \Delta ^6+7223 \Delta ^5+18331 \Delta ^4+33808 \Delta ^3+42732 \Delta ^2+33696 \Delta +12960\right) \Gamma (\Delta +2)\right)}\\
&\times \{1728 (18 \Delta ^7+177 \Delta ^6+831 \Delta ^5+2334 \Delta ^4+4645 \Delta ^3+6783 \Delta ^2+5732 \Delta +1680) C_{0,0,3}+(3 \Delta ^{12}+42 \Delta ^{11}+219 \Delta ^{10}\\
&+206 \Delta ^9-3651 \Delta ^8-24138 \Delta ^7-81903 \Delta ^6-183990 \Delta ^5-316308 \Delta ^4-452936 \Delta ^3-445512 \Delta ^2-140544 \Delta +126144) C_{0,1,0}\\
&-2 (3 \Delta ^{12} C_{1,1,0}+42 \Delta ^{11} C_{1,1,0}+432 \Delta ^{10} C_{0,2,1}+285 \Delta ^{10} C_{1,1,0}+720 \Delta ^9 C_{0,2,1}+1448 \Delta ^9 C_{1,1,0}+5184 \Delta ^9 C_{1,1,2}\\
&-31608 \Delta ^8 C_{0,2,1}+6519 \Delta ^8 C_{1,1,0}+62208 \Delta ^8 C_{1,1,2}-264672 \Delta ^7 C_{0,2,1}+24066 \Delta ^7 C_{1,1,0}+340416 \Delta ^7 C_{1,1,2}\\
&-1033008 \Delta ^6 C_{0,2,1}+67035 \Delta ^6 C_{1,1,0}+1107648 \Delta ^6 C_{1,1,2}-2495520 \Delta ^5 C_{0,2,1}+140208 \Delta ^5 C_{1,1,0}+2474496 \Delta ^5 C_{1,1,2}\\
&-4233192 \Delta ^4 C_{0,2,1}+220446 \Delta ^4 C_{1,1,0}+4206816 \Delta ^4 C_{1,1,2}-5473296 \Delta ^3 C_{0,2,1}+264508 \Delta ^3 C_{1,1,0}+5894208 \Delta ^3 C_{1,1,2}\\
&-5511264 \Delta ^2 C_{0,2,1}+234480 \Delta ^2 C_{1,1,0}+6862752 \Delta ^2 C_{1,1,2}+432 (15 \Delta ^8+172 \Delta ^7+888 \Delta ^6+2690 \Delta ^5+5447 \Delta ^4\\
&+8078 \Delta ^3+7918 \Delta ^2+3304 \Delta -624) C_{0,1,2}-12 (9 \Delta ^{10}+141 \Delta ^9+1011 \Delta ^8+4350 \Delta ^7+12601 \Delta ^6+26427 \Delta ^5+43243 \Delta ^4\\
&+58314 \Delta ^3+53728 \Delta ^2+15312 \Delta -14112) C_{0,1,1}-3664512 \Delta  C_{0,2,1}+129600 \Delta  C_{1,1,0}+5173632 \Delta  C_{1,1,2}-967680 C_{0,2,1}\\
&+34560 C_{1,1,0}+1347840 C_{1,1,2}) \}\ ,\\
t_4&=\frac{\pi ^{7/2} 4^{-\Delta } (2 \Delta -3) \Gamma \left(\Delta -\frac{3}{2}\right)}{3 (\Delta -1) \Delta  \left(3 \Delta ^9+51 \Delta ^8+414 \Delta ^7+2078 \Delta ^6+7223 \Delta ^5+18331 \Delta ^4+33808 \Delta ^3+42732 \Delta ^2+33696 \Delta +12960\right) \Gamma (\Delta +2)}\\
&\times \{-1728 \Delta  \left(9 \Delta ^7+105 \Delta ^6+550 \Delta ^5+1797 \Delta ^4+4019 \Delta ^3+5976 \Delta ^2+4704 \Delta +1152\right) C_{0,0,3}\\
&-2 (-144 (33 \Delta ^9+482 \Delta ^8+3220 \Delta ^7+13428 \Delta ^6+39443 \Delta ^5+84574 \Delta ^4+129300 \Delta ^3+133632 \Delta ^2+88992 \Delta +31104) C_{0,1,2}\\
&+12 (15 \Delta ^{11}+300 \Delta ^{10}+2836 \Delta ^9+16806 \Delta ^8+70033 \Delta ^7+217146 \Delta ^6+511924 \Delta ^5+913140 \Delta ^4+1197048 \Delta ^3+1090080 \Delta ^2+634176 \Delta \\
&+186624) C_{0,1,1}+\Delta (-288(9 \Delta ^9+135 \Delta ^8+902 \Delta ^7+3736 \Delta ^6+10842 \Delta ^5+22703 \Delta ^4+33325 \Delta ^3+32796 \Delta ^2+17496 \Delta +1728) C_{1,1,2}\\
&-24 (30 \Delta ^{10}+411 \Delta ^9+2444 \Delta ^8+8520 \Delta ^7+19136 \Delta ^6+25089 \Delta ^5+1406 \Delta ^4-65772 \Delta ^3-129792 \Delta ^2-107712 \Delta -27648) C_{0,2,1}\\
&+(9 \Delta ^{12}+166 \Delta ^{11}+1543 \Delta ^{10}+9146 \Delta ^9+38267 \Delta ^8+119030 \Delta ^7+280469 \Delta ^6+495754 \Delta ^5+634144 \Delta ^4+536256 \Delta ^3\\
&+238752 \Delta ^2-41472) C_{1,1,0}))+(9 \Delta ^{13}+208 \Delta ^{12}+2517 \Delta ^{11}+20148 \Delta ^{10}+116751 \Delta ^9+511632 \Delta ^8+1737543 \Delta ^7\\
&+4628948 \Delta ^6+9669660 \Delta ^5+15584136 \Delta ^4+18714816 \Delta ^3+15761088 \Delta ^2+8439552 \Delta +2239488) C_{0,1,0} \}\ ,\\
t_5&=\frac{\pi ^{7/2} 4^{-\Delta } (2 \Delta -3) \Gamma \left(\Delta -\frac{3}{2}\right)}{3 (9 \Delta ^6+87 \Delta ^5+370 \Delta ^4+951 \Delta ^3+1667 \Delta ^2+1980 \Delta +1008) \Gamma (\Delta +1)}\{(15 \Delta ^8+125 \Delta ^7+636 \Delta ^6+2162 \Delta ^5+5397 \Delta ^4\\
&+9413 \Delta ^3+12150 \Delta ^2+10062 \Delta +4212) 
C_{0,1,0}-2 (15 \Delta ^8 C_{1,1,0}+80 \Delta ^7 C_{1,1,0}+258 \Delta ^6 C_{1,1,0}+583 \Delta ^5 C_{1,1,0}\\
&+2592 \Delta ^5 C_{1,1,2}+1130 \Delta ^4 C_{1,1,0}+9936 \Delta ^4 C_{1,1,2}+1317 \Delta ^3 C_{1,1,0}+12096 \Delta ^3 C_{1,1,2}+1333 \Delta ^2 C_{1,1,0}-6480 \Delta ^2 C_{1,1,2}\\
&+6 (9 \Delta ^6+84 \Delta ^5+400 \Delta ^4+988 \Delta ^3+1387 \Delta ^2+1036 \Delta +384) C_{0,1,1}-12 (18 \Delta ^6+285 \Delta ^5+1136 \Delta ^4+1817 \Delta ^3+752 \Delta ^2\\
&-400 \Delta -168) C_{0,2,1}+252 \Delta  C_{1,1,0}-6912 \Delta  C_{1,1,2}-720 C_{1,1,0}+8208 C_{1,1,2}) \}\ ,\\
t_6&=\frac{-\pi ^{7/2} 2^{1-2 \Delta } (\Delta +1) (2 \Delta -3) \Gamma \left(\Delta -\frac{3}{2}\right)}{(\Delta -1) \left(9 \Delta ^6+87 \Delta ^5+370 \Delta ^4+951 \Delta ^3+1667 \Delta ^2+1980 \Delta +1008\right) \Gamma (\Delta +1)}\{(3 \Delta ^8+28 \Delta ^7+160 \Delta ^6+603 \Delta ^5\\
&+1622 \Delta ^4+3005 \Delta ^3+4191 \Delta ^2+3564 \Delta +1296) C_{0,1,0}-2 (3 \Delta ^8 C_{1,1,0}+19 \Delta ^7 C_{1,1,0}+73 \Delta ^6 C_{1,1,0}+173 \Delta ^5 C_{1,1,0}\\
&+327 \Delta ^4 C_{1,1,0}-864 \Delta ^4 C_{1,1,2}+354 \Delta ^3 C_{1,1,0}-2016 \Delta ^3 C_{1,1,2}+263 \Delta ^2 C_{1,1,0}-4320 \Delta ^2 C_{1,1,2}-12 (6 \Delta ^4+47 \Delta ^3\\
&+104 \Delta ^2+131 \Delta +72) \Delta ^2 C_{0,2,1}+6 (3 \Delta ^6+28 \Delta ^5+112 \Delta ^4+244 \Delta ^3+393 \Delta ^2+372 \Delta +144) C_{0,1,1}\\
&+12 \Delta  C_{1,1,0}-576 \Delta  C_{1,1,2}-144 C_{1,1,0}+3456 C_{1,1,2})\}\ ,\\
t_7&=\frac{\pi ^{7/2} 4^{-\Delta -1} \left(2 \Delta ^2-7 \Delta +6\right) \Gamma \left(\Delta -\frac{3}{2}\right)}{9 \left(3 \Delta ^6+8 \Delta ^5+16 \Delta ^4+15 \Delta ^3+11 \Delta ^2+\Delta -9\right) \Gamma (\Delta )}\{\left(15 \Delta ^6+64 \Delta ^5-4 \Delta ^4-130 \Delta ^3+244 \Delta ^2+270 \Delta +243\right) C_{0,1,0}\\
&-12 \left(\left(12 \Delta ^5+9 \Delta ^4-31 \Delta ^3+13 \Delta ^2+34 \Delta +24\right) C_{0,1,1}+2 \left(-24 \Delta ^5+27 \Delta ^4+47 \Delta ^3-38 \Delta ^2-17 \Delta +15\right) C_{0,2,1}\right)\}\ ,\\
t_8&=\frac{-\pi ^{7/2} 4^{-\Delta -1} \Delta  (2 \Delta -3) \Gamma \left(\Delta -\frac{3}{2}\right)}{3 (\Delta -1) \left(3 \Delta ^6+8 \Delta ^5+16 \Delta ^4+15 \Delta ^3+11 \Delta ^2+\Delta -9\right) \Gamma (\Delta )}\\
&\times\{\left(3 \Delta ^7+9 \Delta ^6-8 \Delta ^5-62 \Delta ^4-30 \Delta ^3-190 \Delta ^2-163 \Delta -207\right) C_{0,1,0}\\
&-12 \left(\left(2 \Delta ^6+\Delta ^5-8 \Delta ^4+2 \Delta ^3-13 \Delta ^2-16 \Delta -22\right) C_{0,1,1}+ 2 \left(-4 \Delta ^6+7 \Delta ^5+4 \Delta ^4-13 \Delta ^3+5 \Delta ^2+8 \Delta -7\right) C_{0,2,1}\right)\}\ ,
\end{align*}
\begin{align*}
t_9&=\frac{\pi ^{7/2} 4^{-2 \Delta -3} C_{0,1,0}}{63 \left(\Delta ^2-\Delta -1\right) \Gamma (\Delta +4)^2}
\{24 \sqrt{\pi } (\Delta +2) (\Delta +3)(16 \Delta ^{12}+112 \Delta ^{11}+802 \Delta ^{10}+2041 \Delta ^9-3583 \Delta ^8-27783 \Delta ^7\\
&-97848 \Delta ^6-361565 \Delta ^5-1046943 \Delta ^4-1943909 \Delta ^3-2130484 \Delta ^2-1182496 \Delta -72840) \Gamma (2 \Delta -2)\\
&-\frac{4^{\Delta }\Gamma \left(\Delta -\frac{1}{2}\right) \Gamma (\Delta +4)}{\Delta  \left(\Delta ^2-1\right)} (48 \Delta ^{12}+560 \Delta ^{11}+2182 \Delta ^{10}+2763 \Delta ^9-7389 \Delta ^8-69237 \Delta ^7-307656 \Delta ^6\\
&-1103735 \Delta ^5-3121789 \Delta ^4-5823663 \Delta ^3-6399516 \Delta ^2-3547488 \Delta -218520) \}\ .
 \end{align*}
 }
 
 \section{Details of Spin-4 Calculation in $D>4$}\label{app4}
 \subsection*{Constraints From Conservation Equation}
 Conservation equation leads to $4$ relations among the OPE coefficients of $\langle X_{\ell=4}X_{\ell=4}T\rangle$:
 \begin{align*}
& \tilde{C}_{0,0,0}= \frac{1}{8} \left((d-2) (d+8) \tilde{C}_{1,1,0}-2 d (d+6) \tilde{C}_{0,2,0}\right)+2 \tilde{C}_{0,1,0}\ , \\
& \tilde{C}_{0,0,1}= \frac{1}{6} \left(-8 d \tilde{C}_{0,2,0}-2 d (d+4) \tilde{C}_{0,2,1}+(d-2) \left((d+6) \tilde{C}_{1,1,1}+3 \tilde{C}_{1,1,0}\right)+12 \tilde{C}_{0,1,1}\right)\ ,\\
& \tilde{C}_{0,0,2}= \frac{1}{4} \left(-6 d \tilde{C}_{0,2,1}-2 d (d+2) \tilde{C}_{0,2,2}+(d-2) \left((d+4) \tilde{C}_{1,1,2}+2 \tilde{C}_{1,1,1}\right)+8 \tilde{C}_{0,1,2}\right)\ ,\\
& \tilde{C}_{0,0,3}= \frac{1}{2} \left(-4 d \tilde{C}_{0,2,2}+(d-2) \left((d+2) \tilde{C}_{1,1,3}+\tilde{C}_{1,1,2}\right)+4 \tilde{C}_{0,1,3}\right)\ .
 \end{align*}
 
  \subsection*{Deriving Constraints from the HNEC}
 The full expression for $\E(\rho)$ is long and not very illuminating, so we will not transcribe it here. Instead we introduce a new basis $\{\t_1,\cdots,\t_{12}\}$  in the space of OPE coefficients $\tilde{C}_{i,j,k}$ and use this new basis to derive constraints. The exact structures of $\t_1,\cdots,\t_{12}$ are not important because  the fact that $\t_1,\cdots,\t_{12}$ are independent linear combinations of $\tilde{C}_{i,j,k}$ is sufficient to rule out the existence of spin-4 operators.

We again start with $\xi=+1$, however, for spin-4, this will be sufficient to rule them out completely. In the limit $\rho\rightarrow 1$, the leading contribution to $\E(\rho)$ goes  as $(1-\rho)^{-(d+5)}$, in particular
\begin{align}
\E_+(\rho)&=\frac{\t_1}{(d-2) d (d+2) (d+4)(1-\rho)^{(d+5)}}\left((d-2) d (d+2) (d+4)\right.\nonumber \\
&\left. -32 d (d+2) (d+4)\lambda ^2+288 (d+2) (d+4) \lambda ^4-768 (d+4) \lambda ^6+384 \lambda ^8  \right)   +\cdots\ .
\end{align}
Positivity of coefficients of each powers of $\lambda^2$ leads to the constraint
\be
\t_1=0\ .
\ee
After imposing this constraint, the next leading term becomes 
\be
\E_+(\rho)=  \frac{\t_2 }{(1-\rho)^{d+4}}\left(1-\frac{24 \lambda ^2}{d-2}+\frac{144 \lambda ^4}{(d-2) d}-\frac{192 \lambda ^6}{d(d^2-4)} \right)+\cdots\ ,
\ee 
where, positivity now implies 
\be
\t_2=0\ .
\ee
After imposing both these constraints the next leading contribution behaves similar to the spin-3 case:
\be
\E_+(\rho)=  \frac{\t_3-(\a_3 \t_3+\a_4 \t_4) \lambda ^2+ \t_4\lambda ^4 +(\b_3 \t_3+\b_4 \t_4)\lambda^6+(\c_3 \t_3+\c_4 \t_4)\lambda^8}{(1-\rho)^{d+3}}+\cdots\ ,
\ee 
where, $\a_3,\a_4,\b_3,\b_4,\c_3,\c_4$ are numerical factors given later in this appendix. Note that $\a_3,\a_4>0$ and hence positivity of coefficients of $\lambda^0, \lambda^2$  and $\lambda^4$ imply that 
\be
\t_3=\t_4=0\ .
\ee
The next order contribution has an identical structure:
 \be
\E_+(\rho)=  \frac{\t_5-(\a_5 \t_5+\a_6 \t_6) \lambda ^2+ \t_6\lambda ^4 +(\b_5 \t_5+\b_6 \t_6)\lambda^6}{(1-\rho)^{d+2}}+\cdots\ ,
\ee 
with $\a_5,\a_6>0$, implying 
\be
\t_5=\t_6=0\ .
\ee
So far, everything is very similar to the spin-3 case. But the next order contribution is somewhat different. In the next order, there are three independent structures
\be
\E_+(\rho)=\frac{\t_7-(\a_7 \t_7+\a_8 \t_8+\a_9 \t_9)\lambda^2+\t_8 \lambda^4+\t_9 \lambda^6+(\b_7 \t_7+\b_8 \t_8+\b_9 \t_9)\lambda^8}{(1-\rho)^{d+1}}+\cdots\ ,
\ee 
where, $\a_7,\a_8,\a_9>0$. Positivity now leads to three constraints 
\be
\t_7=\t_8=\t_9=0\ .
\ee
However, after imposing these constraints, in the next order we get only two new structures mainly because a lot of contributions vanish after imposing the previous constraints. In particular, we obtain
\be
\E_+(\rho)=  \frac{\t_{10}-(\a_{10} \t_{10}+\a_{11} \t_{11}) \lambda ^2+ \t_{11}\lambda ^4 -(\b_{10} \t_{10}+\b_{11} \t_{11})\lambda^6}{(1-\rho)^{d}}+\cdots\ 
\ee 
with either $\a_{10},\a_{11}>0$ or $\b_{10},\b_{11}>0$ which again implies
\be
\t_{10}=\t_{11}=0\ .
\ee
Finally, in the next order we get
\be
\E_+(\rho)=\frac{\t_{12}}{(1-\rho)^{d-1}}\left(1+\a_{12}\lambda^2+\b_{12}\lambda^4-\c_{12}\lambda^6\right)+\cdots\ ,
\ee
where, $\a_{12},\b_{12},\c_{12}>0$ as shown later in this appendix. Note that unlike the spin-3 case, signs of coefficients of different powers of $\lambda^2$ switch sign. Therefore, we can conclude that 
\be
\t_{12}=0\ .
\ee
$\{\t_1,\cdots,\t_{12}\}$ forms a complete basis in the space of OPE coefficients and hence the constraints $\t_1,\cdots,\t_{12}=0$ necessarily require that all OPE coefficients $\tilde{C}_{i,j,k}$ must vanish implying 
\be
\langle X_{\ell=4}X_{\ell=4}T\rangle=0\ .
\ee

 \subsection*{$\a$, $\b$ and $\c$ Coefficients}
$\a$, $\b$ and $\c$ coefficients are given by
 \begin{align*}
 \a_3&=\frac{2 (d (41 \Delta +73)-4 (\Delta +11))}{(d-2) (d (6 \Delta +11)-4 (\Delta +3))}\ ,\qquad \a_4=\frac{(d-3) d (\Delta +2)}{3 (d (6 \Delta +11)-4 (\Delta +3))}\ ,\\
 \b_3&=\frac{48 (d (27 \Delta +43)+52 \Delta +60)}{d \left(d^2-4\right) (d (6 \Delta +11)-4 (\Delta +3))} \ ,\quad \b_4=-\frac{8 (d (3 \Delta +5)+5 \Delta +6)}{(d+2) (d (6 \Delta +11)-4 (\Delta +3))}\ ,\\
 \c_3&=-\frac{192 (5 \Delta +7)}{d \left(d^2-4\right) (d (6 \Delta +11)-4 (\Delta +3))}\ , \quad \c_4=\frac{8 (2 \Delta +3)}{(d+2) (d (6 \Delta +11)-4 (\Delta +3))}\ ,\\
 \a_5&=\frac{d (38 \Delta +24)-4 (\Delta +8)}{(d-2) (d (4 \Delta +3)-2 (\Delta +2))}\ , \qquad \a_6=\frac{(d-3) d (\Delta +1)}{3 (d (4 \Delta +3)-2 (\Delta +2))}\ ,\\
 \b_5&=\frac{144 \Delta }{(d-2) d (d (4 \Delta +3)-2 (\Delta +2))}\ , \qquad \b_6=\frac{8 \Delta +4}{-4 d \Delta -3 d+2 \Delta +4} \ ,\\
 \a_7&=\frac{12 (3 d+1) (\Delta -1) (2 \Delta +1)}{(d-2) \left(d \left(13 \Delta ^2-3\right)+\Delta ^2-1\right)}\ , \b_7=\frac{24 \Delta  \left(\Delta ^2-1\right)}{d \left(d^2-3 d+2\right) (\Delta +2) \left(d \left(13 \Delta ^2-3\right)+\Delta ^2-1\right)}\ ,\\
 \a_8&=\frac{(d-3) d \Delta  (2 \Delta +1)}{d \left(13 \Delta ^2-3\right)+\Delta ^2-1}\ , \qquad \b_8=\frac{2 \left(-4 \Delta ^3-4 \Delta ^2+\Delta +1\right)}{(d-1) (\Delta +2) \left(d \left(13 \Delta ^2-3\right)+\Delta ^2-1\right)}\ ,\\
 \a_9&=\frac{(d-3)^2 d \Delta  (\Delta +1)}{4 \left(d \left(13 \Delta ^2-3\right)+\Delta ^2-1\right)}\ , \quad \b_9=-\frac{(2 \Delta +1) (d (7 \Delta  (\Delta +1)-5)-\Delta  (\Delta +1))}{(d-1) (\Delta +2) \left(d \left(13 \Delta ^2-3\right)+\Delta ^2-1\right)}\ ,\\
 \a_{10}&=-\frac{2 \left(-((d-9) d+26) \Delta ^3-6 ((d-7) d+4) \Delta ^2-(d-1) (11 d+2) \Delta +6 (d-2) (d+1)\right)}{(d-2) (d-2 \Delta ) (d (\Delta  (\Delta +4)-3)-\Delta  (7 \Delta +4)+6)} \ , \\
 \b_{10}&=-\frac{24 \Delta  (\Delta +1) (\Delta +2)}{(d-2) (d-2 \Delta ) \left(d \left(\Delta ^2+4 \Delta -3\right)-7 \Delta ^2-4 \Delta +6\right)}\ ,\\
 \a_{11}&=\frac{(d-3) (\Delta -1) (d-2 (\Delta +1))}{2 (d (\Delta  (\Delta +4)-3)-\Delta  (7 \Delta +4)+6)} \ ,\quad \b_{11}=\frac{2 (\Delta +2) (2 \Delta -1)}{d \left(\Delta ^2+4 \Delta -3\right)-7 \Delta ^2-4 \Delta +6} \ ,\\
 \a_{12}&=\frac{2 \left(2 (5-2 d) \Delta ^2-3 d \Delta +d+2\right)}{(d-2) (2 \Delta -1) (d-2 \Delta +2)}\ , \quad b_{12}=\frac{4 \Delta  (d (\Delta +3) (2 \Delta +1)-2 (\Delta  (4 \Delta +5)+3))}{(d-2) (2 \Delta -1) (d-2 \Delta ) (d-2 \Delta +2)}\ ,\\
 \c_{12}&=\frac{8 \Delta  (\Delta +1)^2}{(d-2) (2 \Delta -1) (d-2 \Delta ) (d-2 \Delta +2)}\ .
 \end{align*}

\section{Details of CFT$_3$ calculations}\label{3d}
In this appendix, we discuss the details of the the parity even structures for spin 3 operators in $d=3$. The full expression for $\E(\rho)$ is rather long and not very illuminating, so we will not transcribe it here. Following the logic of the higher $d$ case, we introduce a new basis $\{t_1,\cdots,t_7\}$  in the space of OPE coefficients $C_{i,j,k}$ and use this new basis to derive constraints. 

In the limit $\rho\rightarrow 1$, the leading parity even contribution to $\E(\rho)$ goes  as $(1-\rho)^{-6}$, in particular
\be
\E(\rho)=  \frac{ j_6(\epsilon^{\mu_1\mu_2\mu_3})}{(1-\rho)^{6}} t_1+\cdots\ ,
\ee
where, $j_6(\epsilon^{\mu_1\mu_2\mu_3})$ is a specific function of the traceless symmetric polarization tensor. $j_6(\epsilon^{\mu_1\mu_2\mu_3})$ has the property that 
\begin{align}
&j_6(\epsilon^{\mu_1\mu_2\mu_3})\sim \epsilon^{000}{\epsilon^{000}}^* \ge 0 \qquad \text{for} \qquad  \epsilon^{\mu_1\mu_2 2}=0\ , \nonumber\\
&j_6(\epsilon^{\mu_1\mu_2\mu_3})\sim -\epsilon^{222}{\epsilon^{222}}^* \le 0 \qquad \text{for} \qquad  \epsilon^{\mu_1\mu_2 0}=0\ .
\end{align}
Therefore, the HNEC implies that
\be
t_1=0\ .
\ee
After imposing this constraint, the next leading term becomes 
\be
\E(\rho)=  \frac{ j_5(\epsilon^{\mu_1\mu_2\mu_3})}{(1-\rho)^{5}} t_2+\cdots\ ,
\ee
where, $j_5(\epsilon^{\mu_1\mu_2\mu_3})$  is another function which has the property that
 \begin{align}
&j_5(\epsilon^{\mu_1\mu_2\mu_3})\sim \text{Re}\left[\epsilon^{000}\left(\epsilon^{001}+\epsilon^{010}+\epsilon^{100} \right)^*\right] \qquad \text{for} \qquad  \epsilon^{\mu_1\mu_2 2}=0\ 
\end{align}
which changes sign as  $\epsilon^{001}\rightarrow -\epsilon^{001}$ implying 
\be
t_2=0\ .
\ee
The next order term has two structures:
\be
\E(\rho)=  \frac{ j_4(\epsilon^{\mu_1\mu_2\mu_3})}{(1-\rho)^{4}} t_3+ \frac{ \tilde{j}_4(\epsilon^{\mu_1\mu_2\mu_3})}{(1-\rho)^{4}} t_4+\cdots\ ,
\ee
where, $j_4$ and $\tilde{j}_4$ are specific functions of the polarization tensors. Now, applying the HNEC for the following set of polarizations:
\begin{align}
&(a)\ \epsilon^{000}=\epsilon^{011}=\epsilon^{101}=\epsilon^{110}=1\ ,\nonumber\\
&(b)\ \epsilon^{012}=1\ ,\nonumber\\
&(c)\ \epsilon^{222}=-\epsilon^{211}=-\epsilon^{121}=-\epsilon^{112}=1\ ,\nonumber\\
&(d)\ \epsilon^{000}=\epsilon^{220}=\epsilon^{202}=\epsilon^{022}=1\ 
\end{align}
we find that both $t_3$ and $t_4$ must vanish. After imposing these constraints, the next order term also has two structures:
\be
\E(\rho)=  \frac{ j_3(\epsilon^{\mu_1\mu_2\mu_3})}{(1-\rho)^{3}} t_5+ \frac{ \tilde{j}_3(\epsilon^{\mu_1\mu_2\mu_3})}{(1-\rho)^{3}} t_6+\cdots\ ,
\ee
where, again we will not transcribe $j_5$ and $\tilde{j}_5$ for simplicity. Now, applying the HNEC for the following set of polarizations:
\begin{align}
&(a)\ \epsilon^{2\mu \nu}=0\ ,\nonumber\\
&(b)\ \epsilon^{012}=\pm 1\ , \qquad \epsilon^{222}=-\epsilon^{211}=-\epsilon^{121}=-\epsilon^{112}=1\ 
\end{align}
we get
\be
t_5=t_6=0\ .
\ee
After imposing all these constraints, we finally obtain
\be
\E(\rho)=  \frac{ j_2(\epsilon^{\mu_1\mu_2\mu_3})}{(1-\rho)^{2}} t_7+\cdots\ .
\ee
We repeat the same procedure by choosing $(a)\ \epsilon^{0\mu \nu}=0$ and $(b)\ \epsilon^{2\mu\nu}=0$ that lead to the final constraint 
\be
t_7=0\ .
\ee
Since, $\{t_1,\cdots,t_7\}$ forms a complete basis in the space of OPE coefficients, the constraints $t_1,\cdots,t_7=0$ necessarily require that all OPE coefficients $C_{i,j,k}$ must vanish. It is interesting to note that the same set of constraints can also be obtained by using the $\lambda$-trick.  We can first impose $C_{1,1,k}=0$ and then use the polarization (\ref{polarization}) to derive constraints in general dimension $d$. Then taking the limit $d\rightarrow 3$ leads to the correct set of constraints at each order.

\end{spacing}

\end{document}